\documentclass[11pt]{article}
\usepackage{}
\usepackage{amssymb}
\usepackage{amsfonts}
\usepackage{mathrsfs}
\usepackage{graphicx}
\usepackage{amsmath}
\usepackage{color}
\usepackage{amsthm}

\usepackage{pifont,bm}
\usepackage{tikz}
\usepackage{enumerate}
\theoremstyle{definition}

\makeatletter
\def\@biblabel#1{[#1]}
\makeatother

\makeatletter \@addtoreset{equation}{section}

\begin{document}
%\begin{CJK*}{GBK}{song}

\begin{titlepage}
\title{\bf{Inverse scattering transform of the general coupled Hirota system with nonzero boundary conditions
\footnote{This work is supported by the National Natural Science Foundation of China under Grant Nos.12201622 and 11975306.\protect\\
\hspace*{3ex}$^{*}$Corresponding authors.\protect\\
\hspace*{3ex} E-mail addresses: xbwang@cumt.edu.cn, xiubinwang@163.com (X. B. Wang) and sftian@cumt.edu.cn (S.F. Tian)}
}}
\author{Xiu-Bin Wang, Shou-Fu Tian$^{*}$\\
%%%%%%%%%%%%%%%%%%%%%%%%%%%%%%%%%%%%%%%%%%%%%%%%%%%%%%%%%%%%%%%%%%%%%%%%%%%%%%%%%%%%%%%%%
%%%%%              以下两行为作者单位
%%%%%%%%%%%%%%%%%%%%%%%%%%%%%%%%%%%%%%%%%%%%%%%%%%%%%%%%%%%%%%%%%%%%%%%%%%%%%%%%%%%%%%%%%
\small \emph{School of Mathematics, China University of Mining and Technology, Xuzhou 221116, } \\
\small \emph{People's Republic of China}\\
\date{}}
\thispagestyle{empty}
\end{titlepage}
\maketitle

\vspace{-0.5cm}
\begin{center}
\rule{15cm}{1pt}\vspace{0.3cm}

\parbox{15cm}{\small
{\bf Abstract}\\
\hspace{0.5cm} The initial value problem for the general coupled Hirota system with nonzero
boundary conditions at infinity is solved by reporting a rigorous theory of the inverse scattering transform.
With the help of a suitable uniformization variable,
both the inverse and the direct problems are analyzed which allows us to develop the inverse scattering transform on the complex $z$-plane.
Firstly, analyticity of the scattering eigenfunctions and scattering
data, properties of the discrete spectrum, symmetries, and asymptotics are discussed in detail.
Moreover,
the inverse problem is posed as a Riemann-Hilbert problem for the eigenfunctions, and the reconstruction formula of the
potential in terms of eigenfunctions and scattering data is presented.
Finally, the main characteristics of these obtained soliton solutions are graphically discussed
in the $2\times2$ self-focusing case. This family of solutions contains novel Akhmediev breather and Kuznetsov-Ma soliton.
These results would be of much importance in understanding and
enriching breather wave phenomena arising in nonlinear and complex systems, especially in Bose-Einstein condensates.
}

\vspace{0.5cm}
\parbox{15cm}{\small{

\vspace{0.3cm} \emph{Key words:} Riemann-Hilbert problem; Solitons; Akhmediev breather; Kuznetsov-Ma soliton; Inverse scattering transform.

 } }
\end{center}
\vspace{0.3cm} \rule{15cm}{1pt} \vspace{0.2cm}

\section{Introduction}
The solutions of nonlinear wave systems with nonzero boundary conditions (NZBCs) are always physically and mathematically interesting subjects.
The inverse scattering transform (IST) was first proposed by Gardner, Greene, Kruskal, and Miura
to exactly analyze the initial-value problems for the famous Korteweg-de Vries (KdV) equation with a Lax pair in 1967 \cite{GCS-1967}.
After that, a mass of attempts were presented
to develop the application of this approach in other integrable nonlinear wave systems with the so-called Lax pairs \cite{Lax-1968}.
For instance,
Zakharov and Shabat investigated the IST of the standard nonlinear Schr\"{o}dinger (NLS) equation in 1972 \cite{AS-1972}.
In addition, Ablowitz, Kaup, Newell and Segur (AKNS) presented a class of new integrable systems, called AKNS systems,
and found a general framework for their ISTs \cite{MJA-1973,MJA-1974}.
Afterwards, many integrable nonlinear wave equations were found to be solved in terms of the IST,
such as the
sine-Gordon equation \cite{MJADJ-1973}, modified KdV equation \cite{MW-1973,MW-1982}, Kadomtsev-Petviashvili equation \cite{MJADB-1983},
Camassa-Holm  equation \cite{ACVS-2006}, Degasperis-Procesi equation \cite{ACBU-1983}, and
Benjamin-Ono Equation \cite{AFMB-1983}, etc.

In recent years,
the ISTs of nonlinear systems with NZBCs have been paid much attention based on the solutions of
the related Riemann-Hilbert (RH) problem \cite{BFMJ-2006}-\cite{wxbJMMA} %\cite{BFMJ-2006,VEZAB-1972,AMJ-2007,NOMJ,FBFLS,BFFBG-2011,BFFBG-2013,GB-19831,BP-2015,XZJLW-2015,DF-2014,GB-2016,GB-1983,IST-2017,IST-JNMP,yzy-cnsns,yzy-2020pd,wxbTHW,wxbJMMA}.
However, the IST for the coupled Hirota equation with NZBCs, to the best of the authors' knowledge, has not been reported before.
The main purpose of this
paper is to study the IST and soliton solutions for the coupled Hirota equation \cite{Hirota-PRE,WSSWL-2020}, whose form is
\begin{equation}\label{mHirota1}
i\mathcal {Q}_{t}+\alpha\left(\mathcal {Q}_{xx}-2\sigma\mathcal {Q}\mathcal {Q}^{\dag}\mathcal {Q}\right)
+i\beta\left(\mathcal {Q}_{xxx}-6\sigma\mathcal {Q}\mathcal {Q}^{\dag}\mathcal {Q}_{x}\right)=0,
\end{equation}
where $\mathcal {Q}=\mathcal {Q}(x,t)$ is a $2\times2$ matrix valued function, and ``\dag'' denotes the Hermitian conjugate.
The choice $\sigma=+1$ and $\sigma=-1$ distinguishes between the self-defocusing and self-focusing regimes, respectively.
Taking the $2\times2$ matrix potential $\mathcal {Q}(x,t)$ is the following symmetric matrix
\begin{equation*}\label{matrix}
\mathcal {Q}=\left(
               \begin{array}{cc}
                 q_{1} & q_{0} \\
                 q_{0} & q_{-1} \\
               \end{array}
             \right),
\end{equation*}
the system \eqref{mHirota1} can be viewed as the matrix Hirota equation.
The matrix Hirota equation \eqref{mHirota1} is completely integrable.
Particularly, when $\alpha=1$, $\beta=0$, system \eqref{mHirota1} can be reduced to the integrable Spin-1 Gross-Pitaevskii equations \cite{YZY-2017}-\cite{wxb-2018} %\cite{YZY-2017,YZY-20171,MU-2017,VS-2009,wxb-2018}
which can be used to describe light transmission in bimodal nonlinear optical fibres.
Here there are
three reasons for choosing the multi-component nonlinear system \eqref{mHirota1} as a model problem:
(I): throughout this paper, we allow constants $\alpha$ and $\beta$ to be arbitrary, so Eq.\eqref{mHirota1} is quite
general and can be used to describe a wide variety of physical processes.
(II): dynamics of two or more components with different modes, frequencies, or polarizations in optical fibers
can be described by the multi-component NLS systems \cite{BFDA-2012}-\cite{BF-2013}. %\cite{BFDA-2012,CSLY-2013,DA-2013,DA-2007,GAA-CMP,BF-2013}.
Such systems allow for energy transfer between their additional degrees of freedom and yield rich families of vector solutions.
(III): in recent years,
many efforts were devoted to studying ISTs for the one component model.
However, there are few research studies on ISTs for multi-component nonlinear equations.
Therefore, more researches about multi-component nonlinear systems are
also inevitable and worthwhile.

In this article we are devoted to extending the IST for Eq.\eqref{mHirota1} in the general case, under NZBCs as $x\rightarrow\pm\infty$,
as a technique to deal with the initial-value problem, and also construct soliton solutions as a byproduct of the IST.
It is necessary to point out that generally the boundary conditions for $\mathcal {Q}$ must be time($t$)-dependent.
However, their time($t$)-independence can be easily obtained by using the gauge transformation
\begin{equation*}%\label{}
\mathcal {Q}(x,t)=\hat{\mathcal {Q}}e^{-2i\sigma k_{0}^2t}.
\end{equation*}
Then Eq.\eqref{mHirota1} reaches to
\begin{equation}\label{mNLSS1}
i\mathcal {Q}_{t}+\alpha\left(\mathcal {Q}_{xx}-2\sigma\left(\mathcal {Q}\mathcal {Q}^{\dag}-k_{0}^2\mathcal {I}_{m}\right)\mathcal {Q}\right)
+i\beta\left(\mathcal {Q}_{xxx}-6\sigma\mathcal {Q}\mathcal {Q}^{\dag}\mathcal {Q}_{x}\right)=0,
\end{equation}
where $\mathcal {I}_{m}$ is an $m\times m$ identity matrix.

In what follows we analyze the matrix Hirota equation \eqref{mNLSS1} under constant NZBCs
\begin{equation}\label{NZBC-1}
\mathcal {Q}(x,t)\rightarrow\mathcal {Q}_{\pm}~~~~\mbox{as}~~~~x\rightarrow\pm\infty.
\end{equation}
Then we suppose that the following conditions on the boundary conditions hold
\begin{equation}\label{NZBC-2}
\mathcal {Q}_{\pm}^{\dag}\mathcal {Q}_{\pm}=\mathcal {Q}_{\pm}\mathcal {Q}_{\pm}^{\dag}=k_{0}^2\mathcal {I}_{m},
\end{equation}
where $k_{0}$ is a positive, real constant.
As $x\rightarrow\pm\infty$, for a $2\times2$ symmetric matrix potential,
the latter is similar to the following constraints
on the boundary values of the individual entries of $\mathcal {Q}(x,t)$
\begin{equation}\label{NZBC-3}
|q_{1,\pm}|^2=|q_{-1,\pm}|^2,~~q_{1,\pm}q_{0,\pm}^{\ast}=k_{0}^2-|q_{1,\pm}|^2\equiv k_{0}^2-|q_{-1,\pm}|^2,~~
q_{1,\pm}q_{0}^{\ast}+q_{0,\pm}q_{-1,\pm}^{\ast}=0.
\end{equation}
Here, without loss of generality, we notice that the boundary condition $\mathcal {Q}_{+}$ can be taken as
$\mathcal {Q}_{+}=k_{0}\mathcal {I}_{m}$.

Recently, there are many investigations on nonlinear wave solutions and long-time asymptotics of nonlinear evolution equations with NZBCs and zero
boundary conditions (ZBCs) \cite{AB111-2013}-\cite{xjfeg-2}.
It is also known that the IST is a powerful approach to derive nonlinear wave solutions.
However, since system \eqref{mNLSS1} admits a $4\times4$ matrix spectral problem,
the IST for Eq.\eqref{mNLSS1} with NZBCs is rather complicated to deal with.
The research work in this paper, to our knowledge, has not been conducted so far.
The aim of the present paper is to derive the soliton solutions of Eq.\eqref{mNLSS1} with NZBCs \eqref{NZBC-1}
by utilizing IST. Additionally, the main characteristics of these solutions are discussed by controlling suitable parameters.

The paper is organized as follows.
In the next section,
we investigate the direct scattering problem for Eq.\eqref{mHirota1} in the general case
with NZBCs satisfying \eqref{NZBC-2}.
In particular,
we derive the uniformization variable, give eigenfunctions and scattering data,
and analyze their analyticity as functions of the uniformization variable.
%Then we present the three symmetries in the scattering data induced by all the symmetries imposed on the potential,
%and analyze in detail the properties of discrete eigenvalues and norming constants.
In section 3, we study the inverse scattering problem for the
eigenfunctions as a RH problem with poles,
provide the formal solution of the latter and the reconstruction formula of the potential in view of
eigenfunctions and scattering data.
In section 4, we construct the soliton solutions for the focusing ($\sigma=-1$) equation in the $m=2$ case.
Then the dynamics of the breather wave solutions are analyzed with some graphics.
The last section summarizes the main results of this paper.

\section{Direct scattering problem}

\subsection{Riemann surface and uniformization coordinate}

Eq.\eqref{mNLSS1} is completely integrable.
Its Lax pair is
\begin{align}\label{LP-1}
\varphi_{x}=U\varphi,~~
\varphi_{t}=V\varphi,
\end{align}
with
\begin{align}\label{LP-2}
&U=-ik\sigma_{3}+\underline{\mathcal {Q}},~~
V=\alpha T_{\mbox{nls}}+\beta T_{\mbox{cmKdV}},\notag\\
&T_{\mbox{nls}}=2kU+i\sigma_{3}\left(\underline{\mathcal {Q}}_{x}-\underline{\mathcal {Q}}^2+\sigma k_{0}^2\mathcal {I}_{2m}\right),\notag\\
&T_{\mbox{cmKdV}}=2k\left(T_{\mbox{nls}}-i\sigma k_{0}^2\mathcal {I}_{2m}\sigma_{3}\right)-\left[\underline{\mathcal {Q}},\underline{\mathcal {Q}}_{x}\right]+2\underline{\mathcal {Q}}^3-\underline{\mathcal {Q}}_{xx},\notag\\
&\sigma_{3}=\left(
              \begin{array}{cc}
                \mathcal {I}_{m} & \textbf{0} \\
                \textbf{0} & -\mathcal {I}_{m} \\
              \end{array}
            \right),~~\underline{\mathcal {Q}}=\left(
                                 \begin{array}{cc}
                                   \textbf{0} & \mathcal {Q} \\
                                   \sigma\mathcal {Q}^{\dag} & \textbf{0} \\
                                 \end{array}
                               \right).
\end{align}
Throughout this work, $\textbf{0}$, $\widetilde{\textbf{0}}$, $\widehat{\textbf{0}}$ are used to represent the $m\times m$, $2m\times 2m$, $2m\times m$ zero matrix, respectively.

The asymptotic problem of the Lax pair \eqref{LP-1} for Eq.\eqref{mHirota1} with NZBCs \eqref{NZBC-1} yields
\begin{equation}\label{LP-3}
\varphi_{x}=U_{\pm}\varphi,~~U_{\pm}=-ik\sigma_{3}+\underline{\mathcal {Q}}_{\pm},
\end{equation}
where                                             $\underline{\mathcal {Q}}_{\pm}=\left(
                                                                       \begin{array}{cc}
                                                                         \textbf{0} & \mathcal {Q}_{\pm} \\
                                                                         \sigma\mathcal {Q}_{\pm}^{\dag} & \textbf{0} \\
                                                                       \end{array}
                                                                     \right)$.
Similar to \eqref{NZBC-2},  we find
\begin{equation}\label{LP-4}
\mathcal {Q}_{\pm}\mathcal {Q}_{\pm}^{\dag}=\mathcal {Q}_{\pm}^{\dag}\mathcal {Q}_{\pm}=k_{0}^2\mathcal {I}_{m}\Leftrightarrow
\underline{\mathcal {Q}}_{\pm}\underline{\mathcal {Q}}_{\pm}^{\dag}=\underline{\mathcal {Q}}_{\pm}^{\dag}\underline{\mathcal {Q}}_{\pm}=k_{0}^2\mathcal {I}_{2m}.
\end{equation}
It is easy to find that the eigenvalues of $U_{\pm}$ are $\pm i\sqrt{k^2-\sigma k_{0}^2}$.
In order to further analyze the branching of the eigenvalues, we next consider the two-sheeted Riemann surface given by
\begin{equation}\label{LP-5}
\lambda^2=k^2-\sigma k_{0}^2,
\end{equation}
in which $\lambda(k)$ represents a single-valued function on this surface.\\

\centerline{\begin{tikzpicture}[scale=0.5]
\tikzstyle{arrow} = [->,>=stealth]
\filldraw (-9,1) -- (-9,9) to (-1,9) -- (-1,1);
\filldraw (9,1) -- (9,9) to (1,9) -- (1,1);
\filldraw (9,-1) -- (9,-9) to (1,-9) -- (1,-1);
\filldraw (-9,-1) -- (-9,-9) to (-1,-9) -- (-1,-1);
\path [fill=white] (1,-5) -- (9,-5) to (9,-9) -- (1,-9);
\filldraw[pink, line width=0.5](-1,-5)--(3,-5) arc (-180:0:2);
\path [fill=pink] (1,-1) -- (9,-1) to (9,-5) -- (1,-5);
\filldraw[white, line width=0.5](3,-5)--(7,-5) arc (0:180:2);
\path [fill=pink] (-9,5)--(-9,9) to (-1,9) -- (-1,5);
\path [fill=white] (-9,1)--(-9,5) to (-1,5) -- (-1,1);
\path [fill=pink] (1,1)--(1,5) to (9,5) -- (9,1);
\path [fill=white] (1,5)--(1,9) to (9,9) -- (9,5);
\path [fill=pink] (-1,-1)--(-1,-5) to (-9,-5) -- (-9,-1);
\path [fill=white] (-1,-5)--(-1,-9) to (-9,-9) -- (-9,-5);
\filldraw[red, line width=0.5] (2,2) to (-2,-2)[arrow];
\filldraw[red, line width=0.5] (-2,-8) to (2,-8)[arrow];
\draw[fill] (-5,5)node[below]{} circle [radius=0.035];
\draw[fill] (5,5)node[below]{} circle [radius=0.035];
\draw[fill] (-5,-5)node[below]{} circle [radius=0.035];
\draw[fill] (5,-5)node[below]{} circle [radius=0.035];
\draw[-][thick](-9,5)--(-8,5);
\draw[-][thick](-8,5)--(-7,5);
\draw[-][thick](-7,5)--(-6,5);
\draw[-][thick](-6,5)--(-5,5);
\draw[-][thick](-5,5)--(-4,5);
\draw[-][thick](-4,5)--(-3,5);
\draw[-][thick](-3,5)--(-2,5);
\draw[-][arrow][thick](-2,5)--(-1,5)[thick]node[right]{$\mbox{Re}(k)$};
\draw[-][thick](-5,1)--(-5,2);
\draw[-][thick](-5,2)--(-5,3);
\draw[-][thick](-5,3)--(-5,4);
\draw[-][thick](-5,4)--(-5,5);
\draw[-][thick](-5,5)--(-5,6);
\draw[-][thick](-5,6)--(-5,7);
\draw[-][thick](-5,7)--(-5,8);
\draw[-][thick](-5,8)--(-5,9)[arrow] [thick]node[above]{$\mbox{Im}(k)$};
\draw[-][thick](1,5)--(2,5);
\draw[-][thick](2,5)--(3,5);
\draw[-][thick](3,5)--(4,5);
\draw[-][thick](4,5)--(5,5);
\draw[-][thick](5,5)--(6,5);
\draw[-][thick](6,5)--(7,5);
\draw[-][thick](7,5)--(8,5);
\draw[-][thick](8,5)--(9,5)[arrow][thick]node[right]{$\mbox{Re}(k)$};
\draw[-][thick](5,1)--(5,2);
\draw[-][thick](5,2)--(5,3);
\draw[-][thick](5,3)--(5,4);
\draw[-][thick](5,4)--(5,5);
\draw[-][thick](5,5)--(5,6);
\draw[-][thick](5,6)--(5,7);
\draw[-][thick](5,7)--(5,8);
\draw[-][thick](5,8)--(5,9);
\draw[-][thick](-9,-5)--(-8,-5);
\draw[-][thick](-8,-5)--(-7,-5);
\draw[-][thick](-7,-5)--(-6,-5);
\draw[-][thick](-6,-5)--(-5,-5);
\draw[-][thick](-5,-5)--(-4,-5);
\draw[-][thick](-4,-5)--(-3,-5);
\draw[-][thick](-3,-5)--(-2,-5);
\draw[-][thick](-2,-5)--(-1,-5)[arrow][thick]node[right]{$\mbox{Re}(\lambda)$};
\draw[-][thick](-5,-1)--(-5,-2);
\draw[-][thick](-5,-2)--(-5,-3);
\draw[-][thick](-5,-3)--(-5,-4);
\draw[-][thick](-5,-4)--(-5,-5);
\draw[-][thick](-5,-5)--(-5,-6);
\draw[-][thick](-5,-6)--(-5,-7);
\draw[-][thick](-5,-7)--(-5,-8);
\draw[-][thick](-5,-8)--(-5,-9);
\draw[-][thick](1,-5)--(2,-5);
\draw[-][thick](2,-5)--(3,-5);
\draw[-][thick](3,-5)--(4,-5);
\draw[-][thick](4,-5)--(5,-5);
\draw[-][thick](5,-5)--(6,-5);
\draw[-][thick](6,-5)--(7,-5);
\draw[-][thick](7,-5)--(8,-5);
\draw[-][thick](8,-5)--(9,-5)[arrow][thick]node[right]{$\mbox{Re}(z)$};
\draw[-][thick](5,-1)--(5,-2);
\draw[-][thick](5,-2)--(5,-3);
\draw[-][thick](5,-3)--(5,-4);
\draw[-][thick](5,-4)--(5,-5);
\draw[-][thick](5,-5)--(5,-6);
\draw[-][thick](5,-6)--(5,-7);
\draw[-][thick](5,-7)--(5,-8);
\draw[-][thick](5,-8)--(5,-9);
\draw[-][thick](5,8)--(5,9)[arrow] [thick]node[above]{$\mbox{Im}(k)$};
\draw[-][thick](-5,-1.35)--(-5,-1.35)[arrow];% [thick]node[above]{$Im\lambda$};
\draw[fill] (-5,-1) node[above]{$\mbox{Im}(\lambda)$};
%\draw[arrow](5,8.5)[thick]node[above]{$Imk$};
%\draw[arrow](-5,-1)[thick]node[above]{$Im\lambda$};
\draw[-][thick](5,-1.35)--(5,-1.35)[arrow];% [thick]node[above]{$Im\lambda$};
\draw[fill] (5,-1) node[above]{$\mbox{Im} (z)$};
%\draw[arrow](5,-1)[thick]node[above]{$Imz$};
\draw[fill] (-5,7) circle [radius=0.055]node[left]{\footnotesize$ik_{0}$};
\draw[fill] (-5,3) circle [radius=0.055]node[left]{\footnotesize$-ik_{0}$};
\draw[fill] (5,7) circle [radius=0.055]node[left]{\footnotesize$ik_{0}$};
\draw[fill] (5,3) circle [radius=0.055]node[left]{\footnotesize$-ik_{0}$};
\draw[fill] (-7,-5) circle [radius=0.055]node[below]{\footnotesize$k_{0}$};
\draw[fill] (-3,-5) circle [radius=0.055]node[below]{\footnotesize$-k_{0}$};
\draw(5,-5) [red, line width=1] circle(2);
\filldraw[red, line width=1.5] (-5,7) to (-5,3);
\filldraw[red, line width=1.5] (5,7) to (5,3);
\filldraw[red, line width=1.5] (-7,-5) to (-3,-5);
\draw[fill][black] (-8,7) [thick]node[right]{\footnotesize$S_{1}$};
\draw[fill][black] (2,7) [thick]node[right]{\footnotesize$S_{2}$};
\draw[fill][black] (-4,7) [thick]node[right]{\footnotesize$\mbox{Im}(k)>0$};
\draw[fill][black] (-4,3) [thick]node[right]{\footnotesize$\mbox{Im}(k)<0$};
\draw[fill][black] (6,7) [thick]node[right]{\footnotesize$\mbox{Im}(k)>0$};
\draw[fill][black] (6,3) [thick]node[right]{\footnotesize$\mbox{Im}(k)<0$};
\draw[fill][black] (-4,-7) [thick]node[right]{\footnotesize$\mbox{Im}(\lambda)<0$};
\draw[fill][black] (-4,-3) [thick]node[right]{\footnotesize$\mbox{Im}(\lambda)>0$};
\draw[fill][black] (7,-7) [thick]node[right]{\footnotesize$D_{-}$};
\draw[fill][black] (7,-3) [thick]node[right]{\footnotesize$D_{+}$};
\draw[fill][red] (0,6) node[]{\footnotesize$+$};
\draw[fill][black] (-2,0) [thick]node[right]
{\footnotesize$\lambda=\sqrt{k^{2}+k_{0}^{2}}$};
\draw[fill][black] (0,-8) [thick]node[below]
{\footnotesize$\lambda=(z+k_{0}^{2}/z)/2$};
\end{tikzpicture}}
\noindent {\small \textbf{Figure 1.}  Transformation relation from $k$ two-sheeted Riemann surface, $\lambda$-plane and z-plane.}

In the $\sigma=-1$ case, taking
$k+ik_{0}=r_{1}e^{i\theta_{1}}$ and $k-ik_{0}=r_{2}e^{i\theta_{2}}$,
we then set
\begin{equation}\label{LP-6}
\left\{ \begin{aligned}
&\lambda(k)=\sqrt{r_{1}r_{2}}e^{i(\theta_{1}+\theta_{2})/2},~~~~\mbox{on sheet}~~\mathbb{C}_{I},\\
&\lambda(k)=-\sqrt{r_{1}r_{2}}e^{i(\theta_{1}+\theta_{2})/2},~~\mbox{on sheet}~~\mathbb{C}_{II},
                  \end{aligned} \right.
\end{equation}
choosing the local angles $\theta_{j}\in\left[-\frac{\pi}{2},\frac{3\pi}{2}\right)$ for $j=1,2$ corresponds to placing the discontinuity of $\lambda$
on the segment $ik_{0}[1,1]$ on the imaginary $k$-axis.
Then we get the Riemann surface by gluing the two copies of the complex plane along the cut.

A similar analysis used in the $\sigma=1$ case, with the branch cut chosen on $k$ (real) axis,
and exactly for $k\in(-\infty,k_{0})\bigcup(k_{0},+\infty)$.
Particularly, one can give the local polar coordinates $k-k_{0}=r_{1}e^{i\theta_{1}}$ and $k+k_{0}=e^{i\theta_{2}}$ on sheet I,
with $r_{1}$, $r_{2}$ uniquely set by the location of $k$, and angles $\theta_{1}\in[0,2\pi)$
and $\theta_{2}\in[-\pi,\pi)$.
Additionally, we can also give \eqref{LP-6}.
In this situation, $(\theta_{1}+\theta_{2})/2$  varies continuously between 0 and $\pi$ both in the lower and in the upper planes,
with a cut on $(-\infty,-k_{0})\bigcup(k_{0},+\infty)$.
The upper branches of the cuts on sheet $\mathbb{C}_{I}$ are then glued with
the lower branches on sheet $\mathbb{C}_{II}$, and vice versa, thus $\lambda(k)$ is again continuous via the cut.

Similar to \cite{BFMJ-2006,AMJ-2007,yzy-cnsns}, we have the following uniformization variable
\begin{equation}\label{LP-7}
z=k+\lambda,
\end{equation}
whose inverse transformation reads
\begin{equation}\label{LP-8}
k=\frac{1}{2}\left(z+\sigma\frac{k_{0}^2}{z}\right),~~\lambda=\frac{1}{2}\left(z-\sigma\frac{k_{0}^2}{z}\right).
\end{equation}
Summarizing the above results, in $\sigma=1$ case the branch cut on either sheet is mapped to the $z$ (real) axis,
the two sheets (i.e., $\mathbb{C}_{I}$ and $\mathbb{C}_{II}$) of
the Riemann surface are mapped to the lower and upper half-planes of the complex $z$-plane, respectively,
a neighborhood of $k=\infty$ on either sheet is mapped onto a neighborhood of $z=\infty$ (or $z=0$) relying on the sign of $\mbox{Im} k$.

In $\sigma=-1$ case the branch cut on either sheet is mapped to the circle $\mathbb{C}_{o}$;
$\mathbb{C}_{I}$ is mapped to the exterior of $\mathbb{C}_{o}$;
$\mathbb{C}_{II}$ is mapped to the interior of $\mathbb{C}_{o}$; $z(\infty_{I})=\infty$ and $z(\infty_{II})=0$.

Therefore, in the $\sigma=1$ case, $\mbox{Im} \lambda<0$ in the lower-half plane and $\mbox{Im} \lambda>0$ in the upper-half plane of $z$
\begin{equation}\label{LP-9}
\sigma=1:~~D^{+}=\left\{z\in\mathbb{C}:\mbox{Im}>0\right\},~~D^{-}=\left\{z\in\mathbb{C}:\mbox{Im}<0\right\}.
\end{equation}
In the $\sigma=-1$ case, $\mbox{Im}\lambda$ is not sign-definite in either half-plane;
but one has $\mbox{Im}\lambda>0$ in $D^{+}$ and  $\mbox{Im}\lambda<0$ in $D^{-}$,
where, for $\sigma=-1$
\begin{align}\label{LP-10}
&D^{+}=\left\{z\in\mathbb{C}:\left(|z|^2-k_{0}^2\right)\mbox{Im} z>0\right\},\notag\\
&D^{-}=\left\{z\in\mathbb{C}:\left(|z|^2-k_{0}^2\right)\mbox{Im} z<0\right\}.
\end{align}
Then we set the focusing ($\sigma=-1$) equation as a case
to demonstrate the transformation between different complex planes in Fig.1.
In addition, the two domains $D^{\pm}$ and the complex $z$ plane are displayed in Fig.2 ($\sigma=1$ case on the below, and $\sigma=-1$ case on the above).
As will be analyzed in subsection 2.2,
we find that the sign of $\mbox{Im} \lambda$ confirms the regions of analyticity of the Jost eigenfunctions.
In the following
we can rewrite all the $k$ dependence as rely on $z$ wherever applicable with some abuse of notation.\\

\centerline{\begin{tikzpicture}
\tikzstyle{arrow} = [->,>=stealth]
\path [fill=pink] (-4,0) -- (0,0) to (0,4) -- (-4,4);
\path [fill=pink] (4,0) -- (0,0) to (0,4) -- (4,4);
\draw [thick][fill][red](-4,0)--(4,0);
\draw[arrow][thick](0,-4)--(0,4);
%\draw[fill] (0,0) circle [radius=2] [fill=white];
%\draw[fill] (2,2) circle [radius=0.035];
\draw[fill] (-2,0) circle [radius=0.035];
\draw[fill] (0,2) circle [radius=0.035];
\draw[fill] (2,0) circle [radius=0.035];
\draw[fill]  (1.414,1.414) circle [radius=0.05] [red] node[above]{$\zeta_{n}$};
\draw[fill] (1.414,-1.414) circle [radius=0.05] [blue] node[below]{$\zeta_{n}^{\ast}$};
\draw[arrow][thick][fill][red](-4,0)--(-3,0);
\draw[arrow][thick][fill][red](0,0)--(0.3,0);
\draw[arrow][thick][fill][red](2,0)--(2.5,0);
%\draw[-][dashed,thick](-2,2)--(-2,-2);
%\draw[fill] (2,2) circle [radius=0.035] node[right]{$a$};
%\draw[fill] (2,-2) circle [radius=0.035] node[right]{$-a$};
%\draw[->][dashed,thick](2,-4)--(2,4);
%\draw[-][dashed,thick](-2,2)--(2,2);
%\draw[-][dashed,thick](-2,-2)--(2,-2);
\draw [-,thick] (0,2)  to [out=180,in=90] (-2,0);
\draw [-,thick] (0,-2)  to [out=180,in=-90] (-2,0);
\draw [-,thick] (0,2)  to [out=0,in=90] (2,0);
\draw [-,thick] (0,-2)  to [out=0,in=-90] (2,0);
%\draw [-,thick] (1,0)  to [out=80,in=-90] (1.75,3.45);
%\draw [-,thick] (1,0)  to [out=-80,in=90] (1.75,-3.45);
%\draw[-][thick](1.75,3.45)--(1.75,4);
%\draw[-][thick](1.75,-3.45)--(1.75,-4);
%\draw[fill] (2,4) node[right]{$\lambda_{2}$};
%\draw[fill] (3,3) node[right]{$\mathbb{D}_{1}$};
%\draw[fill] (3,-3) node[right]{$\mathbb{D}_{4}$};
\draw[fill] (3.6,0) node[below]{$\mbox{Re(z)}$};
\draw[fill] (0,3.8) node[left]{$\mbox{Im(z)}$};
\draw[fill] (-2.3,0) node[below]{$-k_{0}$};
\draw[fill] (2.3,0) node[below]{$k_{0}$};
\draw[fill] (-1,2.3) node[below]{$\mathbb{C}_{o}$};
%\draw[fill] (1.2,2.5) node[right]{$\Gamma$};
%\draw[fill] (1.2,-2.5) node[right]{$\bar{\Gamma}$};
\draw[fill] (-0.2,0) node[below]{$O$};
%\draw[fill] (0,2) node[above]{$\mathbb{D}_{2}$};
%\draw[fill] (0,-2) node[below]{$\mathbb{D}_{3}$};
%\draw[fill] (0.75,0) node[above]{$\lambda_{+}$};
%\draw[fill] (-0.75,0) node[above]{$\lambda_{-}$};
%\draw[fill] (-2.5,0) node[below]{$-b$};
%\draw[fill] (-1.75,1) node[above]{$\gamma$};
%\draw[fill] (-1.75,-1) node[below]{$\bar{\gamma}$};
%\draw[fill] (0,-3) node[below]{$E=-b+ia$};
%\draw[fill] (0,-3.5) node[below]{$E=-b-ia$};
\end{tikzpicture}}

\centerline{\begin{tikzpicture}
\tikzstyle{arrow} = [->,>=stealth]
\path [fill=pink] (-4,0) -- (0,0) to (0,4) -- (-4,4);
\path [fill=pink] (4,0) -- (0,0) to (0,4) -- (4,4);
\filldraw[white, line width=1](2,0) arc (0:180:2);
\filldraw[pink, line width=1](-2,0) arc (-180:0:2);
\draw [thick][fill][red](-4,0)--(4,0);
\draw[arrow][thick](0,-4)--(0,4);
%\draw[fill] (0,0) circle [radius=2] [fill=white];
\draw[fill] [red] (3,3) circle [radius=0.035];
\draw[fill] [blue] (3,-3) circle [radius=0.035];
\draw[fill] [blue] (-0.8,0.6) circle [radius=0.035];
\draw[fill] [red] (-0.8,-0.6) circle [radius=0.035];
\draw[fill] (-2,0) circle [radius=0.035];
\draw[fill] (0,2) circle [radius=0.035];
\draw[fill] (2,0) circle [radius=0.035];
\draw[arrow][thick][fill][red](-4,0)--(-3,0);
\draw[arrow][thick][fill][red](0.01,0)--(0,0);
\draw[arrow][thick][fill][red](2,0)--(2.5,0);
\draw[arrow][thick][fill][red](-0.1,2)--(0,2);
\draw[arrow][thick][fill][red](-0.1,-2)--(0,-2);
\draw [-,thick] [red] (0,2)  to [out=180,in=90] (-2,0);
\draw [-,thick] [red] (0,-2)  to [out=180,in=-90] (-2,0);
\draw [-,thick] [red] (0,2)  to [out=0,in=90] (2,0);
\draw [-,thick] [red] (0,-2)  to [out=0,in=-90] (2,0);
\draw[fill] (3.6,0) node[below]{$\mbox{Re(z)}$};
\draw[fill] (0,3.8) node[left]{$\mbox{Im(z)}$};
\draw[fill] (-2.3,0) node[below]{$-k_{0}$};
\draw[fill] (2.3,0) node[below]{$k_{0}$};
\draw[fill] (-1,2.3) node[below]{$\mathbb{C}_{o}$};
\draw[fill] (-0.2,0) node[below]{$O$};
\draw[fill] (0.3,2) node[above]{$ik_{0}$};
\draw[fill] (0.4,-2) node[below]{$-ik_{0}$};
\draw[fill] (3,3) node[above]{$z_{n}$};
\draw[fill] (3,-3) node[below]{$z_{n}^{\ast}$};
\draw[fill] (-0.8,0.6) node[above]{$-k_{0}^2/z_{n}$};
\draw[fill] (-0.8,-0.6) node[below]{$-k_{0}^2/z_{n}^{\ast}$};
\end{tikzpicture}}
\noindent {\small \textbf{Figure 2.}  The complex $z$-plane, showing the regions $D^{\pm}$ in which $\mbox{Im}\lambda>0$ and $\mbox{Im}\lambda<0$, respectively, in the $\sigma=\pm1$ case. Also seen in the figures are the oriented contours for the RH problem (red), and the symmetries of the discrete spectrum of the scattering problem.}

\subsection{Jost solutions and analyticity}

%The Jost solutions are usually defined in terms of the asymptotic eigenvectors of the equation defining the scattering problem.
It follows from \eqref{LP-3} that the asymptotic eigenvector matrix yields
\begin{equation}\label{JSA-1}
X_{\pm}(k)=\mathcal {I}_{2m}-\frac{i}{k+\lambda}\sigma_{3}\underline{\mathcal {Q}}_{\pm}\equiv\mathcal {I}_{2m}
-\frac{i}{z}\sigma_{3}\underline{\mathcal {Q}}_{\pm},~~U_{\pm}X_{\pm}=-i\lambda X_{\pm}\sigma_{3},
\end{equation}
where $\mathcal {I}_{2m}$ is a $2m\times2m$ identity matrix. %such that
%\begin{equation}\label{JSA-2}
%U_{\pm}X_{\mp}=-i\lambda X_{\pm}\sigma_{3}.
%\end{equation}
Notice that
\begin{align}\label{JSA-3}
&\det X_{\pm}(z)=\left(\frac{2\lambda}{\lambda+k}\right)^{m}=\gamma^{m}(z),~~\gamma(z)=1-\sigma\frac{k_{0}^2}{z^2},\notag\\
&X_{\pm}^{-1}(z)=\frac{1}{\gamma(z)}\left(\mathcal {I}_{2m}+\frac{i}{z}\sigma_{3}\underline{\mathcal {Q}}_{\pm}\right),
\end{align}
where $X_{\pm}^{-1}$ are determined for all values of $z$ and $\gamma(z)\neq0$,
i.e., away from the branch points $z\neq\pm i k_{0}$ in the $\sigma=-1$ case,
and $z\neq\pm k_{0}$ in the $\sigma=1$ case.

The continuous spectrum $\Sigma_{k}$ contain all values of $k$  such that $\lambda(k)\in\mathbb{R}$;
i.e., $\Sigma_{k}=\mathbb{R}\bigcup ik_{0}[-1,1]$ in the $\sigma=-1$ case,
and $\Sigma_{k}=\mathbb{R}\setminus k_{0}(-1,1)$  in the $\sigma=1$ case.
The corresponding sets in the complex z-plane are $\Sigma_{z}=\mathbb{R}\bigcup \mathbb{C}_{o}$ and $\Sigma_{z}=\mathbb{R}$, respectively.
$\mathbb{C}_{o}$ being the circle of radius $k_{0}$ centered at the origin (see Fig.1).
In what follows we omit the subscripts on $\Sigma$, since the result will be found from the context.
For $\forall z\in\Sigma$,
we now conssider the Jost eigenfunctions $\Phi(x,t;z)$ and $\Psi(x,t;z)$ as the simultaneous solutions
of both parts of the Lax pair, we thus have
\begin{align}\label{JSA-4}
\Phi(x,t;z)=\left(\phi(x,t;z),\bar{\phi}(x,t;z)\right)=X_{-}(z)e^{i\theta(x,t;z)\sigma_{3}}+O(1),~~x\rightarrow-\infty,\notag\\
\Psi(x,t;z)=\left(\psi(x,t;z),\bar{\psi}(x,t;z)\right)=X_{+}(z)e^{i\theta(x,t;z)\sigma_{3}}+O(1),~~x\rightarrow+\infty,
\end{align}
where
\begin{equation}\label{JSA-5}
\theta(x,t;z)=\lambda(z)\left\{-x-\left[\beta\left(4k^2-2k_{0}^2\right)+2\alpha k(z)\right]t\right\},
\end{equation}
and $\phi(x,t;z)$, $\bar{\phi}(x,t;z)$ ($\bar{\psi}(x,t;z)$, $\psi(x,t;z)$)
are four $2m\times m$ matrices which group the last $m$ column and the first $m$ vectors of the
$2m\times2m$ matrix solutions $\Phi(x,t;z)$ ($\Psi(x,t;z)$).
For the sake of convenience, we present the following modified eigenfunctions
\begin{align}\label{JSA-6}
&\left(M(x,t;z),\bar{M}(x,t;z)\right)=\Phi(x,t;z)e^{-i\theta(x,t;z)\sigma_{3}},\notag\\
&\left(\bar{N}(x,t;z),N(x,t;z)\right)=\Psi(x,t;z)e^{-i\theta(x,t;z)\sigma_{3}}.
\end{align}
Similar to \cite{GB-1983}, the following integral equations can be obtained
\begin{align}\label{JSA-7}
&\left(M(x,t;z),\bar{M}(x,t;z)\right)=X_{-}\notag\\
&~~+\int_{-\infty}^{x}X_{-}e^{i\lambda\sigma_{3}(\xi-x)}X_{-}^{-1}
\left(\underline{\mathcal {Q}}-\underline{\mathcal {Q}}_{-}\right)\left(M(\xi,t;z),\bar{M}(\xi,t;z)\right)e^{i\lambda\sigma_{3}(x-\xi)}d\xi,\notag\\
&\left(\bar{N}(x,t;z),N(x,t;z)\right)=X_{-}\notag\\
&~~+\int_{-\infty}^{x}X_{-}e^{i\lambda\sigma_{3}(\xi-x)}X_{-}^{-1}
\left(\underline{\mathcal {Q}}-\underline{\mathcal {Q}}_{+}\right)\left(\bar{N}(\xi,t;z),N(\xi,t;z)\right)e^{i\lambda\sigma_{3}(x-\xi)}d\xi,
\end{align}
where the modified eigenfunctions $\bar{M}(x,t;z)$ and $\bar{N}(x,t;z)$ can be analytically
extended in the complex $z$-plane ($\mbox{Im}\lambda(z)<0$),
and $M(x,t;z)$ and $N(x,t;z)$ can be analytically extended in the complex $z$-plane ($\mbox{Im}\lambda(z)>0$).
Let us define by $L^{1,s}(\mathbb{R})$ the complex Banach space of all measurable functions
$f(x)$ for which $(1+|x|)^{s}f(x)\in L^{1}(\mathbb{R})$ for $s=0,1$.
Then we have the following two theorems (i.e., Theorem 1 and Theorem 2).\\

\noindent
\textbf{Theorem 1.}  Assume that $\mathcal {Q}(x,t)-\mathcal {Q}_{+}\in L^{1}([x_{+},+\infty])$
and $\mathcal {Q}(x,t)-\mathcal {Q}_{-}\in L^{1}([-\infty,x_{-}])$ hold for $x_{\pm}\in\mathbb{R}$, all $t\geq0$,
and that the matrix potential function $\mathcal {Q}(x,t)$
is the symmetric matrix as well as admits the boundary conditions \eqref{LP-4}.
For $x_{\pm}\in\mathbb{R}$, $D^{\pm}$ are given by \eqref{LP-9} and \eqref{LP-10}
for ($\sigma=\pm1$ (the defocusing/focusing cases, respectively), i.e.,
\begin{equation}\label{JSA-8}
D^{+}=\left\{z\in\mathbb{C}:\left(|z|^2+\sigma k_{0}^2\right)\mbox{Im}z>0\right\},~~
D^{-}=\left\{z\in\mathbb{C}:\left(|z|^2+\sigma k_{0}^2\right)\mbox{Im}z<0\right\}.
\end{equation}
Let us suppose that the boundary conditions $\mathcal {Q}_{\pm}$ meet \eqref{LP-4}.
Then for the modified eigenfunctions of the problem \eqref{LP-1} expressed by \eqref{JSA-4} and \eqref{JSA-7},
we find: $M(x,t;z)$ and $N(x,t;z)$ are analytic functions of $z$ for $z\in D^{+}$,
and they are continuous up to $\partial D^{+}\backslash \left\{\pm\sqrt{\sigma}k_{0}\right\}$.
Likewise, $\bar{M}(x,t;z)$ and $\bar{N}(x,t;z)$ are analytic functions of $z$ for $z\in D^{-}$,
and they are continuous up to $\partial D^{-}\backslash \left\{\pm\sqrt{\sigma}k_{0}\right\}$.\\

\noindent
\textbf{Theorem 2.} Assume that $\mathcal {Q}(x,t)-\mathcal {Q}_{+}\in L^{1,1}([x_{+},+\infty])$
and $\mathcal {Q}(x,t)-\mathcal {Q}_{-}\in L^{1,1}([-\infty,x_{-}])$
hold for any $x_{\pm}\in \mathbb{R}$,  and the potential function $\mathcal {Q}(x,t)$ satisfies \eqref{LP-4}.
%Also suppose that the boundary conditions $\mathcal {Q}_{\pm}$ admit \eqref{LP-4}.
Then for the modified eigenfunctions of the scattering problem \eqref{LP-1} determined by \eqref{JSA-4} and \eqref{JSA-7},
we see: $M(x,t;z)$ and $N(x,t;z)$ are analytic functions of $z$ for $z\in D^{+}$, and they are continuous up to $\partial D^{+}$;
$\bar{M}(x,t;z)$ and $\bar{N}(x,t;z)$ are analytic functions of $z$ for $z\in D^{-}$, and they are continuous up to $\partial D^{-}$.
i.e., the eigenfunctions are also continuous up to $\pm\sqrt{\sigma}k_{0}$.

Similar to \cite{GB-1983},
the above two theorems can be easily solved by using standard Neumann series representations for the solutions of \eqref{JSA-7}.

\subsection{Scattering coefficients}

In view of Jacobi's formula, we see that any matrix solution $\varphi(x,t;z)$ of \eqref{LP-1} satisfies
$\partial_{x}(\det\varphi)=\partial_{t}(\det\varphi)=0$, and $\mbox{tr}(U)=0$ and $\mbox{tr}(V)=0$ in \eqref{LP-1}.
Consequently, we have
\begin{equation}\label{SCS-1}
\lim_{x\rightarrow-\infty}\Phi(x,t;z)e^{-i\theta\sigma_{3}}=X_{-},~~
\lim_{x\rightarrow+\infty}\Psi(x,t;z)e^{-i\theta\sigma_{3}}=X_{+},~~z\in\Sigma.
\end{equation}
It follows from \eqref{SCS-1} that
\begin{equation}\label{SCS-2}
\det\Phi(x,t;z)=\det\Psi(x,t;z)=\det X_{\pm}(z)=\gamma^{m},~~x,t\in\mathbb{R},~~z\in\Sigma.
\end{equation}
Since $\Sigma_{0}=\Sigma\backslash\left\{\pm\sqrt{\sigma}k_{0}\right\}$,
we then have that all $z\in\Sigma_{0}$, both $\Phi$ and $\Psi$
are two fundamental matrix solutions of the scattering problem.
As a consequence, the $2m \times 2m$ scattering matrix $S(z)$ admits
\begin{equation}\label{SCS-3}
\Phi(x,t;z)=\Psi(x,t;z)\mathcal {S}(z),~~z\in\Sigma_{0},
\end{equation}
with
\begin{equation}\label{SCS-4}
\mathcal {S}(z)=\left(
       \begin{array}{cc}
         a(z) & \bar{b}(z) \\
         b(z) & \bar{a}(z) \\
       \end{array}
     \right).
\end{equation}
In view of the analytic groups of columns given in \eqref{JSA-4}, one can get
\begin{align}\label{SCS-5}
&\phi(x,t;z)=\psi(x,t;z) b(z)+\bar{\psi}(x,t;z)a(z),\notag\\
&\bar{\phi}(x,t;z)=\psi(x,t;z) \bar{b}(z)+\bar{\psi}(x,t;z)\bar{a}(z),
\end{align}
where $a(z)$, $b(z)$, $\bar{a}(z)$, $\bar{b}(z)$ represent the $m \times m$ blocks of the scattering matrix.

Notice that since $\Phi$ and $\Psi$ are simultaneous solutions of both parts of the Lax pair,
the entries of $\mathcal {S}(z)$ are independent of t.
In addition, from Eqs.\eqref{SCS-2} and \eqref{SCS-3}, we have $\det\mathcal {S}(z)=1$.

It follows from \eqref{SCS-3} that
\begin{equation}\label{SCS-6}
\det a(z)=\frac{\mbox{Wr}\left(\phi,\psi\right)}{\mbox{Wr}\left(\bar{\psi},\psi\right)}\equiv\frac{\det\left(\phi,\psi\right)}{\gamma^{m}},~~
\det \bar{a}(z)=\frac{\mbox{Wr}\left(\bar{\phi},\bar{\psi}\right)}{\mbox{Wr}\left(\bar{\psi},\psi\right)}
\equiv\frac{\det\left(\bar{\phi},\bar{\psi}\right)}{\gamma^{m}},
\end{equation}
where $\mbox{Wr}(f,g)$ represents the Wronskian determinant of the $2m\times m$ matrices $f$ and $g$.
As mentioned in \cite{BFFBG-2013}, we have the following integral representations for the scattering matrix
\begin{align}\label{SCS-7}
\mathcal {S}(z)=&\int_{0}^{\infty}e^{i\lambda(z)\xi\sigma_{3}}X_{+}^{-1}(z)\left(\underline{\mathcal {Q}}-\underline{\mathcal {Q}}_{+}\right)\Phi(\xi,t;z)d\xi\notag\\
&+X_{+}^{-1}(z)X_{-}(z)\left\{\mathcal {I}_{2m}+\int_{-\infty}^{0}e^{i\lambda(z)y\sigma_{3}}X_{-}^{-1}(z)\left(\underline{\mathcal {Q}}-\underline{\mathcal {Q}}_{-}\right)\Phi(\xi,t;z)d\xi\right\}.
\end{align}
Then the following theorem 3 holds.\\

\noindent
\textbf{Theorem 3.} Assume $\mathcal {Q}-\mathcal {Q}_{+}\in L^{1}([x_{+},+\infty))$ and $\mathcal {Q}-\mathcal {Q}_{-}\in L^{1}((-\infty,x_{-}])$
as matrix functions of x for all $t\geq0$, for some $x_{\pm}\in\mathbb{R}$, and let $D^{\pm}$ be defined as in \eqref{LP-9} and \eqref{LP-10}
for the $\sigma=\pm1$ cases.
Also suppose that the boundary conditions $Q_{\pm}$ admit \eqref{LP-4}.
For the scattering matrix $\mathcal {S}(z)$ given in view of the eigenfunctions of the scattering problem by \eqref{SCS-3},
we see:
the upper diagonal block $a(z)$  is continuous up to $\Sigma_{0}=\partial D^{+}\backslash\left\{\pm\sqrt{\sigma}k_{0}\right\}$,
and is analytic in $D^{+}$,
and continuous up to $\Sigma_{0}=\partial D^{-}\backslash\left\{\pm\sqrt{\sigma}k_{0}\right\}$,
and the lower diagonal block $\bar{a}(z)$ is analytic in $D^{-}$.
The off-diagonal blocks of the scattering matrix $\mathcal {S}(z)$, i.e., $b(z)$ and $\bar{b}(z)$,
are only expressed for $z\in\Sigma_{0}$, in which they are continuous, However, generally they do not admit analytic continuation off $\Sigma_{0}$.

Theorem 3 is a direct consequence of Theorem 1 and of the integral representation \eqref{SCS-7}.
Another proof of the analyticity of $a(z)$ and $\bar{a}(z)$ that applies the symmetries in the scattering data will be presented
in the next subsection 2.4.

In the end, for $z\in\Sigma_{0}$, it follows from \eqref{JSA-6} and \eqref{SCS-5} that
\begin{align}\label{SCS-8}
&M(x,t;z)a^{-1}(z)=\bar{N}(x,t;z)+e^{-2i\theta(x,t;z)}N(x,t;z)\rho(z),\notag\\
&\bar{M}(x,t;z)\bar{a}^{-1}(z)=N(x,t;z)+e^{2i\theta(x,t;z)}\bar{N}(x,t;z)\bar{\rho}(z),
\end{align}
where $M(x,t;z)a^{-1}(z)$ and $\bar{M}(x,t;z)\bar{a}^{-1}(z)$ are meromorphic in $D^{+}$ and $D^{-1}$,
and we now introduce reflection coefficients
\begin{equation}\label{SCS-9}
\rho(z)=b(z)a^{-1}(z),~~\bar{\rho}(z)=\bar{b}(z)\bar{a}^{-1}(z),~~z\in\Sigma_{0}.
\end{equation}

\subsection{Symmetries}
If the IST can be used to solve an initial-value problem, we must analyze the symmetry of the potential function.
This is because the symmetries of the eigenfunctions can be constructed by making use of the symmetries of the potential function.
The symmetries for the IST with NZBCs are more complicated,
since $\lambda(k)$ changes sign from one sheet of the Riemann surface to the other,
i.e.,$\lambda_{II}(k)=-\lambda_{I}(k)$.
According to the uniformization variable $z$, we need know the following results:\\
(1): (same sheet) $z\mapsto z^{\ast}$ means $(k,\lambda)\mapsto(k^{\ast},\lambda^{\ast})$ ;\\
(2): (opposite sheets) $z\mapsto\sigma k_{0}^2/z$  (outside/inside $\mathbb{C}_{o}$) means $(k,\lambda)\mapsto (k,-\lambda)$.\\

Both these transformation correspond
to symmetries of the scattering problem.
The previous one is the conjugate symmetry in the potential,
$\underline{\mathcal {Q}}^{\dag}=\sigma \underline{\mathcal {Q}}$.
Then
the second one is a straightforward consequence of the branching of
the scattering parameter k-plane.
Besides, we must introduce a third symmetry that corresponds to assuming
$\mathcal {Q}^{T}=\mathcal {Q}$,  which in view of $\underline{\mathcal {Q}}$ admits
\begin{equation}\label{SYMM-1}
\underline{\mathcal {Q}}=-\sigma_{2}\underline{\mathcal {Q}}^{T}\sigma_{2},~~\sigma_{2}=\left(
                                            \begin{array}{cc}
                                              \textbf{0} & i\mathcal {I}_{m} \\
                                              -i\mathcal {I}_{m} & \textbf{0} \\
                                            \end{array}
                                          \right),
\end{equation}
$\sigma_{2}$ being a $2m\times 2m$ generalization of the $2\times2$ Pauli matrix $\sigma_{2}$.

%In the remainder of this section we will discuss all three symmetries,
%taking into account that: (i) unlike the case of ZBCs, after removing the asymptotic oscillations, the Jost eigenfunctions do not tend to the
%identity matrix; (ii) the matrix nature of the equation implies that in some cases the symmetries one obtains for the eigenfunctions are
%bilinear symmetry relations.

\subsubsection{First symmetry}

Following the same process as in \cite{yzy-cnsns,MJBP-2004}
we consider the relationship of the scattering data and eigenfunctions for the matrix
equation with ZBCs when the above involution is investigated.
Let us introduce for $z\in\Sigma$
\begin{equation*}\label{FSYMM-1}
f(x,t;z)=\Phi^{\dag}(x,t;z^{*})J_{\sigma}\Phi(x,t;z),~~g(x,t;z)=\Psi^{\dag}(x,t;z^{\ast})J_{\sigma}\Psi(x,t;z),
\end{equation*}
with
\begin{equation}\label{FSYMM-2}
J_{\sigma}=\left(
             \begin{array}{cc}
               \mathcal {I}_{m} & \textbf{0} \\
               \textbf{0} & -\sigma \mathcal {I}_{m}\\
             \end{array}
           \right),
\end{equation}
where $J_{\sigma}$ is the $2m\times 2m$ identity in the $\sigma=-1$ case.
%and it coincides with $\sigma_{3}$ in the $\sigma=-1$ case.
Since $\Phi$, $\Psi$ are solutions
of the scattering problem \eqref{LP-1},  it is not hard to check that $f$, $g$ are independent of $x$.
Then by evaluating the limits as $x\rightarrow\pm\infty$, we find
\begin{equation}\label{FSYMM-3}
\Phi^{\dag}(x,t;z^{\ast})J_{\sigma}\Phi(x,t;z)=\Psi^{\dag}(x,t;z^{\ast})J_{\sigma}\Psi(x,t;z)=\gamma(z)J_{\sigma}.
\end{equation}
On the one hand, we write the above relations as
\begin{equation}\label{FSYMM-4}
\Psi^{-1}(x,t;z)=\frac{1}{\gamma(z)}J_{\sigma}\Psi^{\dag}(x,t;z^{\ast})J_{\sigma},
~~\Phi^{-1}(x,t;z)=\frac{1}{\gamma(z)}J_{\sigma}\Phi^{\dag}(x,t;z^{\ast})J_{\sigma}.
\end{equation}
Then the following representation for the scattering matrix can be otained
\begin{equation}\label{FSYMM-5}
\mathcal {S}(z)=\Psi^{-1}(x,t;z)\Phi(x,t;z)=\frac{1}{\gamma(z)}J_{\sigma}\Psi^{\dag}(x,t;z^{\ast})J_{\sigma}\Phi(x,t;z).
\end{equation}

For the sake of convenience, we introduce the following notation for the upper/lower blocks of the eigenfunctions
\begin{equation*}\label{FSYMM-6}
\Phi(x,t;z)=\left(
              \begin{array}{cc}
                \phi^{\mbox{up}} & \bar{\phi}^{\mbox{up}} \\
                \phi^{\mbox{dn}} & \bar{\phi}^{\mbox{dn}} \\
              \end{array}
            \right),~~\Psi(x,t;z)=\left(
                                    \begin{array}{cc}
                                      \bar{\psi}^{\mbox{up}} & \psi^{\mbox{up}} \\
                                      \bar{\psi}^{\mbox{dn}} & \psi^{\mbox{dn}} \\
                                    \end{array}
                                  \right),
\end{equation*}
where each block $^{\mbox{up}}$, $^{\mbox{dn}}$ represents an $m\times m$ matrix.
Then solving the $m\times m$ blocks of $\mathcal {S}(z)$ in \eqref{FSYMM-5} and comparing with \eqref{SCS-3} yields
\begin{align}\label{FSYMM-7}
&\gamma(z)a(z)=\left(\bar{\psi}^{\mbox{up}}\left(x,t;z^{\ast}\right)\right)^{\dag}\phi^{\mbox{up}}(x,t;z)
-\sigma\left(\bar{\psi}^{\mbox{dn}}(x,t;z^{\ast})\right)^{\dag}\phi^{\mbox{dn}}(x,t;z),\notag\\
&\gamma(z)\bar{a}(z)=\left(\psi^{\mbox{dn}}\left(x,t;z^{\ast}\right)\right)^{\dag}\bar{\phi}^{\mbox{dn}}(x,t;z)
-\sigma\left(\psi^{\mbox{up}}(x,t;z^{\ast})\right)^{\dag}\bar{\phi}^{\mbox{up}}(x,t;z),\notag\\
&\gamma(z)b(z)=\left(\psi^{\mbox{dn}}\left(x,t;z^{\ast}\right)\right)^{\dag}\phi^{\mbox{dn}}(x,t;z)
-\sigma\left(\psi^{\mbox{up}}(x,t;z^{\ast})\right)^{\dag}\phi^{\mbox{up}}(x,t;z),\notag\\
&\gamma(z)\bar{b}(z)=\left(\bar{\psi}^{\mbox{up}}\left(x,t;z^{\ast}\right)\right)^{\dag}\bar{\phi}^{\mbox{up}}(x,t;z)
-\sigma\left(\bar{\psi}^{\mbox{dn}}(x,t;z^{\ast})\right)^{\dag}\bar{\phi}^{\mbox{dn}}(x,t;z).
\end{align}
From the above expressions, we find that $a(z)$ can be analytically continued in $D^{+}$,
and $\bar{a}(z)$ can be analytically continued in $D^{-}$.
Consequently, the following theorem can be easily established.\\

\noindent
\textbf{Theorem 4.}
Assume that $\mathcal {Q}-\mathcal {Q}_{+}\in L^{1,1}([x_{+},+\infty))$ and  $\mathcal {Q}-\mathcal {Q}_{-}\in L^{1,1}([-\infty,x_{-}))$
as matrix functions of $x$ for all $t\geq 0$, for some $x_{\pm}\in\mathbb{R}$,
then $\gamma(z)\mathcal {S}(z)$ is continuous for all $z\in\Sigma$, containing the branch points.
The functions $a(z)$, $\bar{a}(z)$, $b(z)$, $\bar{b}(z)$
have simple poles at the branch points $z=\pm\sqrt{\sigma}k_{0}$,  the following residue conditions can be obtained
\begin{align*}\label{FSYMM-9}
&\mathop{\mbox{Res}}\limits_{z=\pm ik_{0}}a(z)=\notag\\
&\pm\frac{k_{0}}{2}\left[\left(\bar{\psi}^{\mbox{up}}\left(x,t;\mp k_{0}\right)\right)^{\dag}\phi^{\mbox{up}}(x,t;\pm k_{0})
-\left(\bar{\psi}^{\mbox{dn}}\left(x,t;\mp k_{0}\right)\right)^{\dag}\phi^{\mbox{dn}}\left(x,t;\mp k_{0}\right)\right],\notag\\
&\mathop{\mbox{Res}}\limits_{z=\pm ik_{0}}\bar{a}(z)=\notag\\
&\pm\frac{k_{0}}{2}\left[\left(\psi^{\mbox{dn}}\left(x,t;\mp k_{0}\right)\right)^{\dag}\bar{\phi}^{\mbox{dn}}(x,t;\mp k_{0})
-\left(\psi^{\mbox{up}}\left(x,t;\mp k_{0}\right)\right)^{\dag}\bar{\phi}^{\mbox{up}}\left(x,t;\mp k_{0}\right)\right],\notag\\
&\lim_{z\rightarrow\pm k_{0}}(z\mp ik_{0})b(z)=\notag\\
&\pm\frac{k_{0}}{2}\left[\left(\psi^{\mbox{dn}}\left(x,t;\mp k_{0}\right)\right)^{\dag}\phi^{\mbox{dn}}(x,t;\pm k_{0})
-\left(\psi^{\mbox{up}}\left(x,t;\mp k_{0}\right)\right)^{\dag}\phi^{\mbox{up}}\left(x,t;\mp k_{0}\right)\right],\notag\\
&\lim_{z\rightarrow\mp k_{0}}(z\mp ik_{0})\bar{b}(z)=\notag\\
&\pm\frac{k_{0}}{2}\left[\left(\bar{\psi}^{\mbox{up}}\left(x,t;\mp k_{0}\right)\right)^{\dag}\bar{\phi}^{\mbox{up}}(x,t;\mp k_{0})
-\left(\bar{\psi}^{\mbox{dn}}\left(x,t;\mp k_{0}\right)\right)^{\dag}\bar{\phi}^{\mbox{dn}}\left(x,t;\mp k_{0}\right)\right],
\end{align*}
in the $\sigma=-1$ case, and
\begin{align*}
&\mathop{\mbox{Res}}\limits_{z=\pm k_{0}}a(z)=\notag\\
&\pm\frac{k_{0}}{2}\left[\left(\bar{\psi}^{\mbox{up}}\left(x,t;\pm k_{0}\right)\right)^{\dag}\phi^{\mbox{up}}(x,t;\pm k_{0})
-\left(\bar{\psi}^{\mbox{dn}}\left(x,t;\pm k_{0}\right)\right)^{\dag}\phi^{\mbox{dn}}\left(x,t;\pm k_{0}\right)\right],\notag\\
&\mathop{\mbox{Res}}\limits_{z=\pm k_{0}}\bar{a}(z)=\notag\\
&\pm\frac{k_{0}}{2}\left[\left(\psi^{\mbox{dn}}\left(x,t;\pm k_{0}\right)\right)^{\dag}\bar{\phi}^{\mbox{dn}}(x,t;\pm k_{0})
-\left(\psi^{\mbox{up}}\left(x,t;\pm k_{0}\right)\right)^{\dag}\bar{\phi}^{\mbox{up}}\left(x,t;\pm k_{0}\right)\right],\notag\\
&\lim_{z\rightarrow\pm k_{0}}(z\mp k_{0})b(z)=\notag\\
&\pm\frac{k_{0}}{2}\left[\left(\psi^{\mbox{dn}}\left(x,t;\pm k_{0}\right)\right)^{\dag}\phi^{\mbox{dn}}(x,t;\pm k_{0})
-\left(\psi^{\mbox{up}}\left(x,t;\pm k_{0}\right)\right)^{\dag}\phi^{\mbox{up}}\left(x,t;\pm k_{0}\right)\right],\notag\\
&\lim_{z\rightarrow\pm k_{0}}(z\mp k_{0})\bar{b}(z)=\notag\\
&\pm\frac{k_{0}}{2}\left[\left(\bar{\psi}^{\mbox{up}}\left(x,t;\pm k_{0}\right)\right)^{\dag}\bar{\phi}^{\mbox{up}}(x,t;\pm k_{0})
-\left(\bar{\psi}^{\mbox{dn}}\left(x,t;\pm k_{0}\right)\right)^{\dag}\bar{\phi}^{\mbox{dn}}\left(x,t;\pm k_{0}\right)\right],
\end{align*}
in the $\sigma=1$ case.
Additionally, under the condition $\det a(z)\neq0$, $\det \bar{a}(z)\neq0$ for $z\in\Sigma$,
the reflection coefficients $\rho(z)$ and $\bar{\rho}(z)$ in \eqref{SCS-9}
are a removable singularity at the branch points $z\pm\sqrt{\sigma}k_{0}$ for all $z\in\Sigma$.
%(thus including the branch points $z\pm\sqrt{\sigma}k_{0}$).

Eq.\eqref{FSYMM-3} claims that the off-diagonal blocks of
\begin{equation*}%\label{}
\Phi^{\dag}(z^{\ast})J_{\sigma}\Phi(z)=\Psi^{\dag}(z^{\ast})J_{\sigma}\Psi(z)
\end{equation*}
are zero, it then follows from \eqref{FSYMM-5} that
\begin{align}\label{FSYMM-10}
&\left(\phi^{\mbox{up}}\left(x,t;z^{\ast}\right)\right)^{\dag}\bar{\phi}^{\mbox{up}}(x,t;z)
=\sigma(\phi^{\mbox{dn}}(x,t;z^{\ast}))^{\dag}\bar{\phi}^{\mbox{dn}}(x,t;z),\notag\\
&\left(\psi^{\mbox{up}}\left(x,t;z^{\ast}\right)\right)^{\dag}\bar{\psi}^{\mbox{up}}(x,t;z)
=\sigma(\psi^{\mbox{dn}}(x,t;z^{\ast}))^{\dag}\bar{\psi}^{\mbox{dn}}(x,t;z),
\end{align}
In addition, Eq.\eqref{FSYMM-3} implies
\begin{equation}\label{FSYMM-11}
\mathcal {S}^{\dag}(z^{\ast})J_{\sigma}\mathcal {S}(z)=J_{\sigma},~~z\in\Sigma.
\end{equation}
Precisely, in view of $\mathcal {S}(z)$ in \eqref{SCS-3}
one has the same symmetries as in the case of ZBCs
\begin{align}\label{FSYMM-12}
&a^{\dag}(z^{\ast})a(z)-\sigma b^{\dag}(z^{\ast})b(z)=\mathcal {I}_{m},\notag\\
&a^{\dag}(z^{\ast})\bar{b}(z)-\sigma b^{\dag}(z^{\ast})\bar{a}(z)=\textbf{0},\notag\\
&\bar{b}^{\dag}(z^{\ast})a(z)-\sigma \bar{a}^{\dag}(z^{\ast})b(z)=\textbf{0},\notag\\
&\bar{b}^{\dag}(z^{\ast})\bar{b}(z)-\sigma \bar{a}^{\dag}(z^{\ast})\bar{a}(z)=\mathcal {I}_{m},
\end{align}
and consequently also the following symmetry %for the reflection coefficients
\begin{equation}\label{FSYMM-13}
\bar{\rho}(z)=\sigma\rho^{\dag}(z^{\ast}),~~z\in\Sigma,
\end{equation}
as well as
\begin{align}\label{FSYMM-14}
&a(z)a^{\dag}(z^{\ast})=\left[\mathcal {I}_{m}-\sigma\rho^{\dag}(z^{\ast})\rho(z)\right]^{-1},\notag\\
&\bar{a}(z)\bar{a}^{\dag}(z^{\ast})=\left[\mathcal {I}_{m}-\sigma\bar{\rho}^{\dag}(z^{\ast})\bar{\rho}(z)\right]^{-1}.
\end{align}
It follows from \eqref{FSYMM-11} that
\begin{equation}\label{FSYMM-15}
\mathcal {S}^{-1}(z)=J_{\sigma}\mathcal {S}^{\dag}(z^{\ast})J_{\sigma},~~S^{-1}(z)=\left(
                                                                                     \begin{array}{cc}
                                                                                       \bar{c}(z) & d(z) \\
                                                                                       \bar{d}(z) & c(z) \\
                                                                                     \end{array}
                                                                                   \right),
\end{equation}
which indicates a relationship between those of its inverse and the blocks of $\mathcal {S}(z)$ for $z\in\Sigma$, i.e.,
\begin{equation}\label{FSYMM-16}
\bar{c}(z)=a^{\dag}(z^{\ast}),~~c(z)=\bar{a}^{\dag}(z^{\ast}),~~d(z)=-\sigma b^{\dag}(z^{\ast}),~~\bar{d}(z)=-\sigma b^{\dag}(z^{\ast}).
\end{equation}
$\bar{c}(z)=a^{\dag}(z^{\ast})$, $c(z)=\bar{a}^{\dag}(z^{\ast})$ can be extended to $D^{+}$ and $D^{-}$ by Schwarz reflection principle, respectively,
while the $d(z)=-\sigma b^{\dag}(z^{\ast})$, $\bar{d}(z)=-\sigma b^{\dag}(z^{\ast})$ are generally only valid for $z \in\Sigma$.

In turn, similar to \eqref{SCS-6}, we have
\begin{align}\label{FSYMM-7}
&\det c(z)=\frac{\mbox{Wr}(\phi,\psi)}{\mbox{Wr}(\phi,\bar{\phi})}\equiv\frac{\det(\phi,\psi)}{\gamma^{m}},\notag\\
&\det \bar{c}(z)=\frac{\mbox{Wr}(\bar{\phi},\bar{\psi})}{\mbox{Wr}(\phi,\bar{\phi})}\equiv\frac{\det(\bar{\phi},\bar{\psi})}{\gamma^{m}},
\end{align}
which allow us conclude that
\begin{equation}\label{FSYMM-8}
\det a(z)=\det c(z)~~\mbox{for}~~ z\in D^{+},~~\det \bar{a}(z)=\det \bar{c}(z)~~\mbox{for}~~ z\in D^{-}.
\end{equation}
Finally, it follows from \eqref{FSYMM-16} that
\begin{equation}\label{FSYMM-9}
\det \bar{a}(z)=\det a^{\dag}(z^{\ast})\equiv(\det a(z^{\ast}))^{\ast}~~\mbox{for}~~z\in D^{-}.
\end{equation}

\subsubsection{Second symmetry}

%Because Jost eigenfunctions and scattering coefficients rely on $\lambda=\left(z-\sigma k_{0}^2/z\right)/2$,
%the symmetry should
%relate their values on opposite sheets of the Riemann surface (inside and outside the circle $\mathbb{C}_{o}$).
Actually, it is easy to check that
\begin{equation*}\label{SSYM1-1}
X_{\pm}(z)=-\frac{i}{z}X_{\pm}\left(\sigma k_{0}^2/z\right)\sigma_{3}\underline{\mathcal {Q}}_{\pm},
\end{equation*}
which, in view of that $\theta\left(\sigma k_{0}^2/z\right)=-\theta(z)$ and $\underline{\mathcal {Q}}_{\pm}e^{i\theta(z)\sigma_{3}}=e^{-i\theta(z)\sigma_{3}}\underline{\mathcal {Q}}_{\pm}$,
yields
\begin{align}\label{SSYM1-2}
&\Phi(x,t;z)=\Phi\left(x,t;\sigma k_{0}^2/z\right)\sigma_{3}\underline{\mathcal {Q}}_{-}/iz,\notag\\
&\Psi(x,t;z)=\Psi\left(x,t;\sigma k_{0}^2/z\right)\sigma_{3}\underline{\mathcal {Q}}_{+}/iz,
\end{align}
with $z\in\Sigma$.
The $2m \times m$ blocks of columns means
\begin{align}\label{SSYM1-3}
&\phi(x,t;z)=\frac{i\sigma}{z}\bar{\phi}\left(x,t;\sigma k_{0}^2/z\right)\mathcal {Q}_{-}^{\dag},~~
\bar{\phi}(x,t;z)=-\frac{i}{z}\phi\left(x,t;\sigma k_{0}^2/z\right)\mathcal {Q}_{-},\notag\\
&\bar{\psi}(x,t;z)=\frac{i\sigma}{z}\psi\left(x,t;\sigma k_{0}^2/z\right)\mathcal {Q}_{+}^{\dag},~~
\psi(x,t;z)=-\frac{i}{z}\bar{\psi}\left(x,t;\sigma k_{0}^2/z\right)\mathcal {Q}_{+},
\end{align}
and for all $z\in\Sigma$, it follows from \eqref{SCS-3} and \eqref{SSYM1-2} that
\begin{equation}\label{SSYM1-4}
\mathcal {S}\left(\sigma k_{0}^2/z\right)=\sigma_{3}\underline{\mathcal {Q}}_{+}\mathcal {S}(z)\underline{\mathcal {Q}}_{-}^{-1}\sigma_{3}
\equiv\frac{\sigma}{k_{0}^2}\sigma_{3}\underline{\mathcal {Q}}_{+}\mathcal {S}(z)\underline{\mathcal {Q}}_{-}\sigma_{3}.
\end{equation}
%where for the last equality we have used \eqref{LP-4},
%so that $Q_{\pm}^{-1}=Q_{\pm}^{\dag}/k_{0}^2=\sigma Q_{\pm}/k_{0}^2$.
In view of \eqref{SCS-3} and  \eqref{SSYM1-2}, we then have
\begin{align}\label{SSYM1-5}
&a\left(\sigma k_{0}^2/z\right)=\frac{1}{k_{0}^2}\mathcal {Q}_{+}\bar{a}(z)\mathcal {Q}_{-}^{\dag},~~
\bar{a}\left(\sigma k_{0}^2/z\right)=\frac{1}{k_{0}^2}\mathcal {Q}_{+}^{\dag}a(z)\mathcal {Q}_{-},\notag\\
&b\left(\sigma k_{0}^2/z\right)=-\frac{\sigma}{k_{0}^2}\mathcal {Q}_{+}^{\dag}\bar{b}(z)\mathcal {Q}_{-}^{\dag},~~
\bar{b}\left(\sigma k_{0}^2/z\right)=-\frac{\sigma}{k_{0}^2}\mathcal {Q}_{+}b(z)\mathcal {Q}_{-},
\end{align}
Finally, we obtain the corresponding symmetry
\begin{equation}\label{SSYM1-6}
\rho\left(\sigma k_{0}^2/z\right)=-\sigma \mathcal {Q}_{+}^{\dag}\bar{\rho}(z)\mathcal {Q}_{+}^{-1}
\equiv-\sigma/k_{0}^2\mathcal {Q}_{+}^{\dag}\bar{\rho}(z)\mathcal {Q}_{+}^{\dag},~~\forall z\in\Sigma.
\end{equation}

\subsubsection{Third symmetry}

Similar to the first symmetry in the potential ($\underline{\mathcal {Q}}=-\sigma_{2}\underline{\mathcal {Q}}^{T}\sigma_{2}$,  corresponding to $\mathcal {Q}^{T}=\mathcal {Q}$),
let us next introduce
\begin{equation*}\label{TSSS-1}
\tilde{f}(x,t;z)=\Phi^{T}(x,t;z)\sigma_{2}\Phi(x,t;z),~~\tilde{g}(x,t;z)=\Psi^{T}(x,t;z)\sigma_{2}\Psi(x,t;z)
\end{equation*}
for $z\in\Sigma$, where $\sigma_{2}$ is given by \eqref{SYMM-1}.
In addition, we can easily check that $\tilde{f}$, $\tilde{g}$ are independent of $x$.
For example, it follows from the scattering problem \eqref{LP-1} that
\begin{equation*}\label{TSSS-2}
\partial_{x}\tilde{f}=\Phi^{T}\left(-k\sigma_{3}\sigma_{2}+\underline{\mathcal {Q}}^{T}\sigma_{2}-ik\sigma_{2}\sigma_{3}
+\sigma_{2}\underline{\mathcal {Q}}\right)\Phi=\textbf{0},
\end{equation*}
since $\sigma_{3}\sigma_{2}=-\sigma_{2}\sigma_{3}$ and $Q^{T}\sigma_{2}=-\sigma_{2}Q$.
Evaluating the limits as $x\rightarrow\pm\infty$, one gets
\begin{equation}\label{TSSS-3}
\Phi^{T}(x,t;z)\sigma_{2}\Phi(x,t;z)=\Psi^{T}(x,t;z)\sigma_{2}\Psi(x,t;z)=\gamma(z)\sigma_{2},
\end{equation}
which indicates
\begin{equation}\label{TSSS-4}
\mathcal {S}^{T}(z)\sigma_{2}\mathcal {S}(z)=\sigma_{2},~~z\in\Sigma.
\end{equation}
It follows from the blocks of the scattering matrix that the latter gives
\begin{equation*}\label{TSSS-5}
b^{T}(z)a(z)=a^{T}(z)b(z),~~\bar{b}^{T}(z)\bar{a}(z)=\bar{a}^{T}(z)\bar{b}(z),~~a^{T}(z)\bar{a}(z)-b^{T}(z)\bar{b}(z)=\mathcal {I}_{m},
\end{equation*}
which then particularly mean
\begin{equation}\label{TSSS-6}
\rho^{T}(z)=\rho(z),~~\bar{\rho}^{T}(z)=\bar{\rho}(z),
\end{equation}
showing that the reflection coefficients should be symmetric matrices themselves, as well as
\begin{equation}\label{TSSS-7}
a(z)a^{T}(z)=\left(\mathcal {I}_{m}-\bar{\rho}(z)\rho(z)\right)^{-1},~~z\in\Sigma.
\end{equation}
At last, it also follows from \eqref{TSSS-4} that $\mathcal {S}^{-1}(z)=\sigma_{2}\mathcal {S}^{T}(z)\sigma_{2}$ for $z\in\Sigma$, i.e.,
\begin{equation}\label{TSSS-8}
\bar{c}(z)=\bar{a}^{T}(z),~~c(z)=a^{T}(z),~~d(z)=-\bar{b}^{T}(z),~~\bar{d}(z)=-b^{T}(z),
\end{equation}
which, in view of  \eqref{FSYMM-16}, reach to
\begin{equation}\label{TSSS-9}
\bar{a}(z)=a^{\ast}(z^{\ast}),~~\bar{b}(z)=\sigma b^{\ast}(z^{\ast}),~~z\in\Sigma.
\end{equation}
Summarizing the results of Section 2.4,  we have following proposition. %regarding the symmetries of the scattering data.
\\

\noindent
\textbf{Proposition 5.}
If $\mathcal {Q}-\mathcal {Q}_{+}\in L^{1}([x_{+},+\infty))$
and $\mathcal {Q}-\mathcal {Q}_{-}\in L^{1}((-\infty,x_{-}])$ as matrix functions of $x$ for all $t\geq 0$,
for some $x_{\pm}\in\mathbb{R}$. For all  $z\in\Sigma_{0}$, the coefficients $\rho(z)$ and $\bar{\rho}(z)$
given in view of the blocks of the scattering matrix $\mathcal {S}(z)$ by \eqref{SCS-9} admit
\begin{equation*}\label{TSSS-10}
\rho(z)=\sigma\rho^{\ast}(z^{\ast}),~~\rho\left(\sigma k_{0}^2/z\right)=-\frac{\sigma}{k_{0}^2}\mathcal {Q}_{+}^{\dag}\bar{\rho}(z)\mathcal {Q}_{+}^{\dag}.
\end{equation*}
Furthermore, for $z\in D^{-}\bigcup\Sigma_{0}$,
the diagonal blocks of the scattering matrix $a(z)$ and $\bar{a}(z)$ satisfy
\begin{align*}\label{TSSS-11}
&\det \bar{a}(z)=\det a^{\dag}(z^{\ast}),\notag\\
&a\left(\sigma k_{0}^2/z\right)=\frac{1}{k_{0}^2}\mathcal {Q}_{+}\bar{a}(z)\mathcal {Q}_{-}^{\dag}\Rightarrow
\det \bar{a}(z)=\frac{k_{0}^{2m}}{\det\mathcal {Q}_{+}\left(\det\mathcal {Q}_{-}\right)^{\ast}}\det a\left(\sigma k_{0}^2/z\right).
\end{align*}
If, moreover, $\mathcal {Q}(x,t)$ is a symmetric matrix, we then have
\begin{equation*}\label{TSSS-12}
\rho^{T}(z)=\rho(z),~~\bar{\rho}^{T}(z)=\bar{\rho}(z),~~z\in\Sigma_{0},
\end{equation*}
and
\begin{equation*}\label{TSSS-13}
\bar{a}(z)=a^{*}(z^{*}),~~z\in D^{-}\bigcup\Sigma_{0}.
\end{equation*}
Assume that $\mathcal {Q}-\mathcal {Q}_{+}\in\left([x_{+},+\infty)\right)$ and $\mathcal {Q}-\mathcal {Q}_{-}\in L^{1,1}\left((-\infty,x_{-}]\right)$
as matrix functions of $x$ for all $t\geq0$, for some $x_{\pm}\in\mathbb{R}$,
the above three symmetries also extend to contain the branch points, and consequently are valid for
$z\in\Sigma$, and $z\in D^{-}\bigcup\Sigma$, respectively.

\subsection{Discrete spectrum and residue conditions}

%The discrete spectrum of the scattering problem is the set of all values $z\in\mathbb{C}\setminus\Sigma$
%such that the scattering problem admits eigenfunctions in $L^2(\mathbb{R})$.
Similar to \cite{IST-2017},
these discrete spectral points are the zeros of the functions $\det a(z)$ and $\det \bar{a}(z)$
in $D^{+}$ and  $D^{-}$, respectively.
Suppose that $\det a(z)$ admits a finite number $\mathcal {N}$ of simple zeros $z_{1},\ldots,z_{\mathcal {N}}$
in $D^{+}\bigcap\{z\in\mathbb{C}:\mbox{Im}z>0\}$.
That is to say, let $\det a(z_{n})=0$ and $(\det a)'(z_{n})\neq 0$, with $|z_{n}|> k_{0}$ and $\mbox{Im}z_{n}>0$ for $n=1,2,\ldots,\mathcal {N}$,
and in which the prime represents differentiation with respect to $z$.
From the symmetries \eqref{FSYMM-9} and \eqref{SSYM1-5}, it follows that
\begin{equation}\label{DSRC-1}
\det a(z_{n})=0\Leftrightarrow \det \bar{a}(z_{n}^{\ast})=0\Leftrightarrow\det \bar{a}\left(\sigma k_{0}^2/z_{n}\right)=0
\Leftrightarrow\det a\left(\sigma k_{0}^2/z_{n}\right)=0.
\end{equation}
For each $n=1,2,\ldots,\mathcal {N}$, we thus have a quartet of discrete eigenvalues,
which indicates that the discrete spectrum is expressed by the set
\begin{equation}\label{DSRC-2}
Z=\left\{z_{n},z_{n}^{\ast},\sigma k_{0}^2/z_{n},\sigma k_{0}^2/z_{n}^{\ast}\right\}_{n=1}^{\mathcal {N}}.
\end{equation}
%However, in the $\sigma=1$ case,
%the self-adjointness of the scattering problem indicates that the discrete eigenvalues are real in the $k$ plane.
%Since spectral singularities are excluded, then eigenvalues must be such that $-k_{0}<k_{j}<k_{0}$
%which means that in the $z$ plane the eigenvalues are all located on the circle of radius $k_{0}$, i.e., $z_{j}\in \mathbb{C}_{o}\setminus\{\pm k_{0}\}$
%for $j=1,2,\ldots,\mathcal {N}$,  and they come in complex conjugate pairs. Therefore, in the defocusing case the second symmetry does not give rise to
%a quartet of eigenvalues, because if $\sigma=1$ and $z_{j}\in \mathbb{C}_{o}$, then $k_{0}^2/z_{n}^{\ast}\equiv z_{n}$
%and, obviously, $k_{0}^2/z_{n}\equiv z_{n}^{\ast}$.
Then,
the discrete spectrum is expressed by
\begin{align}\label{DSRC-3}
&\sigma=-1~~(\mbox{focusing case}):~~Z=\left\{z_{n},-k_{0}^2/z_{n}^{\ast},z_{n}^{\ast},-k_{0}^2/z_{n}\right\}_{1}^{\mathcal {N}},\notag\\
&\sigma=1~~~~(\mbox{defocusing case}):~~~Z=\{\zeta_{n},\zeta_{n}^{\ast}\}_{1}^{\mathcal {N}},
\end{align}
where in
the defocusing ($\sigma=1$) case the eigenvalues are on the circle $\mathbb{C}_{o}$;
in the focusing ($\sigma=-1$) case each first pair is in $D^{+}$ %(and we assume without loss of generality $\mbox{Im}z_{n}>0$)
and each second pair is in $D^{-}$;.

Now suppose that $\det a(z)$ admits $\mathcal {N}$ simple zeros $z_{n}$ $(n=1,2,\ldots,N)$,
namely, $\det a(z_{n})=0$,
which implies that from Eq.\eqref{SCS-6} the Jost eigenfunctions $\psi(x,t;z_{n})$ and
$\phi(x,t;z_{n})$ are linearly dependent.
Therefore there is a nonzero constant $b_{n}$ that admits the following equation
\begin{equation}\label{DSRC-4}
\phi\left(x,t;z_{n}\right)=\psi\left(x,t;z_{n}\right)b_{n},~~\bar{\phi}\left(x,t;z_{n}^{\ast}\right)=\bar{\psi}\left(x,t;z_{n}^{\ast}\right)\bar{b}_{n},
\end{equation}
where $b_{n}$, $b_{n}$ are $m\times m$ non-zero constant matrices.

In the following we construct the residue conditions that will be required for the inverse problem.
In view of \eqref{DSRC-4}, we have $M(x,t;z_{n})=e^{2i\theta(x,t;z_{n})}N(x,t;z_{n})b_{n}$.
As a result, we have the following residue condition in the context of a simple zero of $\det a(z)$
\begin{equation}\label{DSRC-5}
\mathop{\mbox{Res}}\limits_{z=z_{n}}\left[M(x,t;z)a^{-1}(z)\right]=e^{-2i\theta(x,t;z_{n})}N(x,t;z_{n})C_{n},~~
C_{n}=\frac{b_{n}\alpha(z_{n})}{(\det a)'(z_{n})},
\end{equation}
where $\alpha(z):=\mbox{cof}a(z)$ is the cofactor (or adjugate) matrix of $a(z)$. %i.e.,
%such that $a(z)\alpha(z)=\alpha(z)a(z)=\det a(z)\mathcal {I}_{m}$.
Following a similar way, if $z_{n}^{\ast}\in D^{-}$  is a simple zero of $\det \bar{a}(z)$  we also get
\begin{equation}\label{DSRC-6}
\mathop{\mbox{Res}}\limits_{z=z_{n}^{\ast}}\left[\bar{M}(x,t;z)\bar{a}^{-1}(z)\right]
=e^{2i\theta(x,t;z_{n}^{\ast})}\bar{N}(x,t;z_{n}^{\ast})\bar{C}_{n},~~
\bar{C}_{n}=\frac{b_{n}\alpha(z_{n}^{\ast})}{(\det a)'(z_{n}^{\ast})},
\end{equation}
where $\bar{\alpha}(z)$ denotes the cofactor matrix of $\bar{a}(z)$.

As is well known, for an $m\times m$  matrix $A$, one obtains $\det(\mbox{cof} A)=(\det A)^{m-1}$,
generally, so
\begin{equation*}\label{DSRC-7}
\det \alpha(z)=(\det a(z))^{m-1},
\end{equation*}
which, particularly, implies
\begin{equation*}\label{DSRC-8}
\det\alpha(z)=\det a(z),
\end{equation*}
for the special case $m=2$.
As a result,  $\det\alpha(z)$ admits a zero of the same order as $\det a(z)$ for each $z_{n}\in D^{+}\bigcap Z$ in the physically relevant case.
The same of course holds for $\det\bar{\alpha}(z)$,  which will have a zero of the same order as $\det \bar{a}(z)$
for each $z_{n}^{\ast}\in D^{-}\bigcap Z$.

For the simple eigenvalues, we also have
\begin{equation}\label{DSRC-9}
\tau_{n}:=\mathop{\mbox{Res}}\limits_{z=z_{n}}a^{-1}(z)=\frac{\alpha(z)}{(\det a)'(z_{n})},~~
\bar{\tau}_{n}:=\mathop{\mbox{Res}}\limits_{z=z_{n}^{\ast}}\bar{a}^{-1}(z)=\frac{\bar{\alpha}(z)}{(\det \bar{a})'(z_{n}^{\ast})},
\end{equation}
and $\det \tau_{n}=\det\bar{\tau}_{n}=0$,
so in the case of simple eigenvalues, the residues are always rank m-1 matrices.
We then show the norming constants presented in \eqref{DSRC-5} and \eqref{DSRC-6} in view of the above residues
\begin{equation}\label{DSRC-10}
C_{n}=b_{n}\tau_{n},~~\bar{C}_{n}=\bar{b}_{n}\bar{\tau}_{n},
\end{equation}
and we think that for simple discrete eigenvalues one knows
\begin{equation}\label{DSRC-11}
\det C_{n}=\det \bar{C}_{n}=0,
\end{equation}
so for simple zeros of $\det a(z)$, the norming constants are rank m-1 matrices.

It is also worth to point out that since for any $z\in D^{+}\setminus Z$,
one has
\begin{equation*}%\label{}
a^{-1}(z)=\alpha(z)/(\det a(z)).
\end{equation*}
Because $\alpha(z)$ is analytic in $D^{+}$,
we find that $a^{-1}(z)$ will be meromorphic in $D^{+}$,
with poles at each of the discrete eigenvalues, and the order of the pole at each $z_{n}$
is at most equal to the order of $z_{n}$ as a zero of $\det a(z)$.
Obviously, the same holds for $\bar{a}^{-1}(z)$ in $D^{-1}$.

If $z_{n}$ is a second order zero of $\det a(z)$,
then $\det \alpha(z)$ admits zero of order $2(m-1)$ at $z_{n}$.
However, in a neighborhood of $z_{n}$ one has
\begin{equation*}\label{DSRC-12}
a^{-1}(z)=\frac{1}{(z-z_{n})^2}\tau_{n,2}+\frac{1}{z-z_{n}}\tau_{n,1}+\tilde{a}(z),
\end{equation*}
where $\tilde{a}(z)$ is analytic at $z_{n}$, and
\begin{align}\label{DSRC-13}
&\tau_{n,2}=\lim_{z\rightarrow z_{n}}(z-z_{n})^2a^{-1}(z)\equiv\frac{2}{(\det a)''(z_{n})},\notag\\
&\tau_{n,1}=\lim_{z\rightarrow z_{n}}\frac{d}{dz}\left[\left(z-z_{n}\right)^2a^{-1}(z)\right]\equiv\frac{2}{(\det a)''(z_{n})}\alpha'(z_{n})-
\frac{2}{3}\frac{(\det a)'''(z_{n})}{((\det a)''(z_{n}))^2}\alpha(z_{n}).
\end{align}
%For a genuine second order zero, we find that $\tau_{n,2}\neq 0_{m}$ and $\det \tau_{n,2}=0$.
%On the other hand, $\tau_{n,1}$ might or might not be zero,  and if it is non-zero it is possible to have
%$\det\tau_{n,1}\neq 0$. On the other hand, it
%is also possible to have $\tau_{n,2}=0_{m}$, which is equivalent to $\alpha(z_{n})=0_{m}$.
%Here one would still have a first order pole for $a^{-1}(z)$,  with
%residue
%\begin{equation}\label{DSRC-14}
%\tau_{n,1}=\frac{2}{(\det a)''(z_{n})}\alpha'(z_{n}).
%\end{equation}
%Importantly, in this case in general $\det(\alpha')(z_{n})$ needs not to be zero, so $\tau_{n,1}$ needs not to be rank 1.
%In this situation, $z_{n}$ is a double zero of $\det a(z)$, and 0 is an eigenvalue of $a(z_{n})$
%with algebraic and geometric multiplicity equal to 2.
Since $\tau_{n,2}=0_{m}$, $a^{-1}(z)$ admits only a pole of first order at $z_{n}$,
and \eqref{DSRC-5}, Eq.\eqref{DSRC-6} reads
\begin{align}\label{DSRC-15}
&\mathop{\mbox{Res}}\limits_{z=z_{n}}\left[M(x,t;z)a^{-1}(z)\right]=e^{2i\theta(x,t;z_{n})}N(x,t;z_{n})C_{n},
~~C_{n}=\frac{2}{(\det a)''(z_{n})}b_{n}\alpha'(z_{n}),\notag\\
&\mathop{\mbox{Res}}\limits_{z=z_{n}^{\ast}}\left[\bar{M}(x,t;z)\bar{a}^{-1}(z)\right]=e^{-2i\theta(x,t;z_{n}^{\ast})}\bar{N}(x,t;z_{n}^{\ast})\bar{C}_{n},
~~\bar{C}_{n}=\frac{2}{(\det \bar{a})''(z_{n}^{\ast})}b_{n}\bar{\alpha}'(z_{n}^{\ast}).
\end{align}
%and in this case $C_{n}$, $\bar{C}_{n}$ need not be rank $m-1$.

The norming constants are related by the above symmetries.
In the following, we analyze the symmetry relationship between these norming constants $C_{n}$ and $\bar{C}_{n}$,
\begin{equation}\label{DSRC-16}
\bar{C}_{n}=\sigma C_{_{n}}^{\dag}.
\end{equation}
Furthermore, the third symmetry also claims that $C_{n}$ and $\bar{C}_{n}$ be symmetric matrices
\begin{equation}\label{DSRC-17}
C_{n}=C_{n}^{T},~~\bar{C}_{n}=\bar{C}_{n}^{T}.
\end{equation}
In the ($\sigma=-1$) focusing case we should discuss the remaining two points of the eigenvalue quartet.
Similar to \eqref{DSRC-4}, we introduce
\begin{align}\label{DSRC-18}
&\phi\left(x,t;\hat{z}_{n}\right)=\psi\left(x,t;\hat{z}_{n}\right)\hat{b}_{n},~~\hat{z}_{n}=\sigma k_{0}^2/z_{n}^{\star},\notag\\
&\bar{\phi}\left(x,t;\hat{z}_{n}^{\ast}\right)=\bar{\psi}\left(x,t;\hat{z}_{n}^{\ast}\right)\hat{\bar{b}}_{n},~~\hat{z}_{n}^{\ast}=\sigma k_{0}^2/z_{n}.
\end{align}
This second symmetry for the discrete eigenvalues and associated norming
constants only applies to the $\sigma=-1$ (focusing) case, thus for the rest of this section one choose $\sigma=-1$.

Using \eqref{SSYM1-3} and \eqref{DSRC-18}, we find
\begin{equation*}\label{DSRC-19}
\phi(x,t;z_{n})=\frac{i\sigma}{z_{n}}\bar{\phi}\left(x,t;\hat{z}_{n}^{\ast}\right)\mathcal {Q}_{-}^{\dag}
=\frac{i\sigma}{z_{n}}\psi\left(x,t;\hat{z}_{n}^{\ast}\right)\hat{\bar{b}}_{n}\mathcal {Q}_{-}^{\dag}.
\end{equation*}
Following the similar idea, we then have
\begin{equation*}\label{DSRC-20}
\phi\left(x,t;z_{n}\right)=\psi\left(x,t;z_{n}\right)b_{n}=-\frac{i}{z_{n}}\bar{\psi}\left(x,t;\hat{z}_{n}^{\ast}\right)\mathcal {Q}_{+}b_{n}.
\end{equation*}
From above two expression, we finally get
\begin{equation}\label{DSRC-21}
\hat{\bar{b}}_{n}=-\sigma\mathcal {Q}_{+}^{\dag}b_{n}\left(\mathcal {Q}_{-}^{\dag}\right)^{-1}\equiv-\frac{\sigma}{k_{0}^2}\mathcal {Q}_{+}b_{n}\mathcal {Q}_{-}.
\end{equation}
In analogous of \eqref{DSRC-21}, it follows from \eqref{SSYM1-3} and \eqref{DSRC-4} that
\begin{equation}\label{DSRC-22}
\bar{b}_{n}=-\sigma\mathcal {Q}_{+}b_{n}\left(\mathcal {Q}_{-}\right)^{-1}
\equiv-\frac{\sigma}{k_{0}^2}\mathcal {Q}_{+}^{\dag}\bar{b}_{n}\mathcal {Q}_{-}^{\dag}.
\end{equation}
Besides, differentiating \eqref{SSYM1-5} with respect to $z$ and evaluating at $z=z_{n}$ (or $z=z_{n}^{\ast}$), we find
\begin{align}\label{DSRC-23}
&(\det a)'\left(\sigma k_{0}^2/z_{n}^{\ast}\right)=-\sigma\left(\frac{z_{n}^{\ast}}{k_{0}}\right)^2\frac{\det\mathcal {Q}_{+}\det\mathcal {Q}_{-}^{\dag}}{k_{0}^{2m}}(\det \bar{a})'\left(z_{n}^{\ast}\right),\notag\\
&(\det \bar{a})'\left(\sigma k_{0}^2/z_{n}\right)=-\sigma\left(\frac{z_{n}}{k_{0}}\right)^2\frac{\det\mathcal {Q}_{+}^{\dag}\det\mathcal {Q}_{-}}{k_{0}^{2m}}\left(\det \bar{a})'(z_{n}\right).
\end{align}
It also follows from \eqref{SSYM1-5} that
\begin{align}\label{DSRC-24}
&\alpha\left(\sigma k_{0}^2/z_{n}^{\ast}\right)=\frac{1}{k_{0}^2}cof\left(\mathcal {Q}_{-}^{\dag}\right)\bar{\alpha}(z_{n}^{\ast})cof\left(\mathcal {Q}_{+}\right),\notag\\
&\bar{\alpha}\left(\sigma k_{0}^2/z_{n}\right)=\frac{1}{k_{0}^2}cof\left(\mathcal {Q}_{+}^{\dag}\right)\alpha(z_{n}^{\ast})cof\left(\mathcal {Q}_{-}\right).
\end{align}
Summarizing these relations, we then obtain
\begin{align}\label{DSRC-25}
&\mathop{\mbox{Res}}\limits_{z=\hat{z}_{n}
\equiv\sigma k_{0}^2/z_{n}^{\ast}}\left[M(x,t;z)a^{-1}(z)\right]=e^{-2i\theta(x,t;\hat{z}_{n})}N(x,t;\hat{z}_{n})\hat{C}_{n},\notag\\
&\mathop{\mbox{Res}}\limits_{z=\hat{z}^{\ast}_{n}
\equiv\sigma k_{0}^2/z_{n}}\left[\bar{M}(x,t;z)\bar{a}^{-1}(z)\right]=e^{2i\theta(x,t;\hat{z}_{n}^{\ast})}\bar{N}(x,t;\hat{z}_{n}^{\ast})\hat{\bar{C}}_{n},
\end{align}
where the norming constants $\hat{C}_{n}$ admit the following relations
\begin{equation}\label{DSRC-26}
\hat{C}_{n}=\frac{1}{\left(z_{n}^{\ast}\right)^2}\mathcal {Q}_{+}^{\dag}\bar{C}_{n}\mathcal {Q}_{+}^{\dag},~~
\hat{\bar{C}}_{n}=\frac{1}{z_{n}^2}\mathcal {Q}_{+}^{\dag}\bar{C}_{n}\mathcal {Q}_{+}.
\end{equation}
Note that $\hat{\bar{C}}_{n}=\sigma \hat{C}_{n}^{\dag}$.

Summarizing the results of subsection 2.5 regarding the discrete scattering data,
the following proposition holds. \\

\noindent
\textbf{Proposition 6.}
The discrete spectrum of the scattering problem \eqref{LP-1} is defined by
\begin{align*}\label{DSRC-27}
&\sigma=-1:~~Z=\left\{z_{n},-k_{0}^2/z_{n}^{\ast}, z_{n}^{\ast},-k_{0}^2/z_{n}\right\}_{n=1}^{\mathcal {N}},\notag\\
&\sigma=1:~~~~Z=\left\{\zeta_{n},\zeta_{n}^{\ast}\right\}_{n=1}^{\mathcal {N}},
\end{align*}
where in the $\sigma=1$ case the eigenvalues are on the circle $\mathbb{C}_{o}$.
In the $\sigma=-1$ case each first pair is in $D^{+}$, and $\mbox{Im}z_{n}>0$ and each second pair is in $D^{-}$.
In the $\sigma=1$ case the eigenvalues are on the circle $\mathbb{C}_{o}$ (Fig.2).
%((and we assume without loss of generality $\mbox{Im}z_{n}>0$) and each second pair is in $D^{-}$.

Discrete eigenvalues that correspond to simple poles of
$a^{-1}(z)$ in $D^{+}$ and $\bar{a}^{-1}(z)$ in $D^{-}$,
one considers a pair or a quartet of norming constants %$\{C_{n},\bar{C}_{n}\}$ and $\{C_{n},\bar{C}_{n},\hat{C}_{n},\hat{\bar{C}}_{n}\}$
such that
\begin{align*}
&\mathop{\mbox{Res}}\limits_{z=z_{n}}\left[M(x,t;z)a^{-1}(z)\right]=e^{-2i\theta\left(x,t;z_{n}\right)}N(x,t;z_{n})C_{n},\notag\\
&\mathop{\mbox{Res}}\limits_{z=z_{n}^{\ast}}\left[\bar{M}(x,t;z)\bar{a}^{-1}(z)\right]
=e^{2i\theta(x,t;z_{n}^{\ast})}\bar{N}\left(x,t;z_{n}^{\ast}\right)\bar{C}_{n},\notag\\
&\mathop{\mbox{Res}}\limits_{z=\hat{z}_{n}
\equiv\sigma k_{0}^2/z_{n}^{\ast}}\left[M(x,t;z)a^{-1}(z)\right]=e^{-2i\theta\left(x,t;\hat{z}_{n}\right)}N(x,t;\hat{z}_{n})\hat{C}_{n},~~
\hat{C}_{n}=\frac{1}{(z_{n}^{\ast})^2}\mathcal {Q}_{+}^{\dag}\bar{C}_{n}\mathcal {Q}_{+}^{\dag},\notag\\
&\mathop{\mbox{Res}}\limits_{z=\hat{z}_{n}^{\ast}
\equiv\sigma k_{0}^2/z_{n}}\left[\bar{M}(x,t;z)\bar{a}^{-1}(z)\right]
=e^{2i\theta\left(x,t;\hat{z}_{n}^{\ast}\right)}\bar{N}\left(x,t;\hat{z}_{n}^{\ast}\right)\hat{\bar{C}}_{n},~~
\hat{\bar{C}}_{n}=\frac{1}{z_{n}^2}\mathcal {Q}_{+}^{\dag}C_{n}\mathcal {Q}_{+}.
\end{align*}

In the $\sigma=1$ case only the first two equations are discussed.
Generally, when the discrete eigenvalues are simple zeros, for $m=2$, which indicates that the norming constants are a matrix with rank 1.

\subsection{Generalized norming constants}

Let us first introduce the $2m \times2m$ matrix solutions of \eqref{LP-1}
\begin{equation}\label{GNCC-1}
P(x,t;z)=(\phi(x,t;z),\psi(x,t;z)),~~\bar{P}(x,t;z)=(\bar{\phi}(x,t;z),\bar{\psi}(x,t;z)).
\end{equation}
where $P(x,t;z)$ is analytic for $z\in D^{+}$ and $\bar{P}(x,t;z)$ is analytic for $z\in D^{-}$.
One then discuss the bilinear combinations
$A_{\sigma}(z)=\bar{P}^{\dag}(x,t;z^{\ast})J_{\sigma}P(x,t;z)$ and
$A^{\dag}_{\sigma}(z^{\ast})=P^{\dag}(x,t;z^{\ast})J_{\sigma}\bar{P}(x,t;z)$,
which are analytic in $D^{+}$ and $D^{-}$, respectively, and independent of $x$.
Computing the $m\times m$ blocks of $P^{\dag}(x,t;z^{\ast})J_{\sigma}\bar{P}(x,t;z)$
and $\bar{P}^{\dag}(x,t;z^{\ast})J_{\sigma}P(x,t;z)$ in view of the blocks of the eigenfunctions,
and it follows from \eqref{FSYMM-7} and \eqref{FSYMM-10} that
\begin{align}\label{GNCC-2}
&A_{\sigma}(z)=\bar{P}^{\dag}(x,t;z^{*})J_{\sigma}P(x,t;z)\equiv\left(
                                                                 \begin{array}{cc}
                                                                   \gamma(z)a(z) & \textbf{0} \\
                                                                   \textbf{0} & -\sigma\gamma^{*}(z^{\ast})\bar{a}^{\dag}(z^{\ast}) \\
                                                                 \end{array}
                                                               \right),~~z\in D^{+},\notag\\
&A^{\dag}_{\sigma}(z^{\ast})=P^{\dag}(x,t;z^{*})J_{\sigma}\bar{P}(x,t;z)\equiv\left(
                                                                 \begin{array}{cc}
                                                                   \gamma^{\ast}(z^{\ast})a^{\dag}(z^{\ast}) & \textbf{0} \\
                                                                   \textbf{0} & -\sigma\gamma(z)\bar{a}(z) \\
                                                                 \end{array}
                                                               \right),~~z\in D^{-}.
\end{align}
Next we discuss that $z_{n}\in D^{+}$ is the simple zero of the $\det a(z)$,
we easily obtain the $\det \bar{a}(z^{\ast}_{n})=0$ with  $(\det \bar{a})'(z^{\ast}_{n})\neq0$
from the fist symmetry.
Now for free $m$ let $\chi_{n}\in\mathbb{C}^{2m}\setminus\{0\}$ be
a right null vector of $P(x,t;z_{n})$, i.e., $\chi_{n}\in \mbox{ker} P(x,t;z_{n})$.
If we show
\begin{equation*}\label{GNCC-3}
\chi_{n}=\left(
           \begin{array}{c}
             \chi_{n}^{\mbox{up}} \\
             \chi_{n}^{\mbox{dn}} \\
           \end{array}
         \right),~~\chi_{n}^{\mbox{up}},\chi_{n}^{\mbox{dn}}\in\mathbb{C}^{m},
\end{equation*}
then it follows from the definition \eqref{GNCC-1} of $P(x,t;z_{n})$ that
\begin{equation*}\label{GNCC-4}
\phi(x,t;z_{n})\chi_{n}^{\mbox{up}}+\psi(x,t;z_{n})\chi_{n}^{\mbox{dn}}=0_{2m\times m},
\end{equation*}
showing that a right null vector of $P(x,t;z_{n})$ arrives at
\begin{equation}\label{GNCC-5}
\phi(x,t;z_{n})\eta_{n}=\psi(x,t;z_{n})\xi_{n},
\end{equation}
with $\eta_{n}=\chi_{n}^{\mbox{up}}$ and $\xi_{n}=-\chi_{n}^{\mbox{dn}}$.
Following the similar way, at $z_{n}^{\ast}\in D^{-}\bigcap Z$
\begin{equation*}\label{GNCC-6}
\bar{\phi}(x,t;z_{n})\bar{\eta}_{n}=\bar{\psi}(x,t;z_{n}^{\ast})\bar{\xi}_{n},
\end{equation*}
for some $\bar{\xi}_{n},\bar{\eta}_{n}\in \mathbb{C}^{m}\setminus\{0\}$.
It is obvious that $\bar{\xi}_{n},\bar{\eta}_{n}$ and $\xi_{n},\eta_{n}$ are not uniquely given.

Since the first $m$ columns of $P(x,t;z_{n})$ are linearly independent,
and so are the last $m$ columns, necessarily
$\eta_{n}=\chi_{n}^{\mbox{up}}\neq 0$ and $\xi_{n}=-\chi_{n}^{\mbox{dn}}\neq 0$.
Given $\xi_{n}$ and $\eta_{n}$ as in \eqref{GNCC-5},
the $2m \times1$ vector $\chi_{n}=(\eta_{n},-\xi_{n})^{T}$  is a right null vector
of $P(x,t;z_{n})$.
Apparently, the analog of all the above discussions can be proved for $z_{n}^{\ast}\in D^{-}\bigcap Z$ and $\bar{P}(x,t;z)$.

If $\xi_{n},\eta_{n}\in\mathbb{C}^{m}\setminus\{0\}$ admit \eqref{GNCC-5},
then $\chi_{n}=(\eta_{n},-\xi_{n})^{T}$ is a right null vector of
\begin{equation*}\label{1}
A_{\sigma}(z_{n})=\bar{P}^{\dag}(x,t;z_{n}^{\ast})J_{\sigma}P(x,t;z_{n})
\end{equation*}
it then follows from \eqref{GNCC-2} that
\begin{equation*}\label{GNCC-7}
a(z_{n})\eta_{n}=\widetilde{\textbf{0}},~~\bar{a}^{\dag}(z_{n}^{\ast})\eta_{n}=\widetilde{\textbf{0}},
\end{equation*}
indicating that $\eta_{n}$ has to be in the right null space of $a(z_{n})$,
and $\xi_{n}$ has to be in the right null space of $\bar{a}^{\dag}(z_{n}^{\ast})$.
On the contrary, right null vectors of $a(z_{n})$ and $\bar{a}^{\dag}(z_{n}^{\ast})$ present vectors that satisfy \eqref{GNCC-5}.
Repeating the same process, we can demonstrate that the same holds for $\bar{\xi}_{n}, \bar{\eta}_{n}$
\begin{equation*}\label{GNCC-8}
a^{\dag}(z_{n})\bar{\xi}_{n}=\widetilde{\textbf{0}},~~\bar{a}(z_{n}^{\ast})\bar{\eta}_{n}=\widetilde{\textbf{0}},
\end{equation*}
so that $\bar{\eta}_{n}$ is in the right null space of $a(z_{n}^{\ast})$ and $\bar{\xi}_{n}$ is in the right null space of $a^{\dag}(z_{n})$.

%Let us recall the definitions of $\alpha(z)$ and $\bar{\alpha}(z)$ as cofactor matrices of $a(z)$ and $\bar{a}(z)$ respectively,
%it follows from $a(z_{n})\alpha(z_{n})=\alpha(z_{n})a(z_{n})=\textbf{0}$ and $\bar{a}(z_{n}^{\ast})\bar{\alpha}(z_{n}^{\ast})
%=\bar{\alpha}(z_{n}^{\ast})\bar{a}(z_{n}^{\ast})=\textbf{0}$ that each of the $m$ columns of $\alpha(z_{n})$
%are both right and left null vectors of $a(z_{n})$, and each of
%the $m$ columns of $\bar{\alpha}(z_{n}^{\ast})$ are both right and left null vectors of $\bar{a}(z_{n}^{\ast})$.
%If $m=2$ this indicates that the two columns of $\alpha(z_{n})$ and $\bar{\alpha}(z_{n}^{\ast})$ are proportional to each other,
%since $\det \alpha(z_{n})=\det\bar{\alpha}(z_{n}^{\ast})=0$.
As shown in \cite{IST-2017}, when $m=2$ we take two right null vectors of $P(x,t;z_{n})$
such that the first two components of each vector coincide with the first and the second columns of $\alpha(z_{n})$, let $-C_{n}$
denote the $2\times2$ matrix that collects column wise the remaining two components of said null vectors
\begin{equation*}\label{GNCC-9}
\widehat{\textbf{0}}=P(x,t;z_{n})\left(
                           \begin{array}{c}
                             \alpha(z_{n}) \\
                             -C_{n} \\
                           \end{array}
                         \right)\Leftrightarrow\phi(x,t;z_{n})\alpha(z_{n})=\psi(x,t;z_{n})C_{n}.
\end{equation*}
%Following the similar idea, we have
%\begin{equation}\label{GNCC-10}
%\widehat{\textbf{0}}=\bar{P}(x,t;z_{n}^{\ast})\left(
%                           \begin{array}{c}
%                             -\bar{C}_{n} \\
%                             \alpha(z_{n}^{\ast}) \\
%                           \end{array}
%                         \right)\Leftrightarrow\phi(x,t;z_{n})\alpha(z_{n})=\psi(x,t;z_{n})C_{n}.
%\end{equation}
If the right null space of $P(x,t;z_{n})$ is 1-dimensional, which happens if $z_{n}$ is a simple zero of $\det a(z)$,
then the two columns of the matrix multiplying $P(x,t;z_{n})$ have to be proportional to each other,
which then means $C_{n}$ is a rank 1 matrix,
as well as, since $\alpha(z)=a^{-1}(z)/\det a(z)$,
in the context of a simple zero the above equation can be written as
\begin{equation*}\label{GNCC-11}
\mathop{\mbox{Res}}\limits_{z=z_{n}}\frac{\phi(x,t;z)\alpha(z)}{\det a(z)}=\psi(x,t;z_{n}),~~\det C_{n}=0,
\end{equation*}
which provides the definition of the norming constant $C_{n}$ for a simple discrete eigenvalue $z_{n}$.
Following the same way, we also have
\begin{equation*}\label{GNCC-12}
\mathop{\mbox{Res}}\limits_{z=z_{n}}\frac{\bar{\phi}(x,t;z)\bar{\alpha}(z)}{\det \bar{a}(z)}=\bar{\psi}(x,t;z_{n}^{\ast}),~~\det \bar{C}_{n}=0.
\end{equation*}

\subsection{Asymptotics as $z\rightarrow\infty$ and $z\rightarrow 0$}

The asymptotic properties of the eigenfunctions and the scattering matrix are used to define the inverse problem.
In addition,
we reconstruct the potential from the solution of the RH problem in terms of the asymptotic behavior of the eigenfunctions.

It is worth mentioning that both limits (the limit $k\rightarrow\infty$  corresponds to $z\rightarrow\infty$ in $\mathbb{C}_{I}$ and to $z\rightarrow\infty$ in  $\mathbb{C}_{II}$).
In view of the uniformization variable
the asymptotic
expansion of the eigenfunctions can be found via the  Wentzel-Kramers-Brillouin (WKB) expansions.
Obviously, the eigenfunctions $\mu=\varphi e^{i\theta\sigma_{3}}$ admit
\begin{equation*}\label{ASAS-1}
\mu_{x}=\left(-ik\sigma_{3}+\underline{\mathcal {Q}}\right)\mu+i\lambda\mu\sigma_{3},
\end{equation*}
which we can write in terms of the uniformization variable $z$ with the help of \eqref{LP-8}.
Then, it follows from $\Phi(x,t;z)e^{i\theta\sigma_{3}}=(M(x,t;z),\bar{M}(x,t;z))$ that
\begin{align*}\label{ASAS-2}
&M^{\mbox{up}}_{x}=-\left(\frac{i\sigma k_{0}^2}{z}\right)M^{\mbox{up}}+\mathcal {Q} M^{\mbox{dn}},
~~M^{\mbox{dn}}_{x}=\sigma \mathcal {Q}^{\dag}M^{\mbox{up}}+iz M^{\mbox{dn}},\notag\\
&\bar{M}^{\mbox{dn}}_{x}=\left(\frac{i\sigma k_{0}^2}{z}\right)\bar{M}^{\mbox{dn}}+\sigma\mathcal {Q}^{\dag} \bar{M}^{\mbox{up}},~~
\bar{M}^{\mbox{up}}_{x}= \mathcal {Q}^{\dag}\bar{M}^{\mbox{dn}}-iz \bar{M}^{\mbox{up}}.
\end{align*}
%where $\mbox{up}$ and $dn$ denote the upper and lower $m\times m$ blocks of the corresponding $2m\times m$ matrices $M$, $\bar{M}$.
We can then write the WKB expansion as
\begin{equation*}\label{ASAS-3}
M^{\mbox{up}}=\mathcal {I}_{m}+\frac{A_{1}}{z}+h.o.t.,~~
M^{\mbox{dn}}=\frac{B_{1}}{z}+\frac{B_{2}}{z^2}+h.o.t..
\end{equation*}
(Here and in the following $h.o.t.$ represents higher order terms)
where $A_{1}, B_{1},\ldots$ are $m\times m$ matrix functions of $x$, $t$ to be known.
Plugging the WKB ansatz into the above differential
equations, and matching equal powers of $z$ leads to
$B_{1}=i\sigma\mathcal {Q}^{\dag}$ and $A_{1,x}=i\sigma\left(\mathcal {Q}\mathcal {Q}^{\dag}-k_{0}^2\mathcal {I}_{m}\right)$,
which in turn yields
\begin{align}\label{ASAS-4}
&M(x,t;z)=\notag\\
&\left(
           \begin{array}{c}
             \mathcal {I}_{m}+\frac{i\sigma}{z}\int_{-\infty}^{x}\left(\mathcal {Q}(x',t)\mathcal {Q}^{\dag}(x',t)-k_{0}^2\mathcal {I}_{m}\right)dx' \\
             \frac{i\sigma}{z}\mathcal {Q}^{\dag}(x,t) \\
           \end{array}
         \right)+O\left(\frac{1}{z^2}\right),~~z\rightarrow\infty,~~z\in D^{+},
\end{align}
where we have taken the boundary conditions for $M$ as $x\rightarrow -\infty$
into account .%(and implicitly assumed that the limits $z\rightarrow\infty$ and
%$x\rightarrow\infty$ commute).

Following a similar way one can see the asymptotic expansion for $\bar{M}$, as well as $N$, $\bar{N}$ as $z\rightarrow\infty$
in the suitable region of analyticity
\begin{align}\label{ASAS-5}
&\bar{M}(x,t;z)=\notag\\
&\left(
                 \begin{array}{c}
                   -\frac{i}{z}\mathcal {Q}(x,t)\\
                   \mathcal {I}_{m}-\frac{i\sigma}{z}\int_{-\infty}^{x}\left(\mathcal {Q}(x',t)\mathcal {Q}^{\dag}(x',t)-k_{0}^2\mathcal {I}_{m}\right)dx' \\
                 \end{array}
               \right)+O\left(\frac{1}{z^2}\right),~~z\rightarrow\infty,~z\in D^{-},
\end{align}
and
\begin{align}\label{ASAS-6}
&\bar{N}(x,t;z)=\notag\\
&\left(
           \begin{array}{c}
             \mathcal {I}_{m}+\frac{i\sigma}{z}\int_{-\infty}^{x}\left(\mathcal {Q}(x',t)\mathcal {Q}^{\dag}(x',t)-k_{0}^2\mathcal {I}_{m}\right)dx' \\
             \frac{i\sigma}{z}\mathcal {Q}^{\dag}(x,t) \\
           \end{array}
         \right)+O\left(\frac{1}{z^2}\right),~~
z\rightarrow\infty,~z\in D^{-},\notag\\
&N(x,t;z)=\notag\\
&\left(
                 \begin{array}{c}
                   -\frac{i}{z}\mathcal {Q}(x,t) \\
                   \mathcal {I}_{m}-\frac{i\sigma}{z}\int_{-\infty}^{x}\left(\mathcal {Q}(x',t)\mathcal {Q}^{\dag}(x',t)-k_{0}^2\mathcal {I}_{m}\right)dx' \\
                 \end{array}
               \right)+O\left(\frac{1}{z^2}\right),
~z\rightarrow\infty,~~z\in D^{+}.
\end{align}
Then asymptotics as $z\rightarrow\infty$ in the appropriate regions $D^{\pm}$ reach to
\begin{align*}\label{ASAS-7}
&M(x,t;z)=\left(
           \begin{array}{c}
             \mathcal {Q}\mathcal {Q}^{\dag}_{-}/k_{0}^2+O(z) \\
             i\sigma\mathcal {Q}^{\dag}_{-}/z+O(1) \\
           \end{array}
         \right),~~\bar{M}(x,t;z)=\left(
                                    \begin{array}{c}
                                      -i\mathcal {Q}_{-}/z+O(1) \\
                                      \mathcal {Q}\mathcal {Q}^{\dag}_{-}/k_{0}^2+O(z) \\
                                    \end{array}
                                  \right),\notag\\
&N(x,t;z)=\left(
            \begin{array}{c}
              \mathcal {Q}\mathcal {Q}_{+}^{\dag}/k_{0}^2+O(z) \\
              i\sigma\mathcal {Q}_{+}^{\dag}/z+O(1) \\
            \end{array}
          \right),~~~~N(x,t;z)=\left(
                                    \begin{array}{c}
                                      -i\mathcal {Q}_{+}/z+O(1) \\
                                      \mathcal {Q}\mathcal {Q}^{\dag}_{+}/k_{0}^2+O(z) \\
                                    \end{array}
                                  \right).
\end{align*}
The above expressions will help us to derive the scattering potential $\mathcal {Q}(x,t)$ from the solution of the inverse problem for the
eigenfunctions.

In the end, substituting the above asymptotic expansions into \eqref{SCS-3},
one can see that, as $z\rightarrow\infty$ in the proper regions of the complex $z$-plane
\begin{equation}\label{ASAS-8}
\mathcal {S}(z)=\mathcal {I}_{2m}+O\left(\frac{1}{z}\right).
\end{equation}
The above asymptotics can obtained with $\mbox{Im} z\geq0$ and $\mbox{Im}z\leq0$ for $a(z)$ and $\bar{a}(z)$, respectively,
and with $z\in\Sigma$ for $b(z)$ and $\bar{b}(z)$.
In a similar way, one also finds that as $z\rightarrow\infty$
\begin{equation}\label{ASAS-9}
\mathcal {S}(z)=\frac{1}{k_{0}^2}\left(
                                   \begin{array}{cc}
                                     \mathcal {Q}_{+}\mathcal {Q}_{-}^{\dag} & 0_{m} \\
                                     0_{m} & \mathcal {Q}_{+}^{\dag}\mathcal {Q}_{-} \\
                                   \end{array}
                                 \right)+O(z),
\end{equation}
where the asymptotics for the off-diagonal blocks hold for $z\in\Sigma$.
while the asymptotics for the block diagonal entries of $\mathcal {S}(z)$ can be extended to $D^{+}$ for $a(z)$
and $D^{-}$ for $\bar{a}(z)$.

\section{Inverse problem}

\subsection{Riemann-Hilbert problem}
As usual, the inverse scattering problem is formulated in terms of a suitable RH problem.
As mentioned above, the starting point for the formulation of the inverse problem is \eqref{SCS-8}, which we now regard as a relationship between eigenfunctions analytic in $D^{-}$ and those analytic in $D^{+}$.

We first introduce the sectionally meromorphic matrices
\begin{equation}\label{RHPRHP-1}
\mu^{+}(x,t;z)=\left(Ma^{-1},N\right),~~\mu^{-}(x,t;z)=\left(\bar{N},\bar{M}\bar{a}^{-1}\right),
\end{equation}
where superscripts $\pm$ differentiate between analyticity in $D^{+}$ and $D^{-}$, respectively.
It follows from \eqref{SCS-8} that
\begin{equation}\label{RHPRHP-2}
\mu^{-}(x,t;z)=\mu^{+}(x,t;z)\left(\mathcal {I}_{2m}-G(x,t;z)\right),~~z\in\Sigma,
\end{equation}
where
\begin{equation}\label{RHPRHP-3}
G(x,t;z)=\left(
           \begin{array}{cc}
             \textbf{0} & -e^{2i\theta(x,t;z)}\bar{\rho}(z) \\
             e^{-2i\theta(x,t;z)}\rho(z) & \rho(z)\bar{\rho}(z) \\
           \end{array}
         \right).
\end{equation}
Eqs.\eqref{RHPRHP-1}-\eqref{RHPRHP-3} give a matrix, multiplicative, homogeneous RH problem.
In order to finish the formulation of the RH problem one needs a normalization condition,
which is the asymptotic behavior of $\mu^{\pm}$ as $z\rightarrow\infty$.
According to the asymptotic behavior of the Jost eigenfunctions and scattering coefficients,
we find
\begin{equation*}\label{RHPRHP-4}
\mu^{\pm}=\mathcal {I}_{2m}+O\left(\frac{1}{z}\right),~~z\rightarrow\infty.
\end{equation*}
In addition,
\begin{equation*}\label{RHPRHP-5}
\mu^{\pm}=-\frac{i}{z}\sigma_{3}Q_{+}+O(1),~~z\rightarrow 0.
\end{equation*}
To solve the RH problem, one needs to regularize it by subtracting out the asymptotic behavior and the pole contributions, which below we
will suppose corresponding to poles of order 1.
Review that in the $\sigma=-1$ focusing case discrete eigenvalues come in symmetric quartets.
It is convenient to give $\zeta_{n}=z_{n}$ and $\zeta_{n+\mathcal {N}}=\sigma k_{0}^2/z_{n}^{\ast}$ for
$n=1,2,\ldots,\mathcal {N}$, as well as $C_{n+\mathcal {N}}=\hat{C}_{n}$ and $\bar{C}_{n+\mathcal {N}}=\hat{\bar{C}}_{n}$
for $n=1,2,\ldots,\mathcal {N}$ and rewrite \eqref{RHPRHP-2} as
\begin{align}\label{RHPRHP-6}
&\mu^{-}-\mathcal {I}_{2m}+(i/z)\sigma_{3}Q_{+}
-\sum_{n=1}^{2\mathcal {N}}\left(\mathop{\mbox{Res}}\limits_{\zeta_{n}^{\ast}}\mu^{-}\right)/(z-\zeta_{n}^{\ast})=\notag\\
&\mu^{+}-\mathcal {I}_{2m}+(i/z)\sigma_{3}Q_{+}-\sum_{n=1}^{2\mathcal {N}}\left(\mathop{\mbox{Res}}\limits_{\zeta_{n}}\mu^{+}\right)/(z-\zeta_{n})\notag\\
&-\sum_{n=1}^{2\mathcal {N}}\left(\mathop{\mbox{Res}}\limits_{\zeta_{n}^{\ast}}\mu^{-}\right)/(z-\zeta_{n}^{\ast})-\mu^{+}G.
\end{align}
The left-hand side of \eqref{RHPRHP-6} is now analytic in $D^{-}$ and is $O(1/z)$ as $z\rightarrow\infty$ from $D^{-1}$,
while the sum of the first four terms of the right-hand side is analytic in $D^{+}$ and is $O(1/z)$ as $z\rightarrow\infty$ from $D^{+}$.
In the end, the asymptotic behavior of the off-diagonal scattering coefficients implies that $G(x,t;z)$ is $O(1/z)$ as $z\rightarrow\pm\infty$
and $O(z)$ as $z\rightarrow\infty$ along the real axis.
We then introduce the analog of Cauchy projectors $P^{\pm}$ over $\Sigma$
\begin{equation*}\label{RHPRHP-7}
P_{\pm}[f](z)=\frac{1}{2\pi i}\int_{\Sigma}\frac{f(\zeta)}{\zeta-(z\pm i0)}d\zeta,
\end{equation*}
where $\int_{\Sigma}$ represents the integral along the oriented contours displayed in Fig.2,
and the notation $z\pm i0$ means that, when $z\in\Sigma$,
the limit is taken from the left/right of it.
Now refer to Plemelj's formulas:
if $f^{\pm}$ are analytic in $D^{\pm}$  and are $O(1/z)$ as $z\rightarrow\infty$, one obtains $P^{\pm}f^{\pm}=\pm f^{\pm}$
and $P^{+}f^{-}=P^{-}f^{+}=0$.
Applying $P^{+}$ and $P^{-}$ to \eqref{RHPRHP-6}  we then obtain
\begin{align}\label{RHPRHP-8}
\mu(x,t;z)=\mathcal {I}_{2m}-&\left(\frac{i}{z}\right)\sigma_{3}Q_{+}+\sum_{n=1}^{2\mathcal {N}}\frac{\mathop{\mbox{Res}}\limits_{\zeta_{n}}\mu^{+}}{z-\zeta_{n}}
+\sum_{n=1}^{2\mathcal {N}}\frac{\mathop{\mbox{Res}}\limits_{\zeta_{n}^{\ast}}\mu^{-}}{z-\zeta_{n}^{\ast}}\notag\\
&+\frac{1}{2\pi i}\int_{\Sigma}\frac{\mu^{+}(x,t;z)}{\zeta-z}G(x,t;\zeta)d\zeta,~~z\in\mathbb{C}\setminus\Sigma.
\end{align}
The expressions for $\mu^{+}$ and $\mu^{-}$
are formally identical, except for the conclusion  that the integral appearing on the right-hand side is a
$P^{+}$ and a $P^{-}$ projector, respectively. In addition, in the $\sigma=1$ (defocusing) case the sums are only for $n=1,2,\ldots\mathcal {N}$.

\subsection{Residue conditions and reconstruction formula}
Eq.\eqref{RHPRHP-8} is an integral equation for $\mu^{\pm}(x,t;z)$ with $z\in D^{\pm}$
which also relies on the residues of $\mu^{\pm}(x,t;z)$ at its poles in $D^{\pm}$.
The residues appearing on the right-hand side of \eqref{RHPRHP-8}
are proportional to the values of the analytic columns of $\mu^{\pm}(x,t;z)$
at the discrete eigenvalues, and therefore the analytic columns of \eqref{RHPRHP-8} reduce to a system of linear algebraic-integral equations.
In fact, it follows from the definition \eqref{RHPRHP-1} that only the first $m$ columns of $\mu^{+}$ admit a pole at
$z=z_{0}$ and $z=\sigma k_{0}^2/z_{n}^{\ast}$ in $D^{+}$,  and only the
last $m$ columns of $\mu^{-}$ have a pole at $z=z^{\ast}$ and $z=\sigma k_{0}^2/z_{n}$ in $D^{-}$.
By using the residue relations \eqref{DSRC-5} and \eqref{DSRC-25}, we have the following relationships %then means that such residues are proportional, respectively,
%to the last $m$ columns of $\mu^{+}$ and the first m columns of $\mu^{-}$. Explicitly:
\begin{align}\label{RCRSF-1}
&\mathop{\mbox{Res}}\limits_{\zeta_{n}}\mu^{+}=\left(e^{-2i\theta(x,t;\zeta_{n})}N(x,t;\zeta_{n})C_{n},\widehat{\textbf{0}}\right),~~n=1,2,\ldots,2\mathcal {N},\notag\\
&\mathop{\mbox{Res}}\limits_{\zeta_{n}^{\ast}}\mu^{+}=\left(\widehat{\textbf{0}},e^{2i\theta(x,t;\zeta_{n}^{\ast})}\bar{N}(x,t;\zeta_{n}^{\ast})\bar{C}_{n}\right),~~n=1,2,\ldots,2\mathcal {N}.
\end{align}
As a result, we can evaluate the last $m$ columns of \eqref{RHPRHP-8} at $z=z_{n}$ and at $z=\sigma k_{0}^2/z_{n}^{*}$,
obtaining
\begin{align*}\label{RCRSF-2}
&N\left(x,t;\zeta_{n}^{\ast}\right)=\notag\\
&~~~~~~\left(
                                     \begin{array}{c}
                                       -i\mathcal {Q}_{+}/\zeta_{n} \\
                                       \mathcal {I}_{m} \\
                                     \end{array}
                                   \right)+\sum_{j=1}^{2\mathcal {N}}\frac{e^{2i\theta(x,t;\zeta_{j}^{\ast})}}{\zeta_{n}-\zeta_{j}^{\ast}}
\bar{N}(x,t;\zeta_{j}^{\ast})\bar{C}_{j}+\frac{1}{2\pi i}\int_{\Sigma}\frac{\left(\mu^{+}G\right)_{2}(x,t;\zeta)}{\zeta-\zeta_{n}}d\zeta
\end{align*}
for $n=1,2,\ldots,2\mathcal {N}$, and where the subscript 2 in $\mu^{+}G$ denotes the last $m$ columns of the product, i.e.,
\begin{equation*}\label{RCRSF-3}
\left(\mu^{+}G\right)_{2}(x,t;\zeta)=-e^{2i\theta(x,t;\zeta)}\bar{N}(x,t;\zeta)\bar{\rho}(\zeta).
\end{equation*}
Similarly, the first $m$ columns of \eqref{RHPRHP-8} $z=z_{n}^{\ast}$ at the points $z=\sigma k_{0}^2/z_{n}^{\ast}$ can arrive at
\begin{align*}\label{RCRSF-4}
&\bar{N}\left(x,t;\zeta_{n}^{\ast}\right)=\left(
                                \begin{array}{c}
                                  \mathcal {I}_{m} \\
                                  i\sigma\mathcal {Q}_{+}^{\dag}/\zeta_{n}^{\ast} \\
                                \end{array}
                              \right)\notag\\
                              &~~~~~~~~~~~~~~~+\sum_{j=1}^{2\mathcal {N}}\frac{e^{-2i\theta(x,t;\zeta_{j})}}{\zeta_{n}^{\ast}-\zeta_{j}}
\bar{N}(x,t;\zeta_{j})\bar{C}_{j}+\frac{1}{2\pi i}\int_{\Sigma}\frac{\left(\mu^{+}G\right)_{2}(x,t;\zeta)}{\zeta-\zeta_{n}^{\ast}}d\zeta,
\end{align*}
where $n=1,2.\ldots,2\mathcal {N}$, and the subscript 1 in $\mu^{+}G$ represents the first $m$ columns of the product, i.e.,
\begin{equation*}\label{RCRSF-5}
\left(\mu^{+}G\right)_{1}(x,t;\zeta)=e^{-2i\theta(x,t;\zeta)}N(x,t;\zeta)\rho(\zeta).
\end{equation*}
In the end, evaluating the first $m$ columns of $\mu^{-}$ and the last m columns of $\mu^{+}(x,t;z)$ through \eqref{RHPRHP-8}
for $z\in\Sigma$ we get
\begin{align}\label{RCRSF-5}
&N(x,t;z)=\left(
           \begin{array}{c}
             -i\mathcal {Q}_{+}/z \\
             \mathcal {I}_{m} \\
           \end{array}
         \right)+\sum_{j=1}^{2\mathcal {N}}\frac{e^{2i\theta(x,t;\zeta_{j}^{\ast})}}{z-\zeta_{j}^{\ast}}\bar{N}(x,t;\zeta_{j}^{\ast})\bar{C}_{j}\notag\\
&~~~~~~~~~~~~~-\frac{1}{2\pi i}\int_{\Sigma}\frac{e^{2i\theta(x,t;\zeta)}\bar{N}(x,t;\zeta)\bar{\rho}(\zeta)}{\zeta-(z+i0)}d\zeta,\notag\\
&\bar{N}(x,t;z)=\left(
           \begin{array}{c}
             \mathcal {I}_{m} \\
             i\sigma\mathcal {Q}_{+}^{\dag}/z \\
           \end{array}
         \right)+\sum_{j=1}^{2\mathcal {N}}\frac{e^{-2i\theta(x,t;\zeta_{j})}}{z-\zeta_{j}}N(x,t;\zeta_{j})C_{j}\notag\\
&~~~~~~~~~~~~~~-\frac{1}{2\pi i}\int_{\Sigma}\frac{e^{-2i\theta(x,t;\zeta)}N(x,t;\zeta)\rho(\zeta)}{\zeta-(z-i0)}d\zeta,
\end{align}
which, together with equations \eqref{RCRSF-5}, lead to a closed system of linear algebraic-integral equations for the solution of the RH problem.

The last task is to derive the potential from the solution of the RH problem.
From \eqref{RHPRHP-8}, one get the asymptotic behavior of $\mu^{\pm}(x,t;z)$ as $z\rightarrow\infty$
\begin{align}\label{RCRSF-6}
&\mu^{\pm}(x,t;z)=\mathcal {I}_{2m}+O\left(\frac{1}{z^2}\right)+\notag\\
&\frac{1}{z}\left\{-i\sigma_{3}Q_{+}+\sum_{n=1}^{2\mathcal {N}}
\left(\mathop{\mbox{Res}}\limits_{\zeta_{n}}\mu^{+}+\mathop{\mbox{Res}}\limits_{\zeta_{n}^{\ast}}\mu^{-}\right)-\frac{1}{2\pi i}\int_{\Sigma}\mu^{+}(x,t;\zeta)G(x,t;\zeta)d\zeta\right\},
\end{align}
where the residues are expressed by \eqref{RCRSF-1}.
Taking $\mu=\mu^{+}$ and comparing the upper right $m\times m$ block of \eqref{RCRSF-6} of this expression with \eqref{ASAS-6} yields
\begin{align}\label{RCRSF-7}
&\mathcal {Q}(x,t)=\mathcal {Q}_{+}\notag\\
&+i\sum_{n=1}^{2\mathcal {N}}e^{2i\theta\left(x,t;\zeta_{n}^{\ast}\right)}\bar{N}^{\mbox{up}}\left(x,t;\zeta_{n}^{\ast}\right)\bar{C}_{n}
+\frac{1}{2\pi}\int_{\Sigma}e^{2i\theta(x,t;\zeta)}\bar{N}^{\mbox{up}}\left(x,t;\zeta\right)\bar{\rho}(\zeta)d\zeta.
\end{align}
Following the similar way, taking $\mu=\mu^{-}$  and comparing the lower left $m\times m$ block of \eqref{RCRSF-6} of this expression with \eqref{ASAS-6},
then we get the reconstruction formula for the potential
\begin{align}\label{RCRSF-8}
&\mathcal {Q}^{\dag}(x,t)=\mathcal {Q}_{+}^{\dag}\notag\\
&-i\sigma\sum_{n=1}^{2\mathcal {N}}e^{-2i\theta\left(x,t;\zeta_{n}\right)}N^{\mbox{dn}}(x,t;\zeta_{n})C_{n}
+\frac{\sigma}{2\pi}\int_{\Sigma}e^{-2i\theta(x,t;\zeta)}N^{\mbox{dn}}(x,t;\zeta)\rho(\zeta)d\zeta.
\end{align}
%Recall that the time dependence of the solution is automatically taken into account by the fact that the Jost eigenfunctions are simultaneous
%solutions of both parts of the Lax pair.

The above formulas help us to determine the symmetries of the norming constants.
Actually, considering that $N^{\mbox{dn}}(x,t;z)\backsim\mathcal {I}_{m}$ as $x\rightarrow\infty$ for any $z\in D^{+}$,
and $\bar{N}^{\mbox{up}}(x,t;z)\backsim\mathcal {I}_{m}$ as $x\rightarrow\infty$ for any $z\in D^{-}$,
comparing the two equations in \eqref{RCRSF-7} and \eqref{RCRSF-8} gives
\begin{equation*}\label{RCRSF-9}
\bar{C}_{n}=\sigma C_{n}^{\dag},~~n=1,2,\ldots,2\mathcal {N}.
\end{equation*}
Because of $\mathcal {Q}^{T}=\mathcal {Q}$ the norming constants must admit the same symmetry, i.e.,
\begin{equation*}\label{RCRSF-10}
C_{n}^{T}=C_{n},~~\bar{C}_{n}^{T}=\bar{C}_{n},~~n=1,2,\ldots,2\mathcal {N}.
\end{equation*}

\subsection{Focusing reflectionless potentials}
In this section, we first discuss potentials $\mathcal {Q}(x,t)$ for which the reflection coefficient $\rho(z)$ fades away identically
for $z\in\Sigma$. In this case there is no jump from $\mu^{+}$ to $\mu^{-}$
across the continuous spectrum, and the inverse problem thus can be reduced to an
algebraic system, whose solution gives the soliton solutions of \eqref{mNLSS1}

In the following we investigate solutions of the focusing equations $\sigma=-1$.
In this focusing case discrete eigenvalues happen in quartets, with $\zeta_{\mathcal {N}+j}=-k_{0}^2/z_{j}^{\ast}$
and $C_{\mathcal {N}+j}=\mathcal {Q}_{+}^{\dag}\bar{C}_{j}\mathcal {Q}_{+}^{\dag}/(z^{\ast})^2$ for all $j=1,2,\ldots \mathcal {N}$.
Furthermore, one can also easily find that $\theta(x,t;z^{\ast})=\theta^{\ast}(x,t;z)$.
According to \eqref{DSRC-16} one has $\bar{C}_{j}=-C_{j}^{\dag}$ for all $j=1,2,\ldots,2\mathcal {N}$.
For convenience, we introduce the quantities
\begin{equation*}\label{FRP-1}
c_{j}(x,t;z)=\frac{C_{j}}{z-\zeta_{j}}e^{-2i\theta(x,t;\zeta_{j})},~~j=1,2,\ldots,2\mathcal {N}.
\end{equation*}
Then the algebraic systems
constructed from the inverse problem for said upper blocks are expressed as
\begin{align}\label{FRP-2}
&N^{\mbox{up}}(\zeta_{j})=-\frac{i}{\zeta_{j}}\mathcal {Q}_{+}-\sum_{\ell=1}^{2\mathcal {N}}\bar{N}^{\mbox{up}}(\zeta_{\ell}^{\ast})c_{\ell}^{\dag}(\zeta_{j}^{\ast}),~~j=1,2,\ldots,2\mathcal {N},\notag\\
&\bar{N}^{\mbox{up}}(\zeta_{n}^{\ast})=\mathcal {I}_{m}+\sum_{j=1}^{2\mathcal {N}}N^{\mbox{up}}(\zeta_{j})c_{j}(\zeta_{n}^{\ast}),~~n=1,2,\ldots,2\mathcal {N},
\end{align}
and substituting \eqref{FRP-2} reaches to
\begin{equation}\label{FRP-3}
\bar{N}^{\mbox{up}}(\zeta_{n}^{\ast})=\mathcal {I}_{m}-i\mathcal {Q}_{+}\sum_{j=1}^{2\mathcal {N}}c_{j}(\zeta_{n}^{\ast})/\zeta_{j}
-\sum_{j=1}^{2\mathcal {N}}\sum_{\ell=1}^{2\mathcal {N}}\bar{N}^{\mbox{up}}\left(\zeta_{\ell}^{\ast}\right)c_{\ell}^{\dag}(\zeta_{j}^{\ast})
c_{j}(\zeta_{n}^{\ast}),
\end{equation}
where $n=1,2,\ldots,2\mathcal {N}$, for simplicity we omitted the $x$ and $t$ dependence.
We now show this system in matrix form.
Introducing $\textbf{X}=(X_{1},X_{2},\ldots,X_{2\mathcal {N}})^{T}$
and $\textbf{B}=(B_{1},B_{2},\ldots,B_{2\mathcal {N}})^{T}$, where
\begin{equation}\label{FRP-4}
X_{n}=\bar{N}^{\mbox{up}}(x,t;\zeta_{n}^{\ast}),~~B_{n}=\mathcal {I}_{m}-i\mathcal {Q}_{+}\sum_{j=1}^{2\mathcal {N}}c_{j}(\zeta_{n}^{\ast})/\zeta_{j},~~
n=1,2,\ldots,2\mathcal {N},
\end{equation}
and defining the block matrix $\Gamma=(\Gamma_{n,\ell})$, in which
\begin{equation}\label{FRP-5}
\Gamma_{n,\ell}=\sum_{j=1}^{2\mathcal {N}}c_{\ell}^{\dag}(\zeta_{j}^{\ast})c_{j}(\zeta_{n}^{\ast}),~~n,\ell=1,2,\ldots,2\mathcal {N},
\end{equation}
the system \eqref{FRP-3} yields simply
\begin{equation}\label{FRP-5}
A\textbf{X}=\textbf{B},~~A=\mathcal {I}_{p}+\Gamma,
\end{equation}
where $\mathcal {I}_{p}$ is the identity matrix of size $p=2m\mathcal {N}$.
Summarizing the above results, we obtain the following theorem.\\

\noindent
\textbf{Theorem 5.}
Substituting $X_{1},\ldots,X_{2\mathcal {N}}$ into the reconstruction formula \eqref{RCRSF-7} and \eqref{RCRSF-8},
one gets the corresponding $\mathcal {N}$ soliton solution for $\mathcal {Q}(x,t)$
\begin{equation*}\label{FRP-6}
\mathcal {Q}(x,t)=\mathcal {Q}_{+}+i\sum_{n=1}^{2\mathcal {N}}e^{2i\theta(x,t;z_{n}^{\ast})}X_{n}\bar{C}_{n},
\end{equation*}
where $X_{n}$ and $\bar{C}_{n}$ are given by \eqref{FRP-4}-\eqref{FRP-5}.\\

%\noindent
%\textbf{Note}. Although the discrete eigenvalues appear in quartets in the NZBC case as opposed to pairs in the case of zero boundary
%conditions, the number of unknowns in the inverse problem is still the same.
%This is because the symmetry \eqref{SSYM1-3} means
%\begin{align*}\label{FRP-7}
%&N^{\mbox{up}}\left(x,t;\zeta_{j}\right)=-\frac{i}{\zeta_{j}}\bar{N}^{\mbox{up}}\left(x,t;\zeta_{j+\mathcal {N}}^{\ast}\right)\mathcal {Q}_{+},\notag\\
%&N^{\mbox{up}}\left(x,t;\zeta_{j+\mathcal {N}}\right)=-\frac{i\sigma\zeta_{j}^{\ast}}{\zeta_{j}}\bar{N}^{\mbox{up}}\left(x,t;\zeta_{j}^{\ast}\right)\mathcal {Q}_{+},
%\end{align*}
%for all $j=1,2,\ldots,\mathcal {N}$.
%Consequently one can equivalently write the linear algebraic system \eqref{FRP-2} in view of just $2\mathcal {N}$ unknowns,
%as in the case of ZBCs.

\section{Soliton solutions for the focusing model}

The focusing and defocusing Hirota equation with NZBCs admit a rich family of soliton solutions \cite{yzy-cnsns},
and in this section we will construct their counterpart in the symmetric matrix Hirota equation \eqref{mNLSS1} with $m=2$.
Let us start by constructing
the one-soliton solution in the $\sigma=-1$ (focusing) case, with one quartet of discrete eigenvalues.
$X_{j}=\bar{N}_{j}^{\mbox{up}}(x,t;\zeta_{j}^{\ast})$  for
$j=1,2$, and utilizing the symmetries \eqref{FRP-2}, we obtain
\begin{equation*}\label{SSF-1}
X_{1}D_{1}=\mathcal {I}_{2}-\frac{i}{\zeta_{1}}X_{2}\mathcal {Q}_{+}c_{1}(\zeta_{1}^{\ast}),~~
X_{2}D_{2}=\mathcal {I}_{2}+\frac{i\zeta_{1}^{\ast}}{k_{0}^2}X_{1}\mathcal {Q}_{+}c_{2}(\zeta_{2}^{\ast}),
\end{equation*}
where
\begin{equation*}\label{SSF-2}
D_{1}=\mathcal {I}_{2}+\frac{i}{(\zeta_{1}^{\ast})^2+k_{0}^2}C_{1}^{\dag}\mathcal {Q}_{+}^{\dag}e^{2i\theta(\zeta_{1}^{\ast})},~~D_{2}=D_{1}^{\dag},
\end{equation*}
where we have used the expression of the matrices $c_{j}(\zeta_{\ell}^{\ast})=C_{j}e^{2i\theta(\zeta_{j})}/(\zeta_{\ell}^{\ast}-\zeta_{j})$,
as well as $\zeta_{2}=-k_{0}^2/\zeta_{1}^{\ast}$,
and the two symmetries for the norming constant $C_{2}=-\mathcal {Q}_{+}^{\dag}C_{1}^{\dag}\mathcal {Q}_{+}^{\dag}/(\zeta_{1}^{\ast})^2$.

This is formally a linear algebraic system of two expressions in the two unknowns $X_{1}$, $X_{2}$,
but both the unknowns and the coefficients are
$2\times2$ matrices. We solve the expression by back substitution, and get
\begin{align*}\label{SSF-3}
&X_{1}=\left[\mathcal {I}_{2}-\frac{i}{\zeta_{1}}D_{2}^{-1}\mathcal {Q}_{+}c_{1}(\zeta_{1}^{\ast})\right]
\left[D_{1}-\frac{\zeta_{1}^{\ast}}{\zeta_{1}k_{0}^2}\mathcal {Q}_{+}c_{2}(\zeta_{2}^{\ast})D_{2}^{-1}\mathcal {Q}_{+}c_{1}(\zeta_{1}^{\ast})\right]^{-1},
\notag\\
&X_{2}=\left[\mathcal {I}_{2}+\frac{i\zeta_{1}^{\ast}}{k_{0}^2}D_{1}^{-1}\mathcal {Q}_{+}c_{2}(\zeta_{2}^{\ast})\right]
\left[D_{2}-\frac{\zeta_{1}^{\ast}}{\zeta_{1}k_{0}^2}\mathcal {Q}_{+}c_{1}(\zeta_{1}^{\ast})D_{1}^{-1}\mathcal {Q}_{+}c_{2}(\zeta_{2}^{\ast})\right]^{-1},
\end{align*}
where
\begin{equation*}\label{SSF-4}
c_{1}(\zeta_{1}^{\ast})=\frac{C_{1}e^{-2i\theta(x,t;\zeta_{1})}}{\zeta_{1}^{\ast}-\zeta_{1}},~~
c_{2}(\zeta_{2}^{\ast})=\frac{\zeta_{1}}{\zeta_{1}k_{0}^2(\zeta_{1}^{\ast}-\zeta_{1})}\mathcal {Q}_{+}^{\dag}C_{1}^{\dag}\mathcal {Q}_{+}^{\dag}
e^{2i\theta(x,t;\zeta_{1}^{\ast})}.
\end{equation*}
We know that in the above expression the eigenvalue $\zeta_{1}\in D^{+}$ is set to be located in the upper half plane,
i.e., such that $|\zeta_{1}|> k_{0}$, $\mbox{Im}\zeta_{1}>0$,
and the associated norming constant $C_{1}$ is a free $2\times2$ symmetric matrix with $\det C_{1}=0$.
Here we can also investigate the case when $C_{1}$ is not rank 1, corresponding to having $a(\zeta_{1})=0$.
Generally, all entries of $C_{1}$ can be complex,
and thus the norming constant is determined in view of two free complex numbers in the rank 1 case,
and three in the rank 2 case
\begin{equation*}\label{SSF-5}
C_{1}=\left(
        \begin{array}{cc}
          \gamma_{1} & \gamma_{0} \\
          \gamma_{0} & \gamma_{-1} \\
        \end{array}
      \right),
\end{equation*}
with $\gamma_{j}\in\mathbb{C}$ for $j=1,0,-1$.
In terms of $X_{1}$, $X_{2}$ via \eqref{RCRSF-7} and \eqref{RCRSF-8},  we have the following one-soliton solution
\begin{equation}\label{SSF-6}
\mathcal {Q}_{[1]}
=\mathcal {Q}_{+}-iX_{1}e^{2i\theta(x,t;\zeta_{1}^{\ast})}C_{1}^{\dag}
+iX_{2}e^{-2i\theta(x,t;\zeta_{1})}\mathcal {Q}_{+}C_{1}\mathcal {Q}_{+}/\zeta_{1}^2.
\end{equation}
The above results can then be used to obtain the various soliton solutions.
In the following, Figs.3-9 present the various breather waves by varying the suitable parameters,
which are useful for understanding the dynamical behaviors of the soliton solutions.

When the discrete eigenvalue $\zeta_{1}\in D^{+}$ is purely imaginary, the corresponding soliton solutions are stationary,
and a solution (analog of Kuznetsov-Ma (KM) breather of the focusing NLS) that is homoclinic in $x$ and periodic in $t$,
shown in Fig.3 for the polar state (rank 2 norming constant).
By comparing Figs.3(a)-(c) and Figs.3(d)-(f), we find, when the free parameter $\beta$ become large,
the direction of the wave propagation can be changed, but its shape remain unchanged.
On the other hand, in the $|q_{0}|$ component, the breather wave emerges without valleys,
and Fig.3(b) and Fig.4(b) reveal that two peaks of the breather wave without valley merge into one peak,
since $\gamma_{0}$ is chosen as 1 in Fig.3 instead of 2 in Fig.4.
If one considers the limit as the discrete eigenvalue $\zeta_{1}$ approaches the circle $\mathbb{C}_{o}$,
one obtains the analog of the Akhmediev breather (AB). The corresponding polar state is plotted in Fig.5,
namely, Fig.5 is plotted for the breather waves with
suitable parameters, which is not the time-periodic breather but the space-periodic breather, thus revealing the
usual AB features.
Note that these solutions are periodic in $x$ and homoclinic in $t$.

A one-soliton solution (analog of the Tajiri-Watanabe soliton for the scalar focusing NLS)
corresponding to $\zeta_{1}\in D^{+}$ in generic position is displayed in Fig.6, and when the norming constant
is a symmetric matrix (rank 2).
From Fig.7, we can easily see that $|q_{1}|$ (or $q_{-1}$) and $|q_{0}|$ have different structures.
In the $|q_{1}|$ (or $q_{-1}$)  component, the dark-breather wave is displayed in Fig.7(a) and Fig.7(a). In addition,
Fig.6(a) and Fig.7(a) reveal that bright-breather wave turns into the dark-breather wave,
since $\gamma_{j}$ is chosen as $\gamma_{1}=i,\gamma_{0}=1+i,\gamma_{-1}=i$ (complex) in Fig.7
instead of $\gamma_{1}=1,\gamma_{0}=1,\gamma_{-1}=2$ (real) in Fig.6.
Note that Fig.8 and Fig.9 are plotted for the breather waves with
suitable parameters, which are homoclinic in $x$ and periodic in $t$, thus revealing the
usual KM breather features.
Besides, we also exhibit a range of
interesting and complicated dynamics, obtained by varying the available parameters (see Figs.10 and 11).
To best of our knowledge, the most types of dynamic patterns presented in Figs.3-11 have never emerged in standard NLS equation so far.

\noindent
{\rotatebox{0}{\includegraphics[width=4.8cm,height=2.5cm,angle=0]{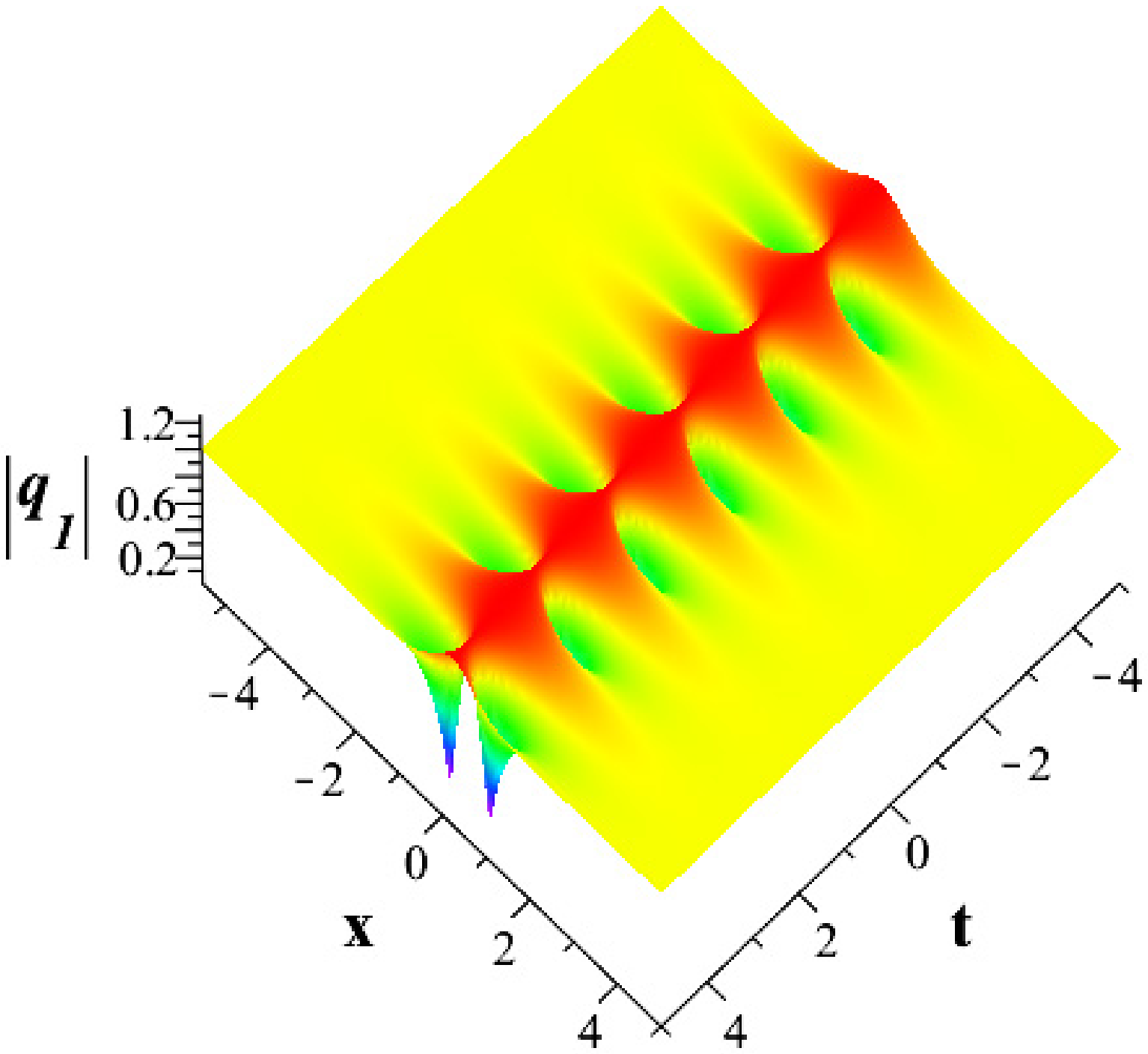}}}
{\rotatebox{0}{\includegraphics[width=4.8cm,height=2.5cm,angle=0]{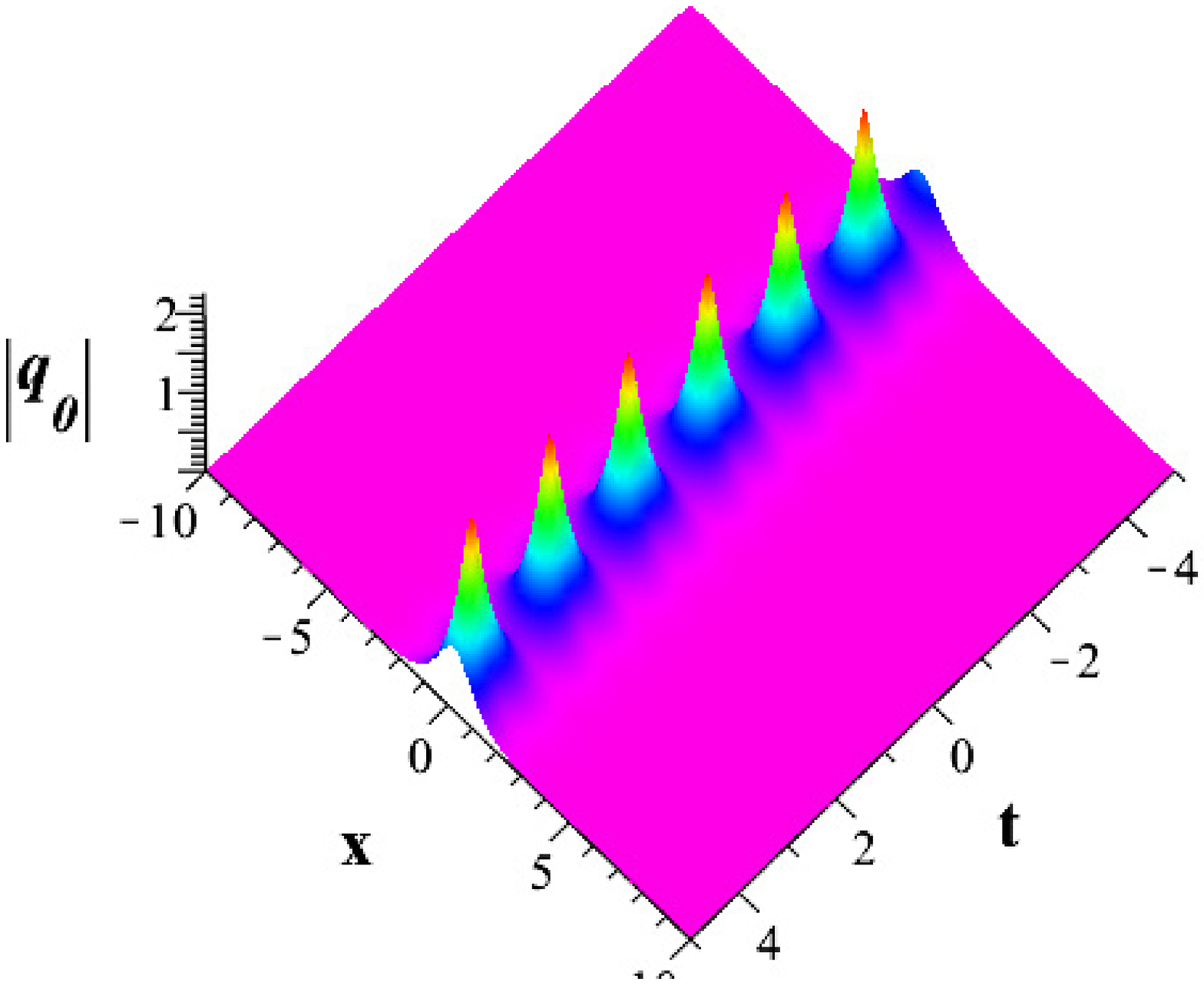}}}
{\rotatebox{0}{\includegraphics[width=4.8cm,height=2.5cm,angle=0]{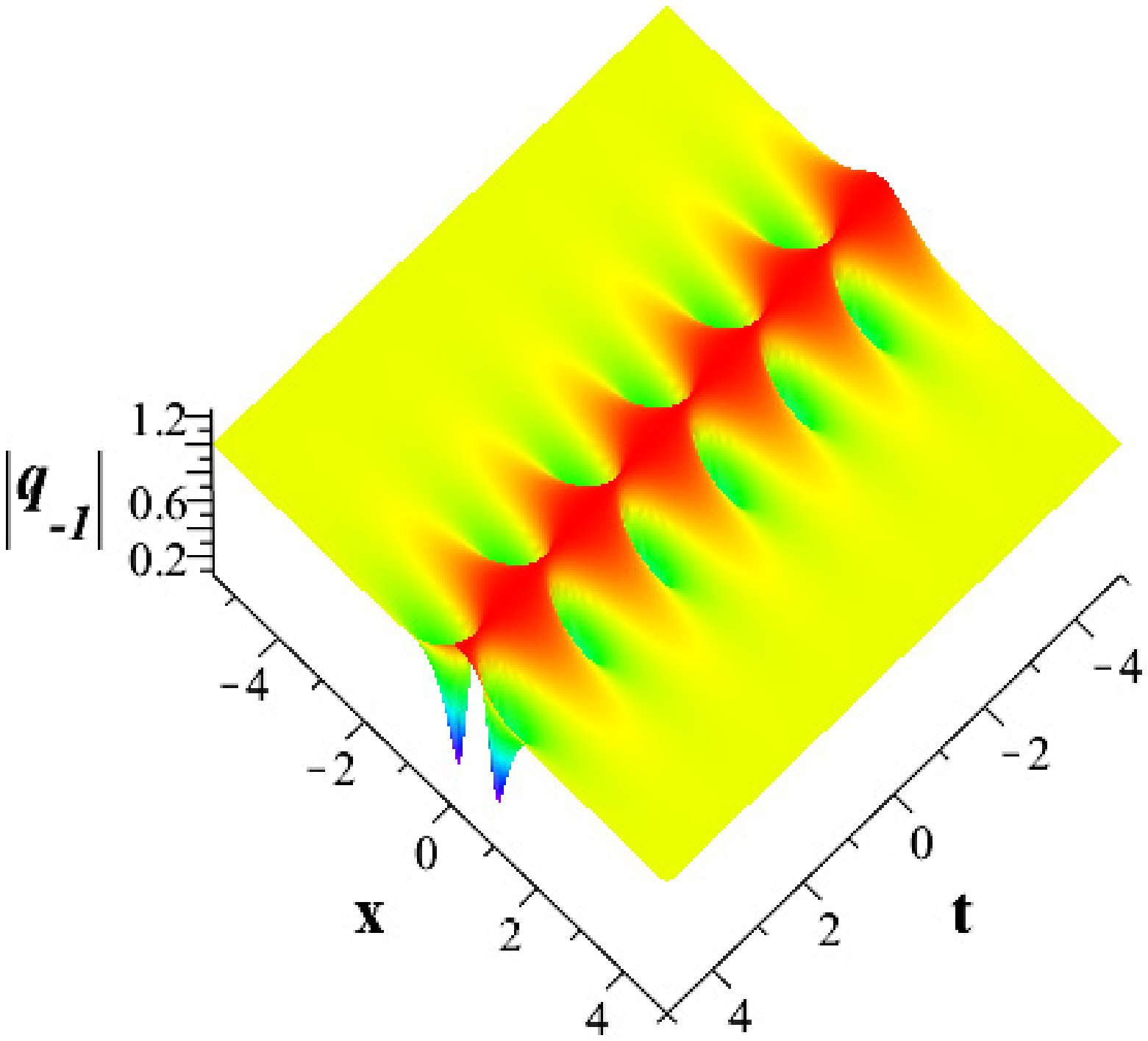}}}

$\qquad\qquad\textbf{(a)}\qquad\qquad\qquad\qquad\qquad\textbf{(b)}
\qquad\qquad\qquad\qquad\qquad\qquad\textbf{(c)}$\\

\noindent
{\rotatebox{0}{\includegraphics[width=4.8cm,height=2.5cm,angle=0]{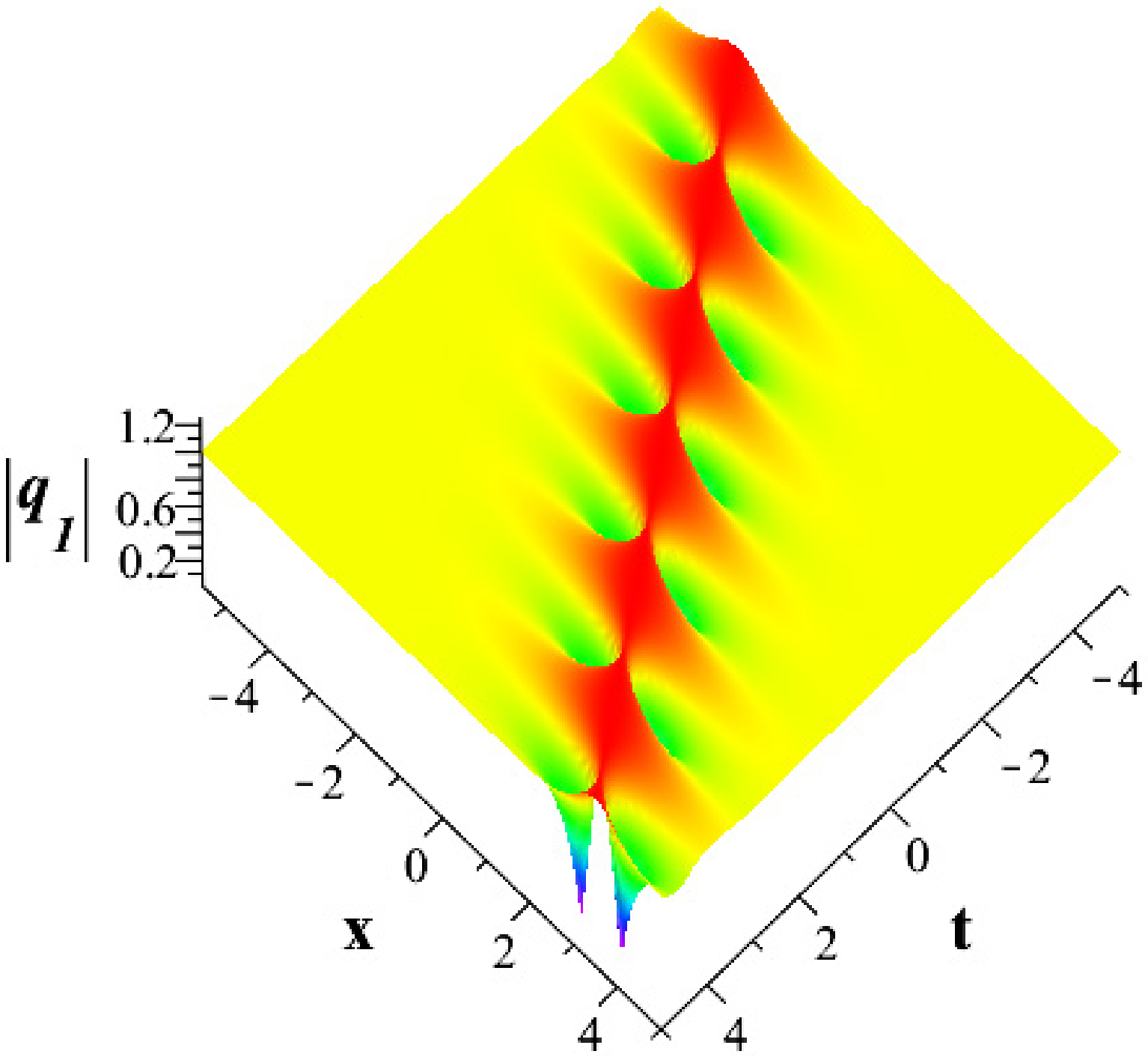}}}
{\rotatebox{0}{\includegraphics[width=4.8cm,height=2.5cm,angle=0]{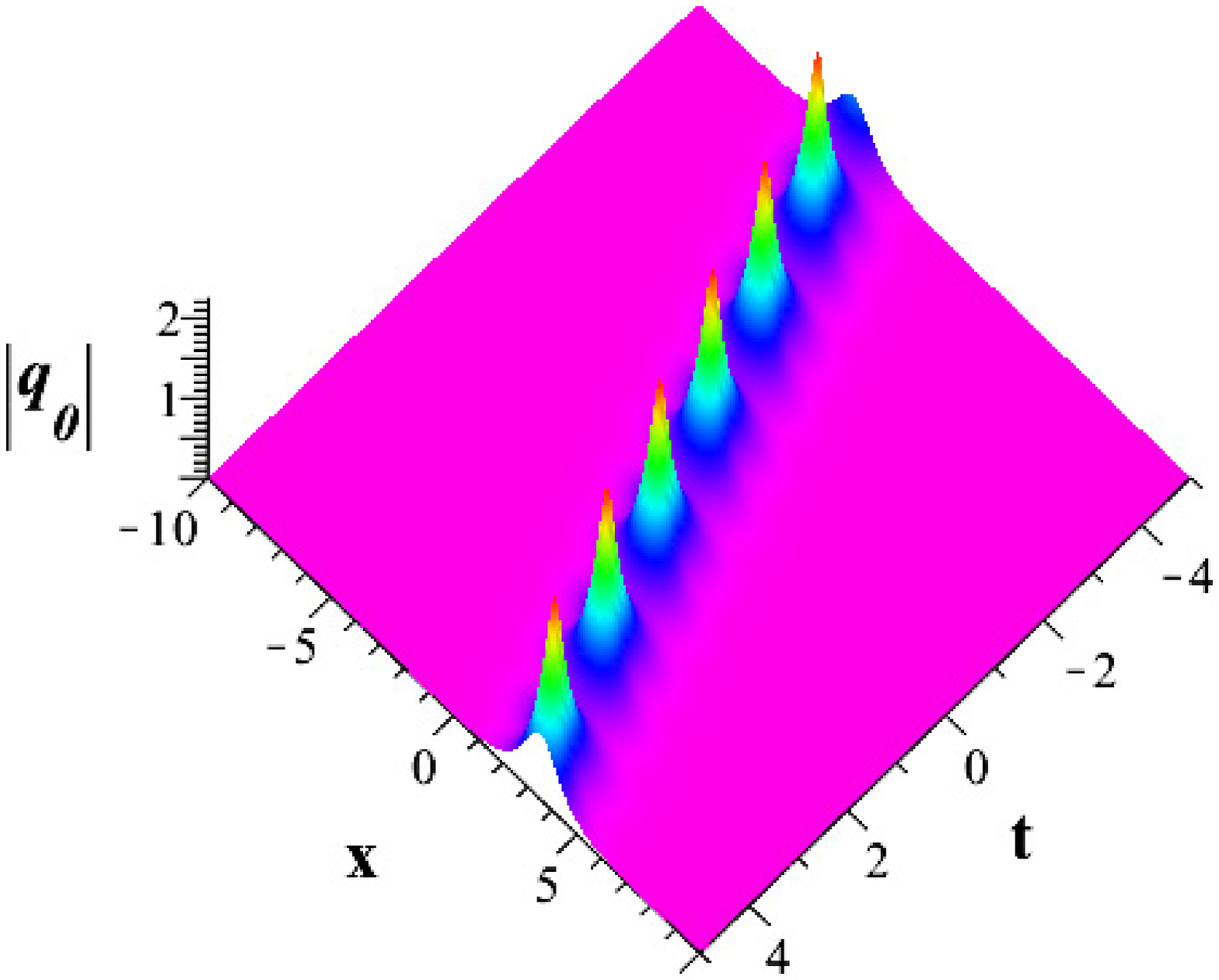}}}
{\rotatebox{0}{\includegraphics[width=4.8cm,height=2.5cm,angle=0]{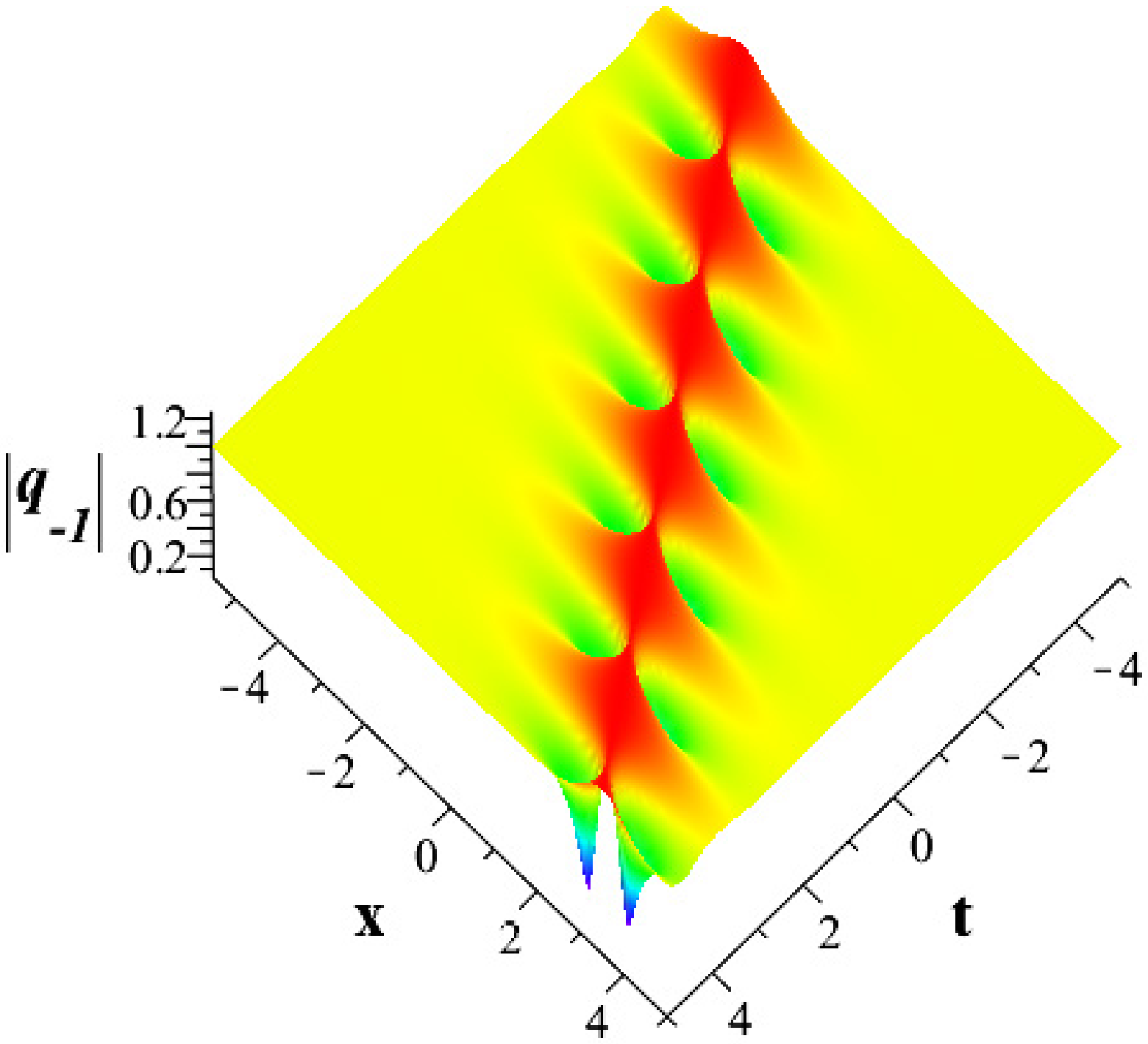}}}

$\qquad\qquad\textbf{(d)}\qquad\qquad\qquad\qquad\qquad\textbf{(e)}
\qquad\qquad\qquad\qquad\qquad\qquad\textbf{(f)}$\\

\noindent { \small \textbf{Figure 3.} (Color online) Breather wave via solution \eqref{SSF-6} with parameters
$\alpha=1, \mathcal {Q}_{+}=\mathcal {I}_{2}, \zeta_{1}=2i, k_{0}=1,\gamma_{1}=1,\gamma_{0}=1,\gamma_{-1}=1$
$(\textbf{a},\textbf{b},\textbf{c})$:  $\beta=0.1$;
$(\textbf{d},\textbf{e},\textbf{f})$:  $\beta=1$.\\}

\noindent
{\rotatebox{0}{\includegraphics[width=4.8cm,height=2.5cm,angle=0]{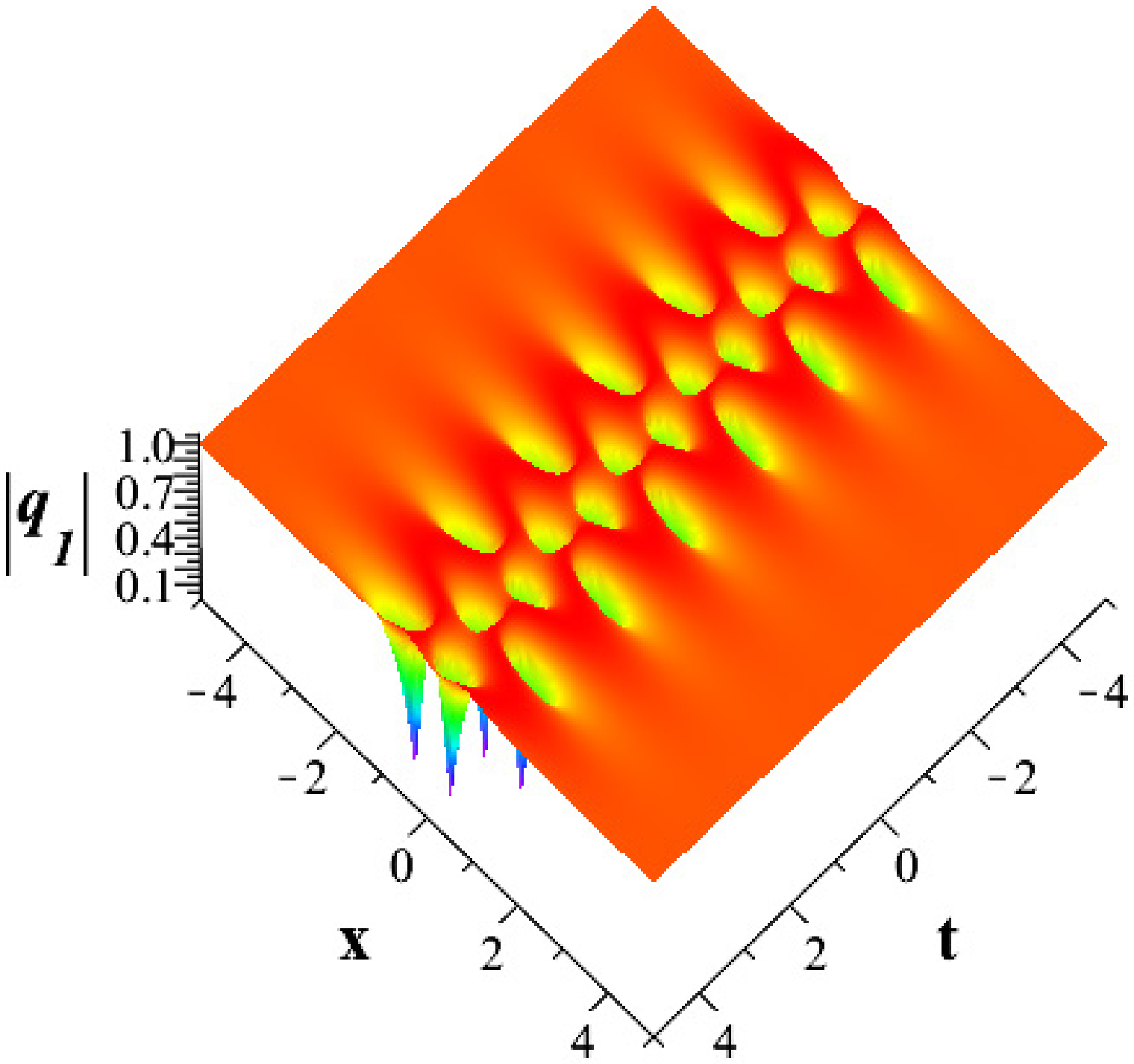}}}
{\rotatebox{0}{\includegraphics[width=4.8cm,height=2.5cm,angle=0]{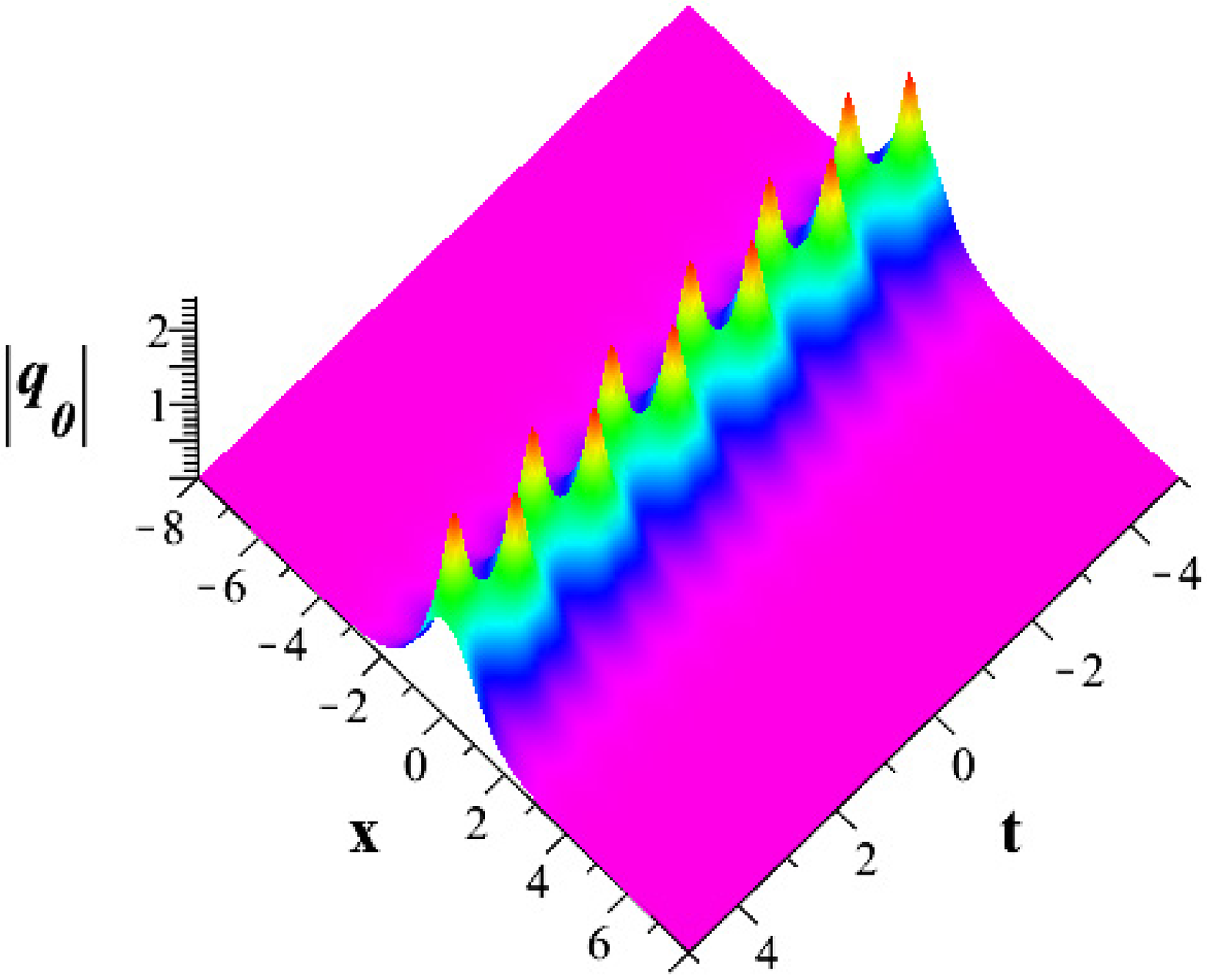}}}
{\rotatebox{0}{\includegraphics[width=4.8cm,height=2.5cm,angle=0]{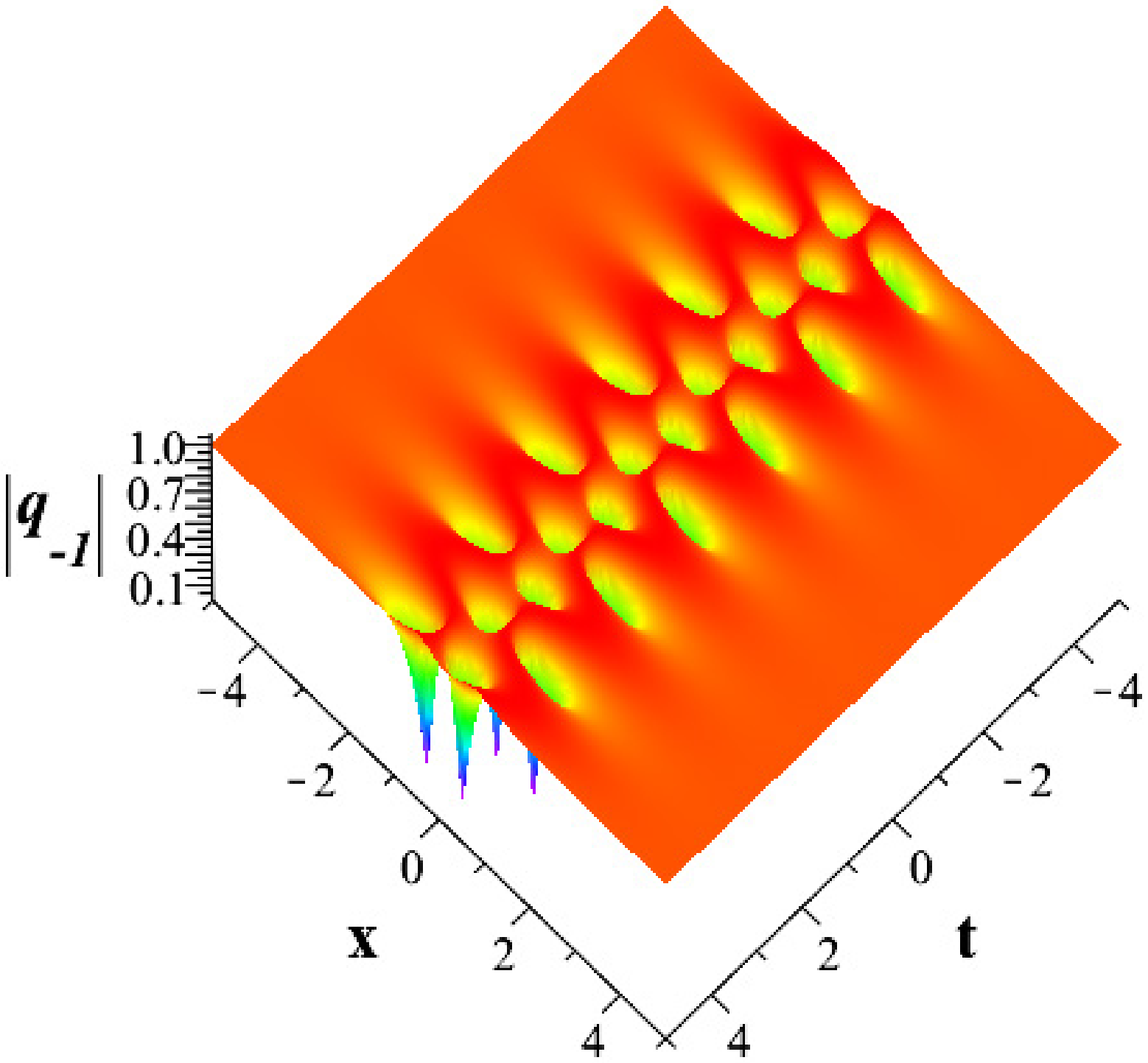}}}

$\qquad\qquad\textbf{(a)}\qquad\qquad\qquad\qquad\qquad\textbf{(b)}
\qquad\qquad\qquad\qquad\qquad\qquad\textbf{(c)}$\\

%\noindent
%{\rotatebox{0}{\includegraphics[width=4.8cm,height=3.9cm,angle=0]{3-4.eps}}}
%~~~
%{\rotatebox{0}{\includegraphics[width=4.8cm,height=3.9cm,angle=0]{3-5.eps}}}
%~~~
%{\rotatebox{0}{\includegraphics[width=4.8cm,height=3.9cm,angle=0]{3-6.eps}}}
%
%$\qquad\qquad\textbf{(d)}\qquad\qquad\qquad\qquad\qquad\qquad\qquad\textbf{(e)}
%\qquad\qquad\qquad\qquad\qquad\qquad\qquad\textbf{(f)}$\\
\noindent { \small \textbf{Figure 4.} (Color online) Breather wave via solution \eqref{SSF-6} with parameters
$\alpha=-1, \beta=0.01, \mathcal {Q}_{+}=\mathcal {I}_{2}, \zeta_{1}=2i, k_{0}=1,\gamma_{1}=1,\gamma_{0}=2,\gamma_{-1}=1$.\\}

\noindent
{\rotatebox{0}{\includegraphics[width=4.8cm,height=2.5cm,angle=0]{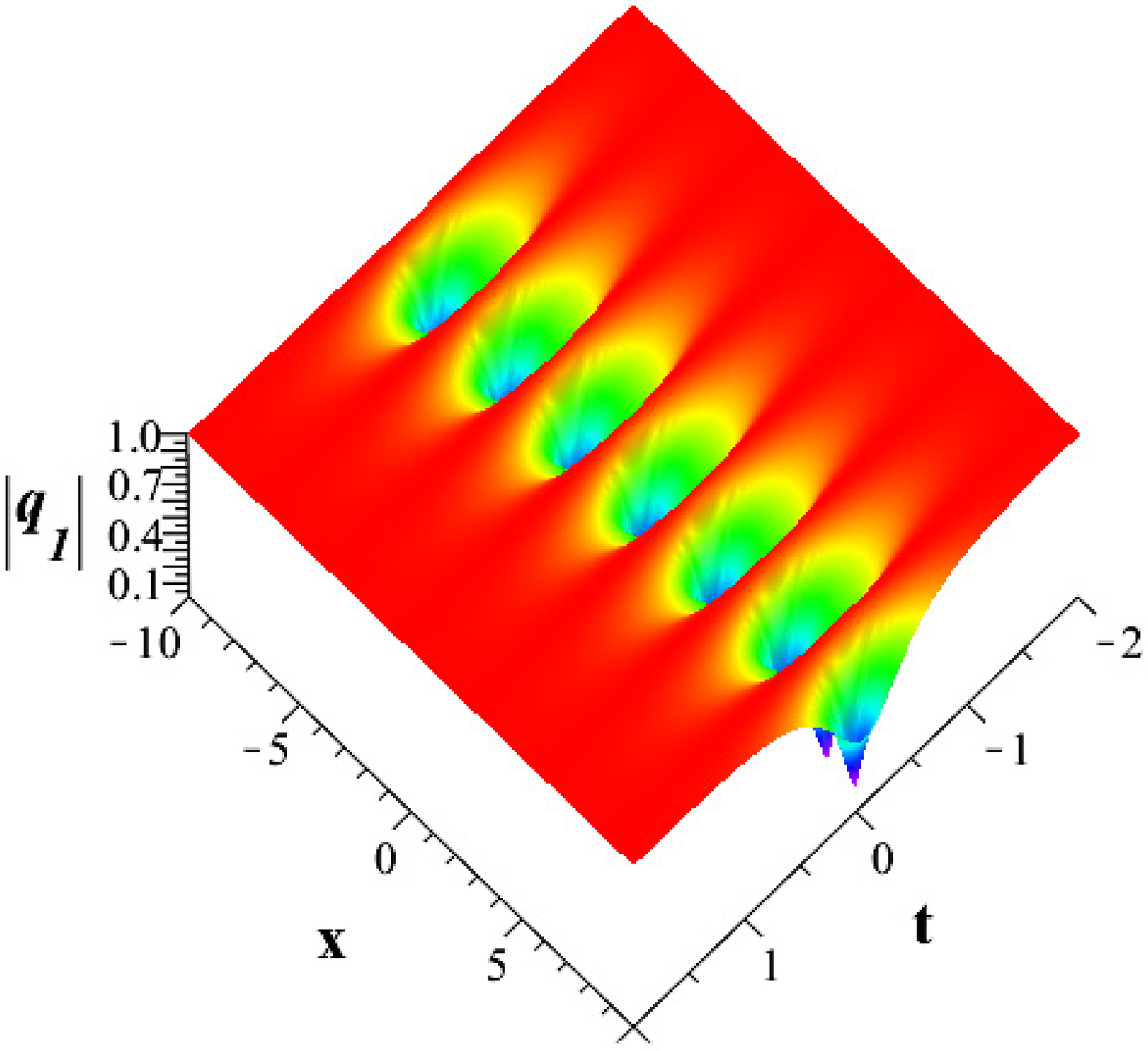}}}
{\rotatebox{0}{\includegraphics[width=4.8cm,height=2.5cm,angle=0]{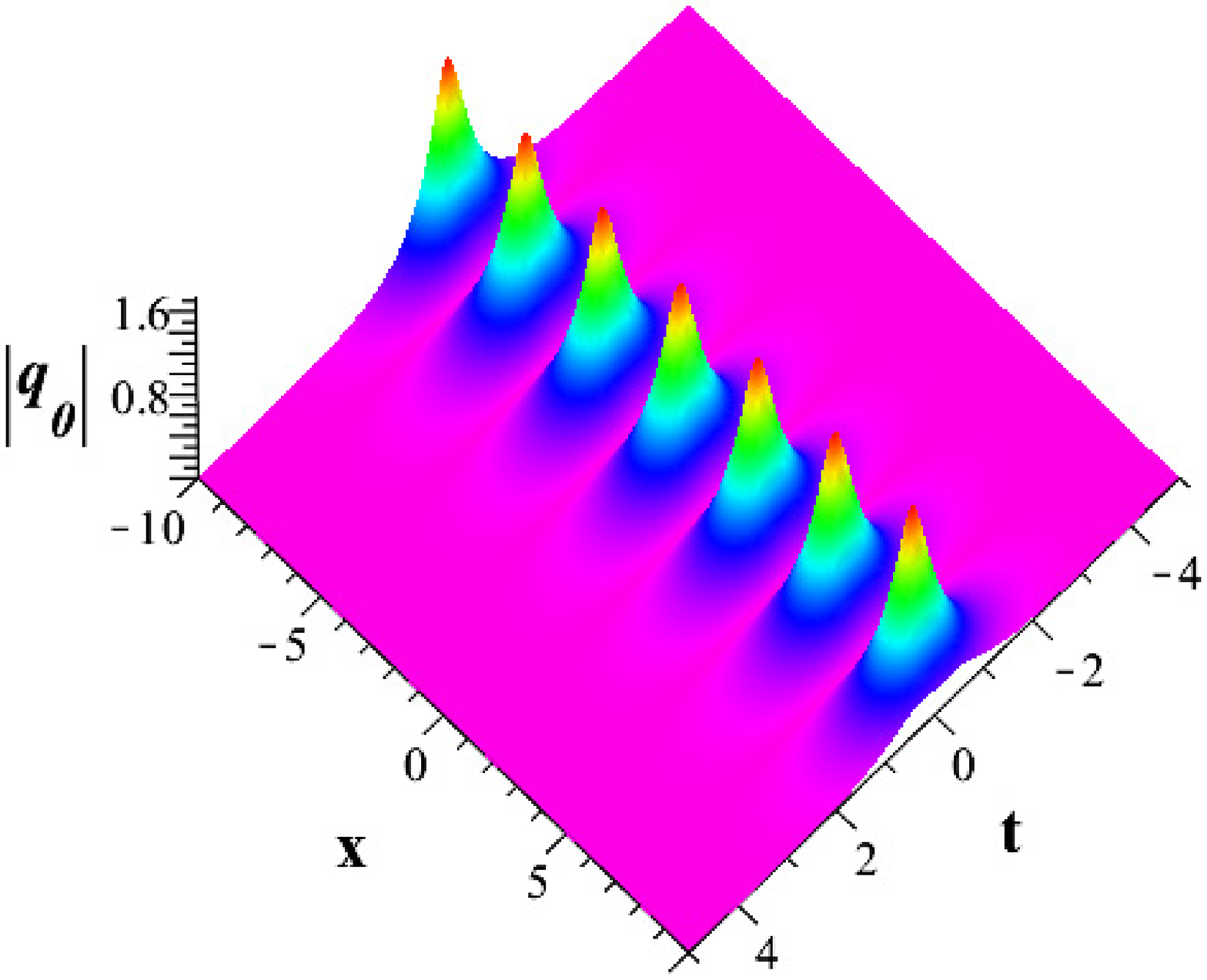}}}
{\rotatebox{0}{\includegraphics[width=4.8cm,height=2.5cm,angle=0]{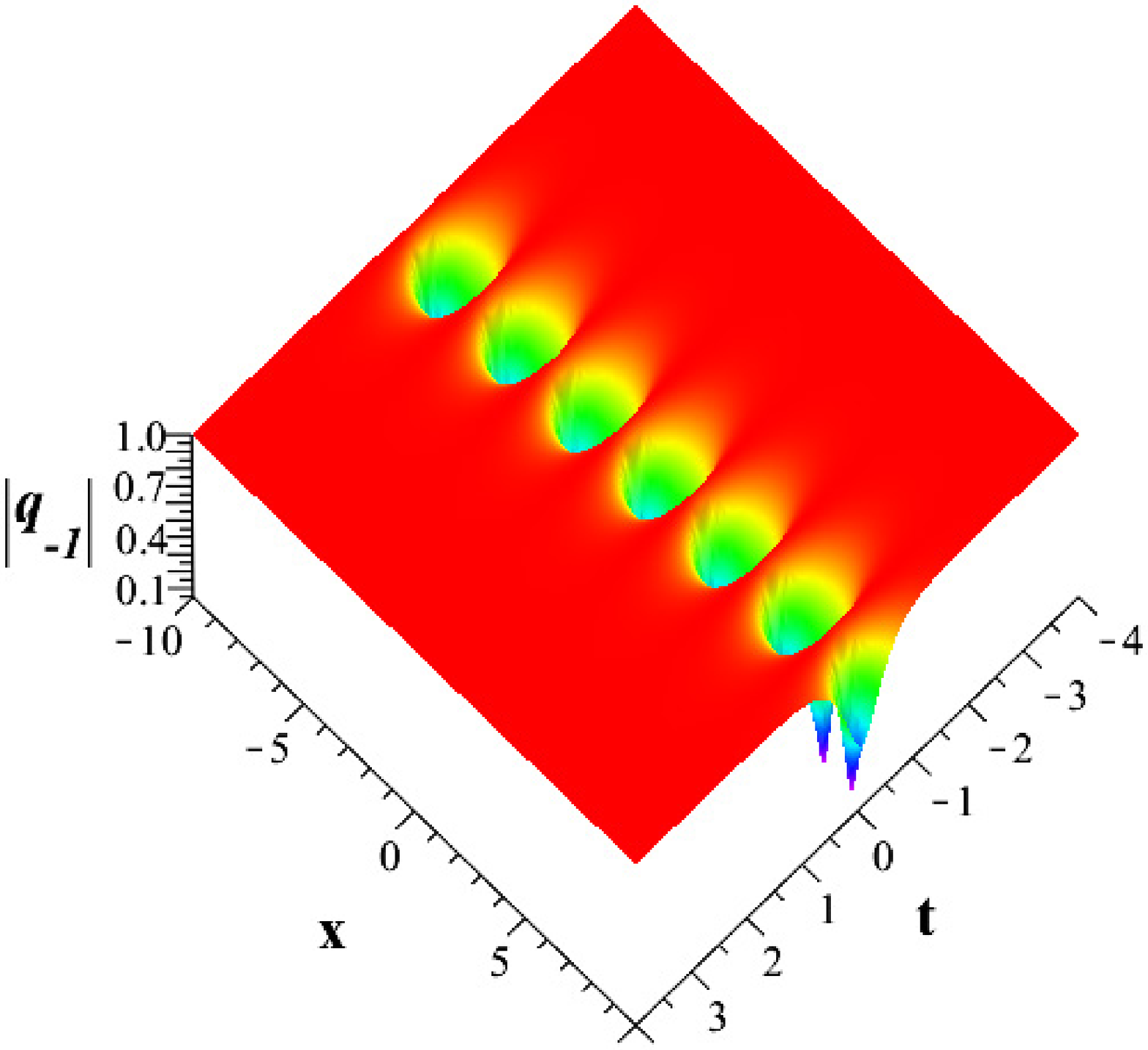}}}

$\qquad\qquad\textbf{(a)}\qquad\qquad\qquad\qquad\qquad\textbf{(b)}
\qquad\qquad\qquad\qquad\qquad\qquad\textbf{(c)}$\\

%\noindent
%{\rotatebox{0}{\includegraphics[width=4.8cm,height=3.9cm,angle=0]{3-4.eps}}}
%~~~
%{\rotatebox{0}{\includegraphics[width=4.8cm,height=3.9cm,angle=0]{3-5.eps}}}
%~~~
%{\rotatebox{0}{\includegraphics[width=4.8cm,height=3.9cm,angle=0]{3-6.eps}}}
%
%$\qquad\qquad\textbf{(d)}\qquad\qquad\qquad\qquad\qquad\qquad\qquad\textbf{(e)}
%\qquad\qquad\qquad\qquad\qquad\qquad\qquad\textbf{(f)}$\\
\noindent { \small \textbf{Figure 5.} (Color online) Breather wave via solution \eqref{SSF-6} with parameters
$\alpha=-1, \beta=0.01, \mathcal {Q}_{+}=\mathcal {I}_{2}, \zeta_{1}=0.5+0.8i, k_{0}=1,\gamma_{1}=0,\gamma_{0}=1,\gamma_{-1}=0$.\\}

\noindent
{\rotatebox{0}{\includegraphics[width=4.8cm,height=2.5cm,angle=0]{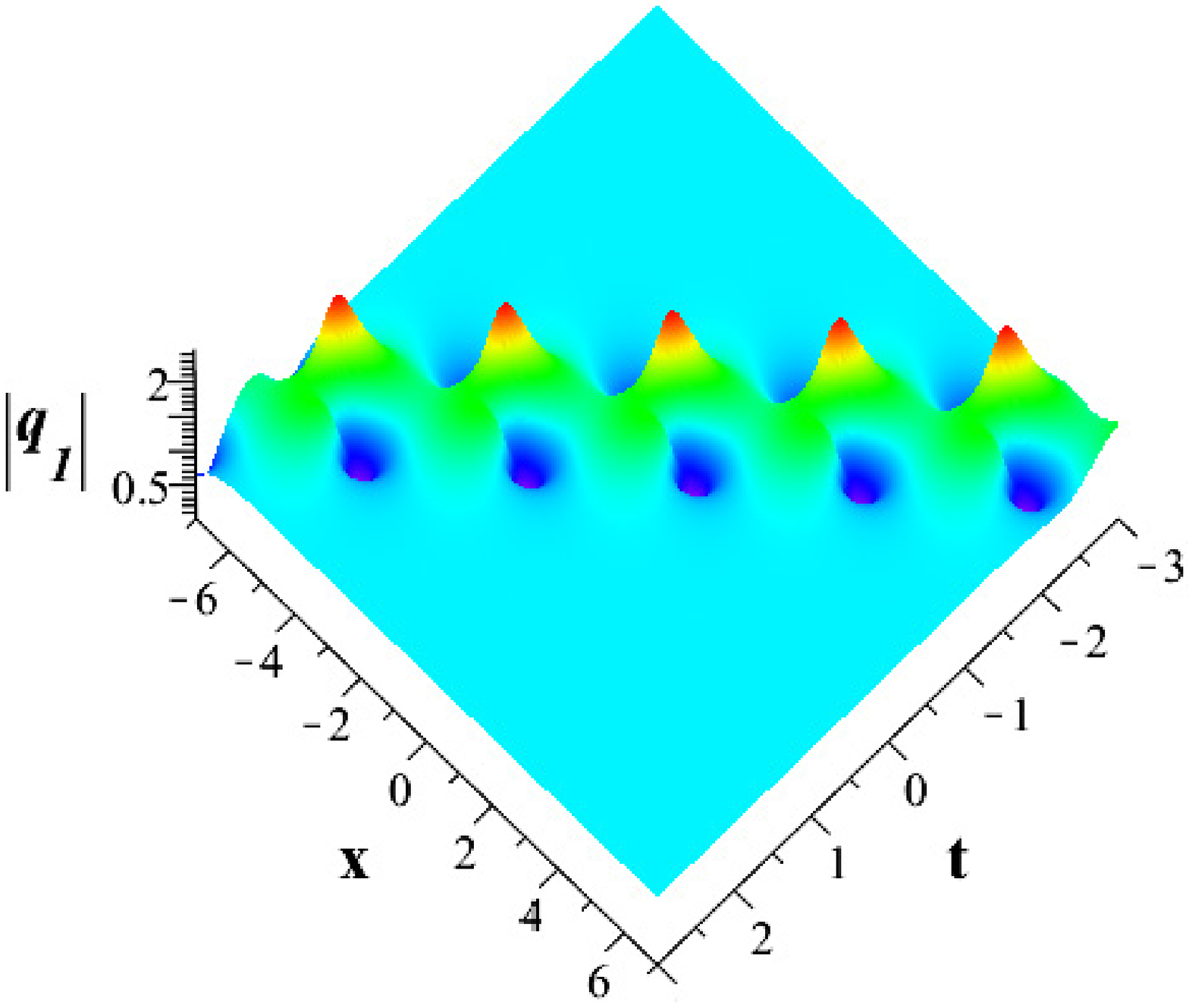}}}
{\rotatebox{0}{\includegraphics[width=4.8cm,height=2.5cm,angle=0]{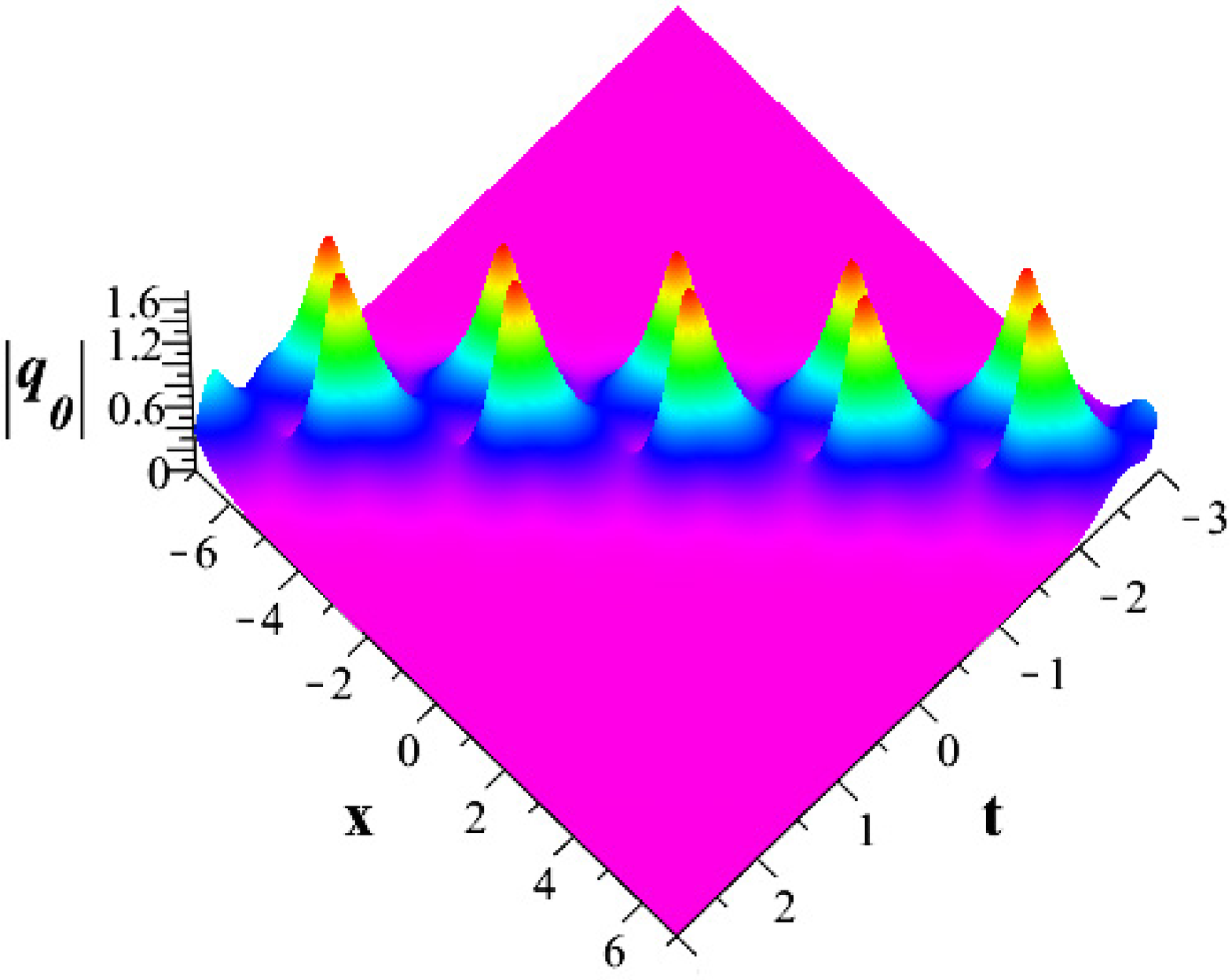}}}
{\rotatebox{0}{\includegraphics[width=4.8cm,height=2.5cm,angle=0]{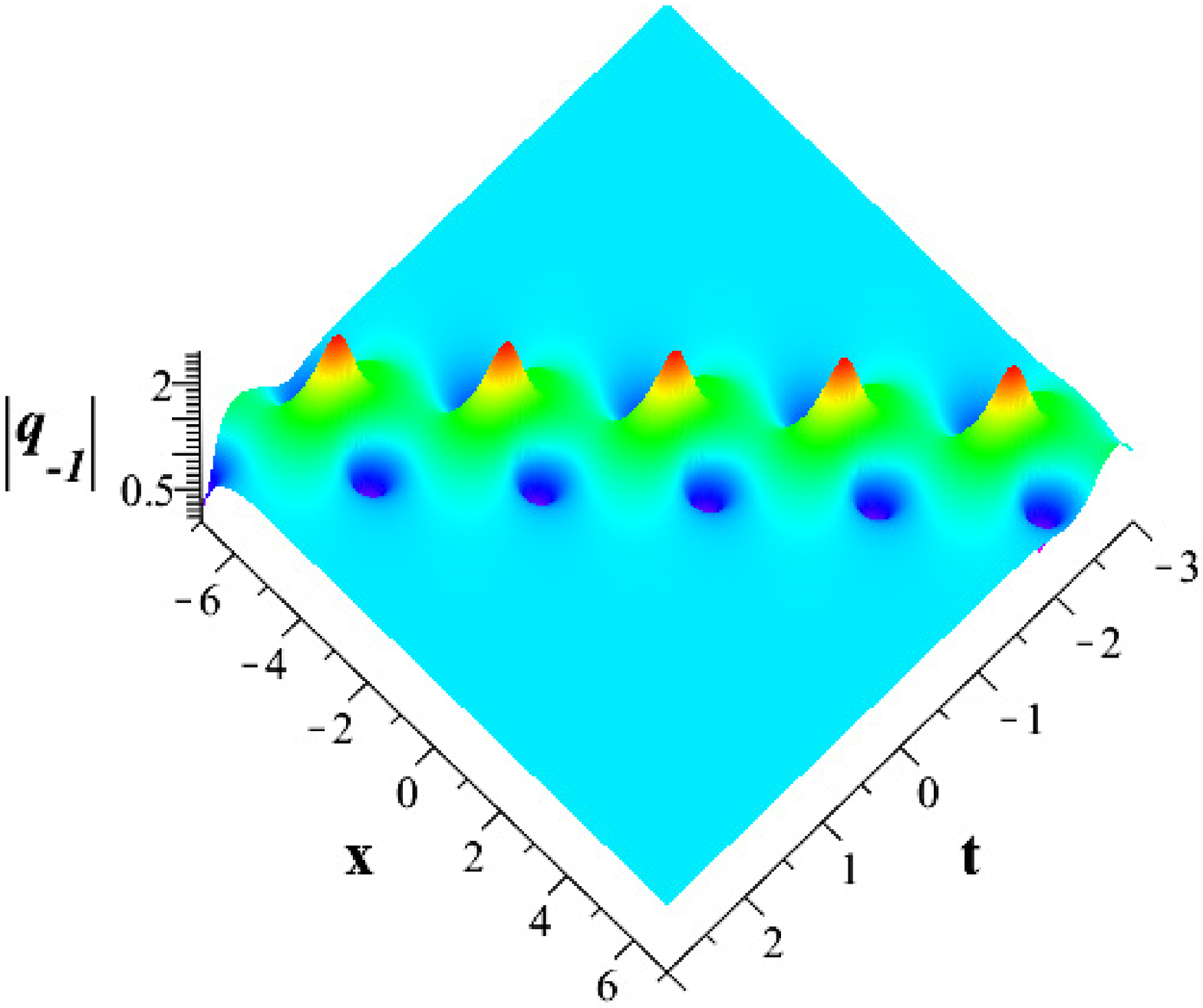}}}

$\qquad\qquad\textbf{(a)}\qquad\qquad\qquad\qquad\qquad\textbf{(b)}
\qquad\qquad\qquad\qquad\qquad\qquad\textbf{(c)}$\\

\noindent { \small \textbf{Figure 6.} (Color online) Breather wave via solution \eqref{SSF-6} with parameters
$\alpha=1, \beta=0.01, \mathcal {Q}_{+}=\mathcal {I}_{2}, \zeta_{1}=1+2i, k_{0}=1,\gamma_{1}=1,\gamma_{0}=1,\gamma_{-1}=2$.\\}

\noindent
{\rotatebox{0}{\includegraphics[width=4.8cm,height=2.5cm,angle=0]{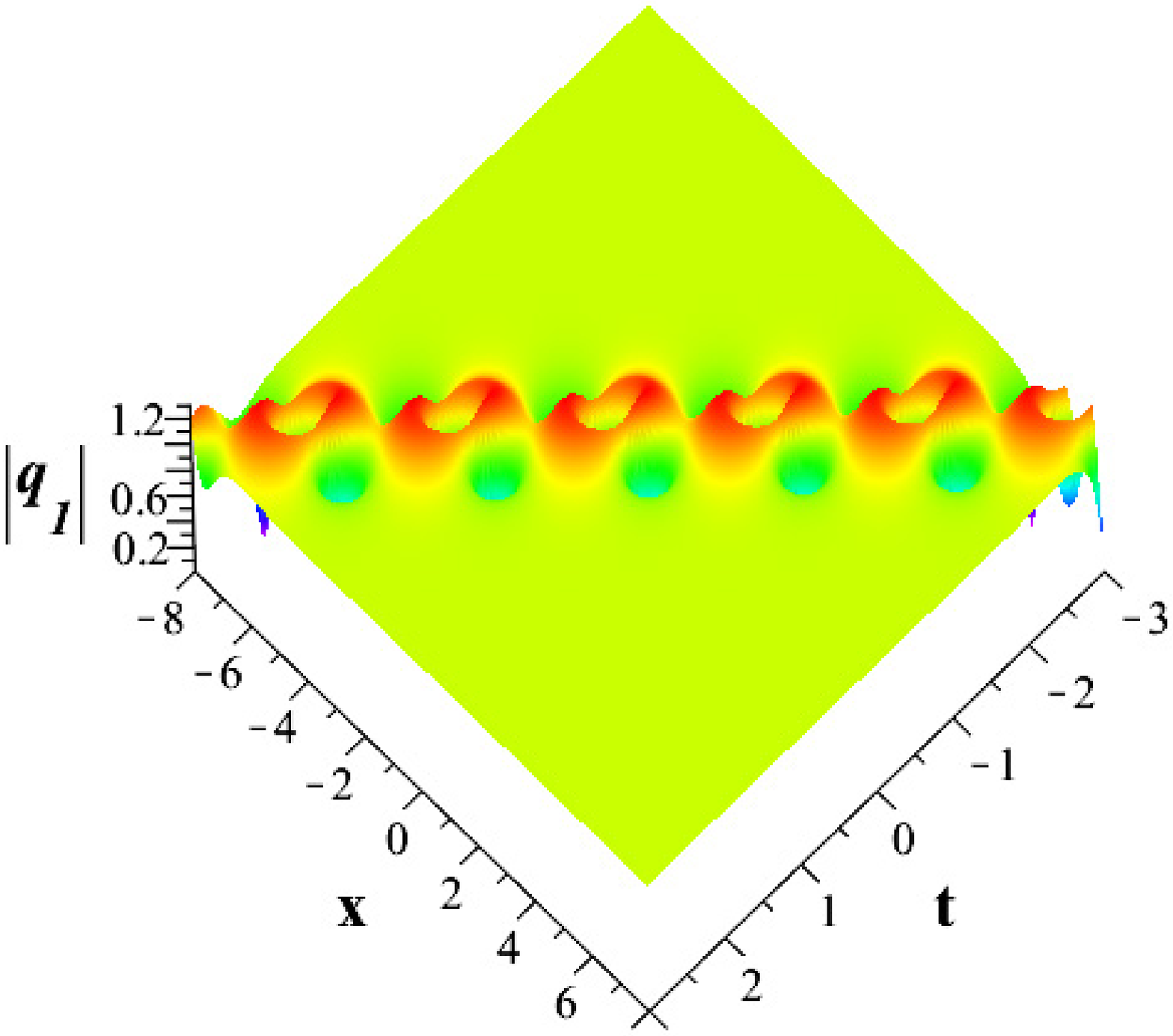}}}
{\rotatebox{0}{\includegraphics[width=4.8cm,height=2.5cm,angle=0]{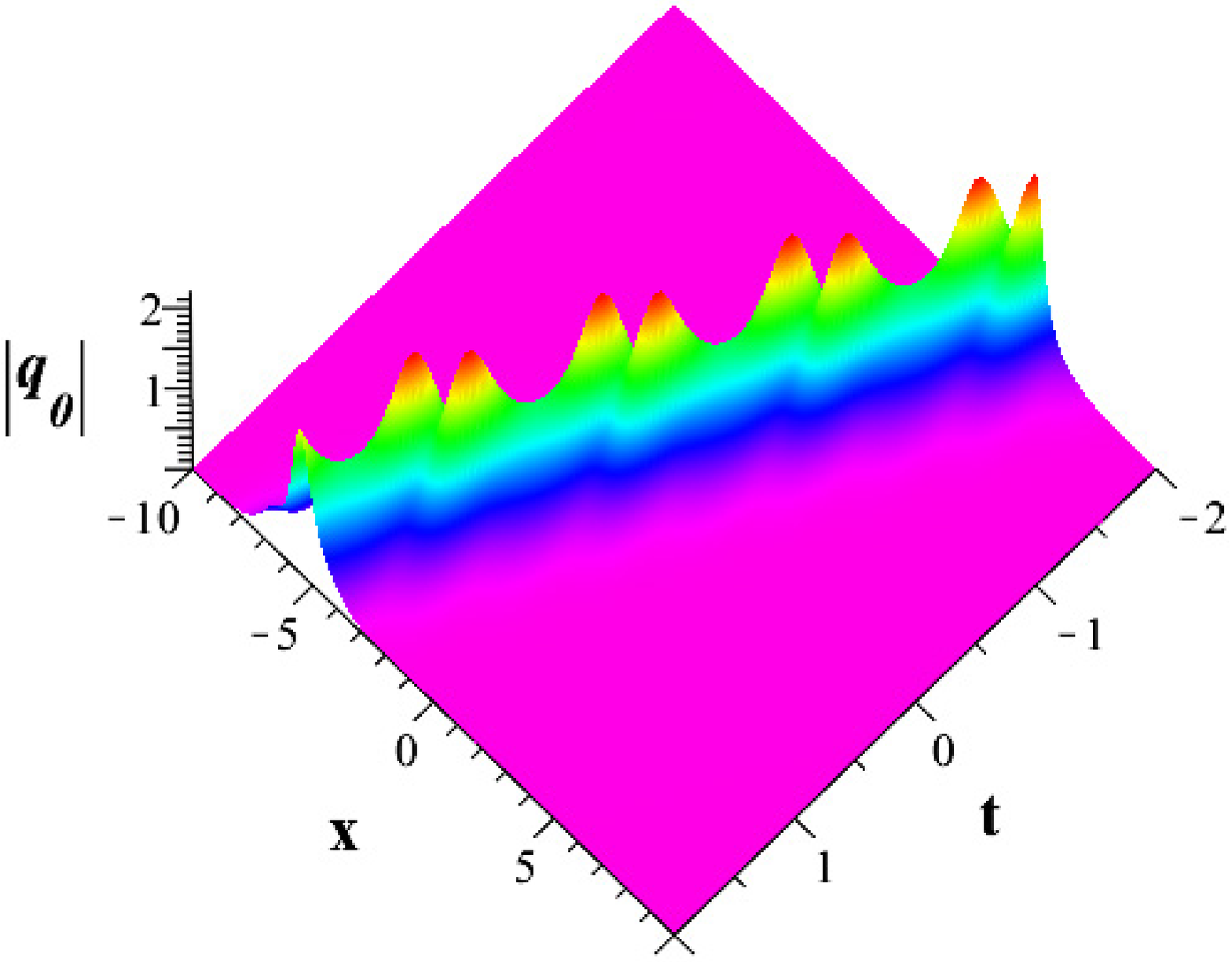}}}
{\rotatebox{0}{\includegraphics[width=4.8cm,height=2.5cm,angle=0]{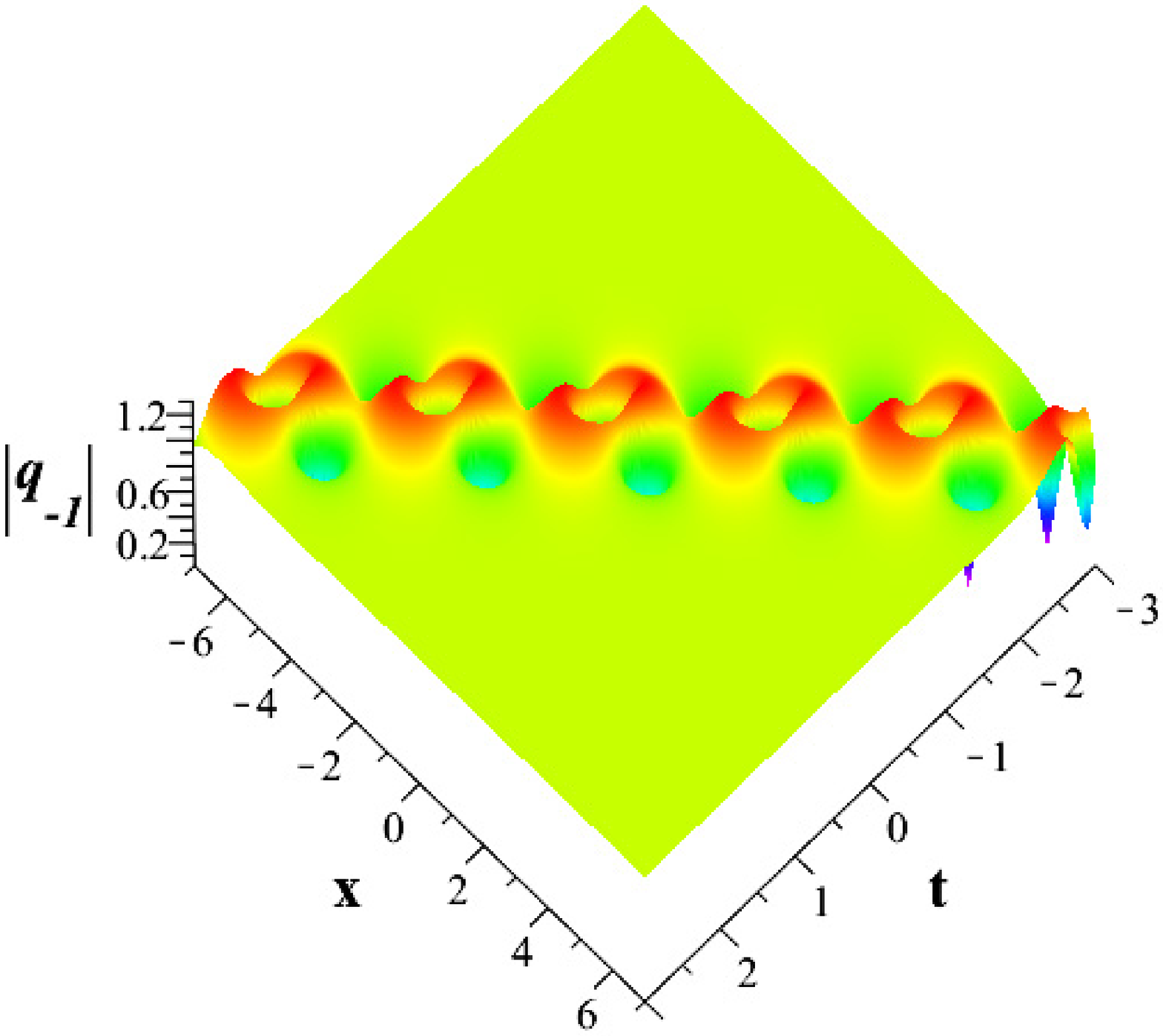}}}

$\qquad\qquad\textbf{(a)}\qquad\qquad\qquad\qquad\qquad\textbf{(b)}
\qquad\qquad\qquad\qquad\qquad\qquad\textbf{(c)}$\\

\noindent { \small \textbf{Figure 7.} (Color online) Breather wave via solution \eqref{SSF-6} with parameters
$\alpha=1, \beta=0.1, \mathcal {Q}_{+}=\mathcal {I}_{2}, \zeta_{1}=1+2i, k_{0}=1,\gamma_{1}=i,\gamma_{0}=1+i,\gamma_{-1}=i$.\\}

\noindent
{\rotatebox{0}{\includegraphics[width=4.8cm,height=2.5cm,angle=0]{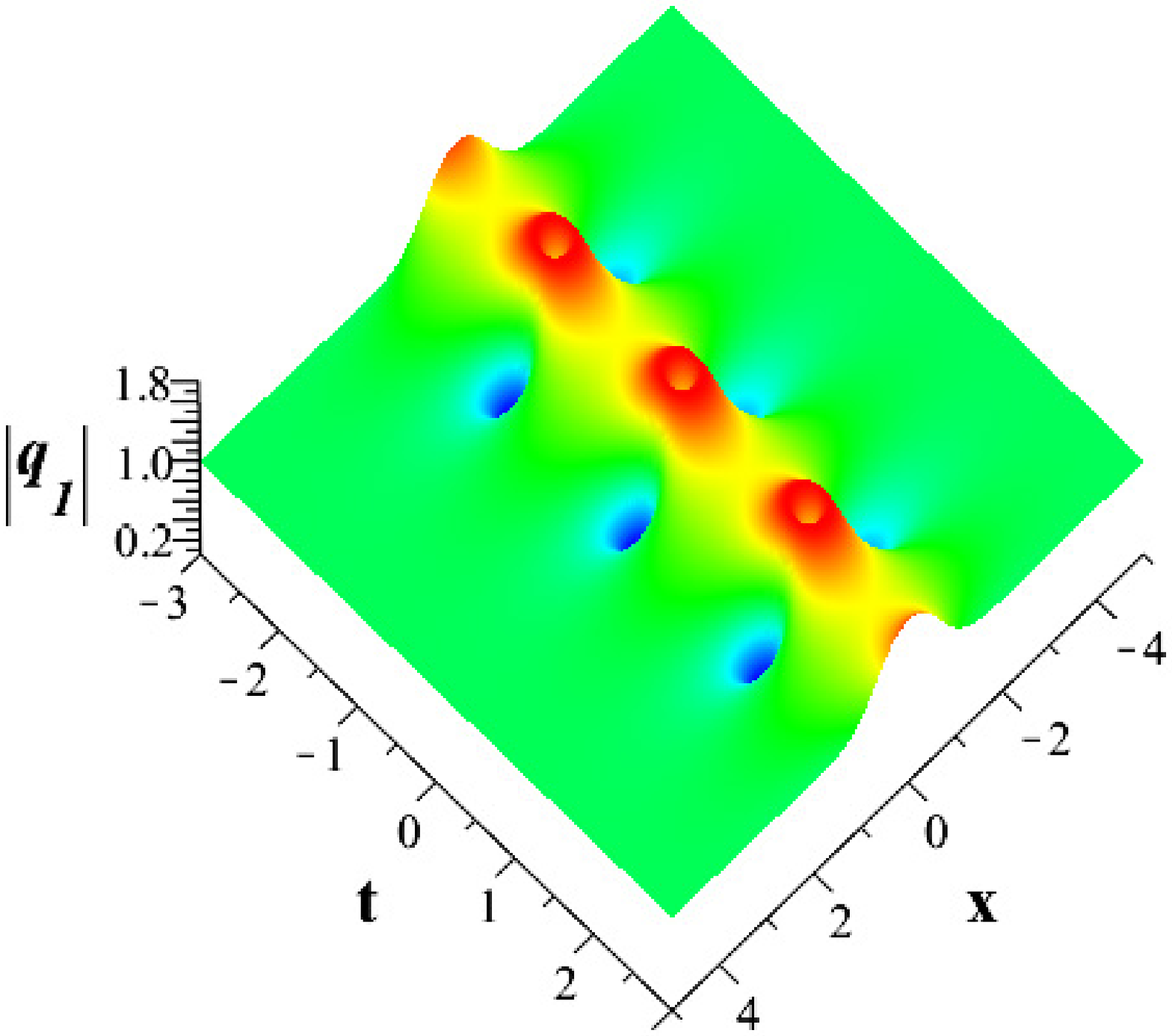}}}
{\rotatebox{0}{\includegraphics[width=4.8cm,height=2.5cm,angle=0]{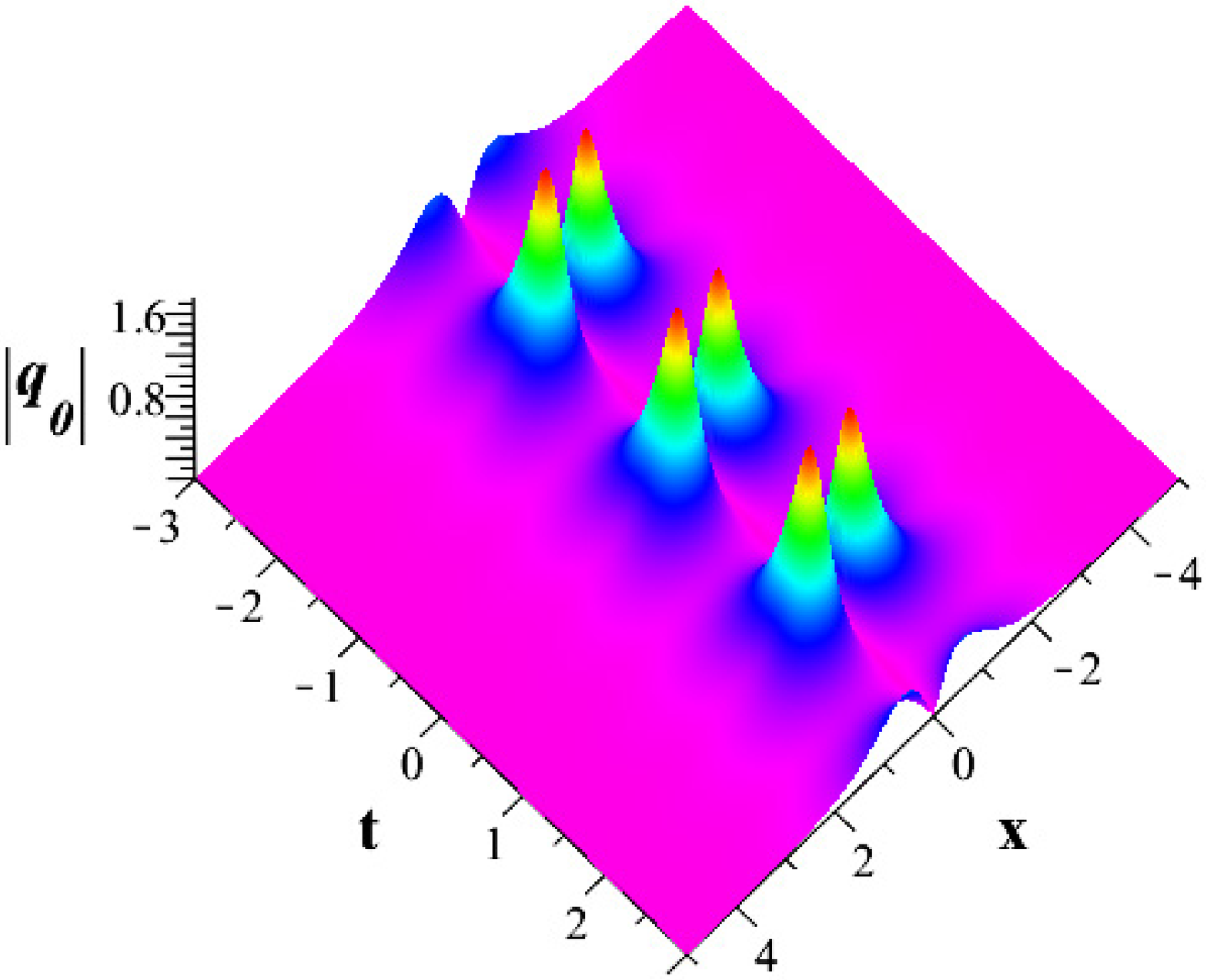}}}
{\rotatebox{0}{\includegraphics[width=4.8cm,height=2.5cm,angle=0]{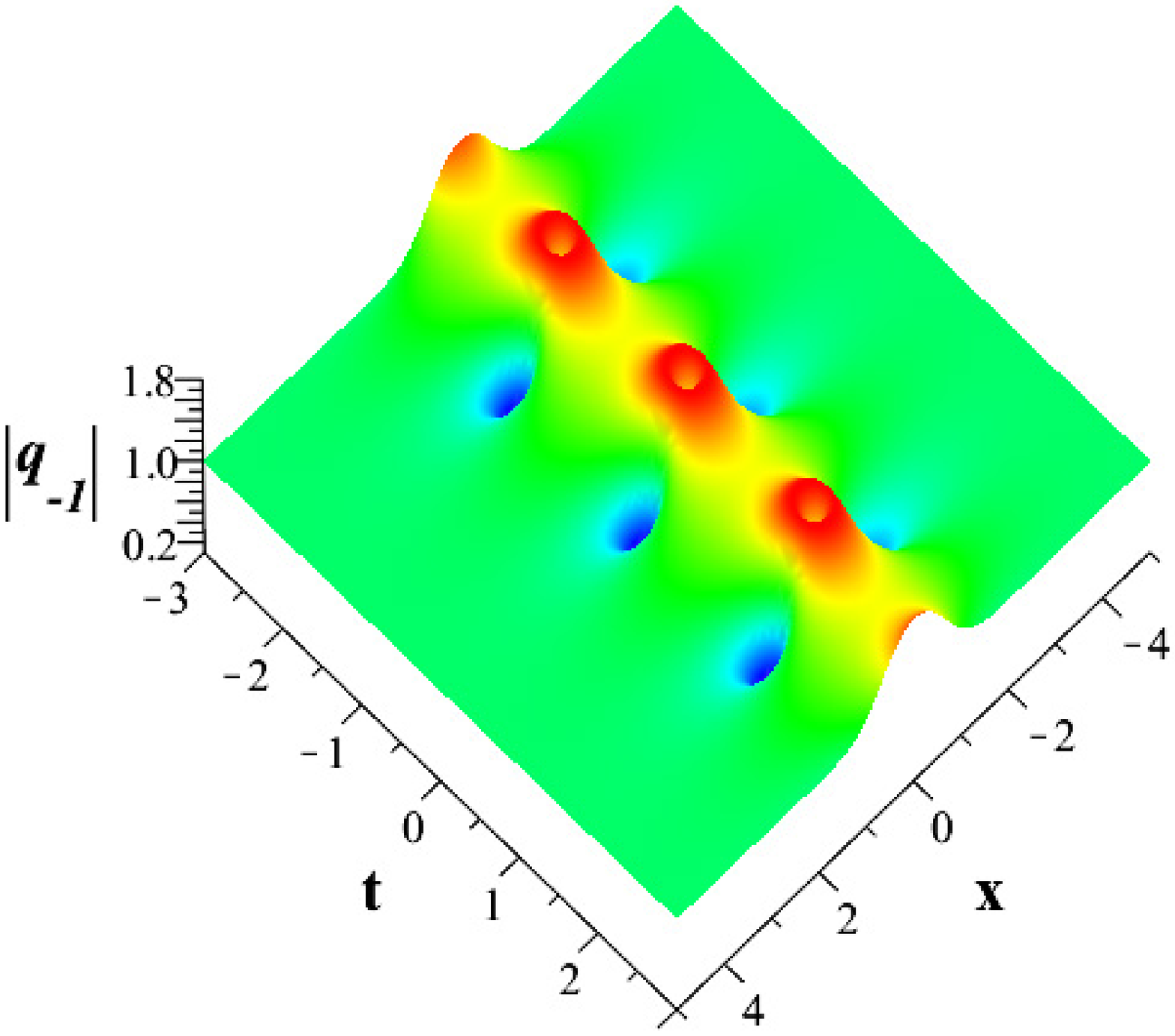}}}

$\qquad\qquad\textbf{(a)}\qquad\qquad\qquad\qquad\qquad\textbf{(b)}
\qquad\qquad\qquad\qquad\qquad\qquad\textbf{(c)}$\\

\noindent { \small \textbf{Figure 8.} (Color online) Breather wave via solution \eqref{SSF-6} with parameters
$\alpha=-1, \beta=0.1, \mathcal {Q}_{+}=\mathcal {I}_{2}, \zeta_{1}=2i, k_{0}=1,\gamma_{1}=2i,\gamma_{0}=i,\gamma_{-1}=2i$.\\}

\noindent
{\rotatebox{0}{\includegraphics[width=4.8cm,height=2.5cm,angle=0]{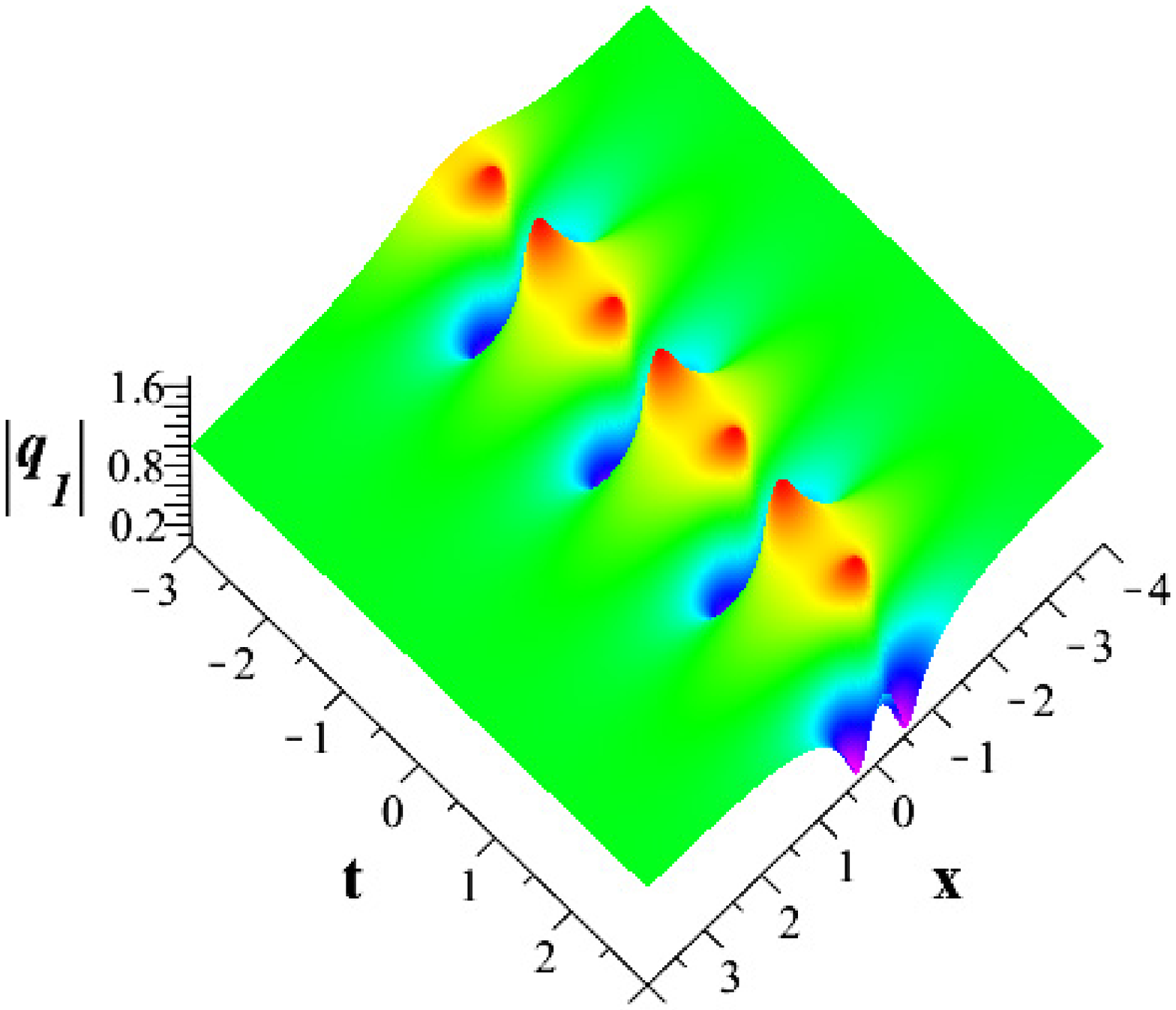}}}
{\rotatebox{0}{\includegraphics[width=4.8cm,height=2.5cm,angle=0]{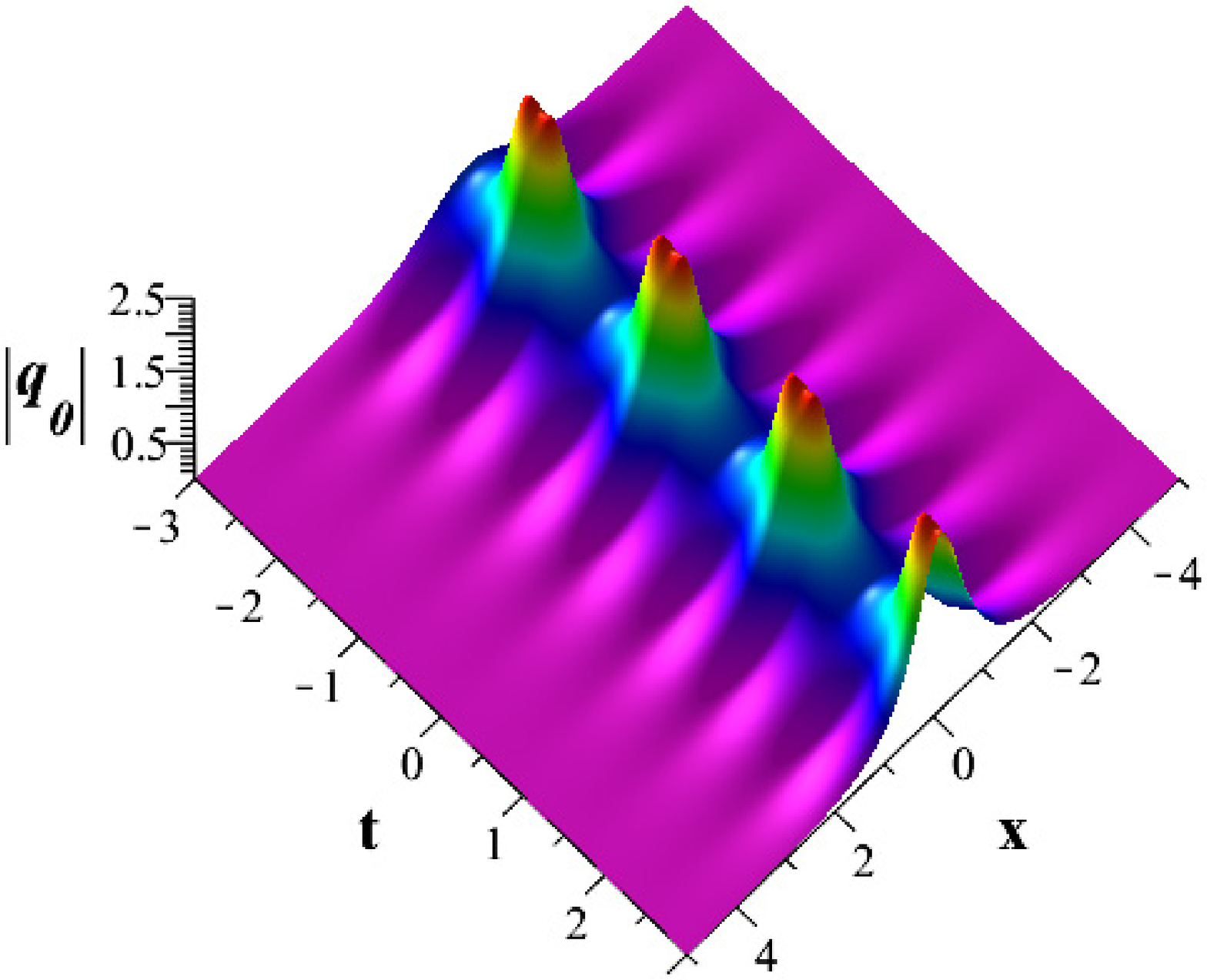}}}
{\rotatebox{0}{\includegraphics[width=4.8cm,height=2.5cm,angle=0]{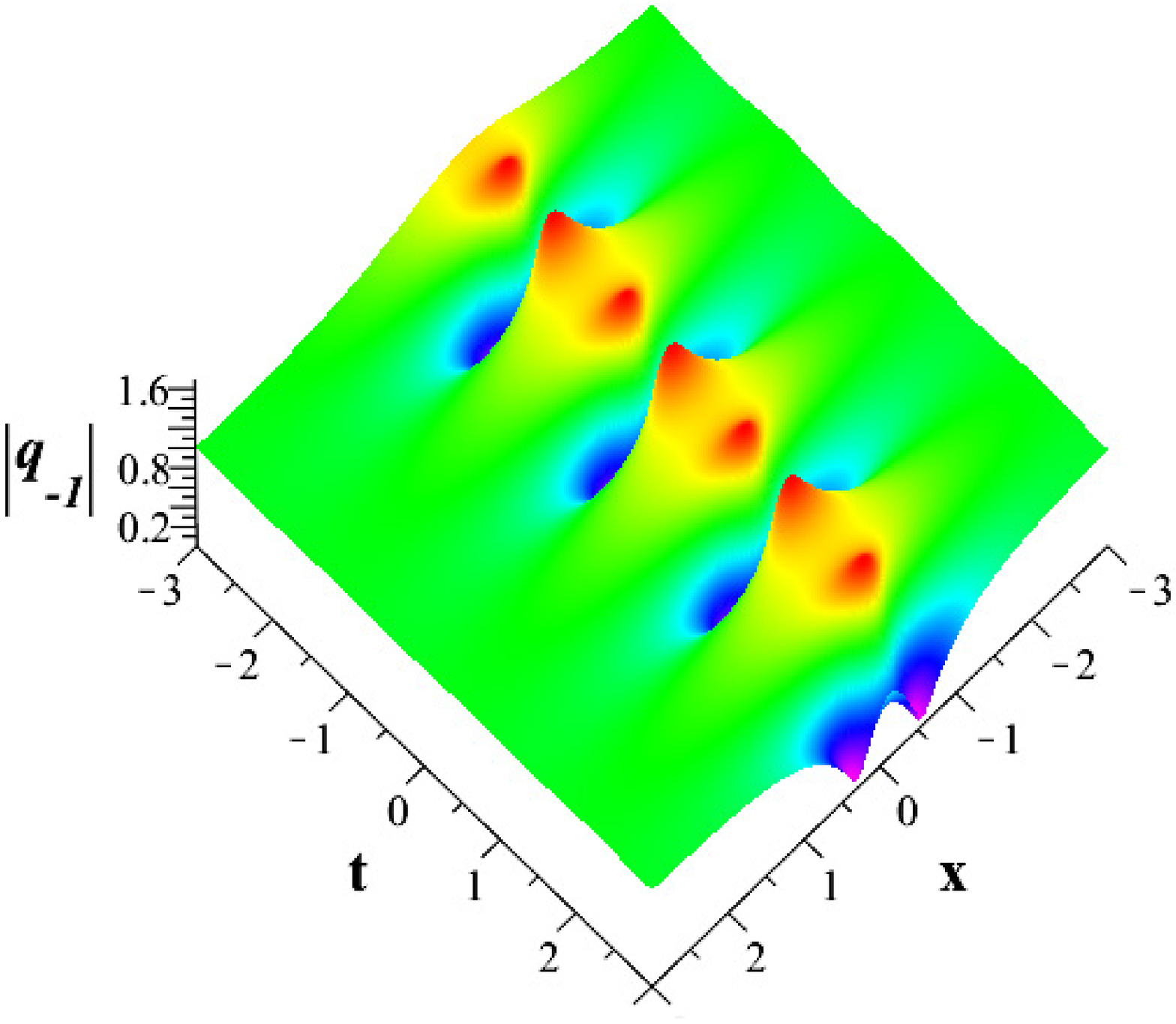}}}

$\qquad\qquad\textbf{(a)}\qquad\qquad\qquad\qquad\qquad\textbf{(b)}
\qquad\qquad\qquad\qquad\qquad\qquad\textbf{(c)}$\\

\noindent { \small \textbf{Figure 9.} (Color online) Breather wave via solution \eqref{SSF-6} with parameters
$\alpha=-1, \beta=0.1, \mathcal {Q}_{+}=\mathcal {I}_{2}, \zeta_{1}=2i, k_{0}=1,\gamma_{1}=1,\gamma_{0}=i,\gamma_{-1}=1$.\\}

\noindent
{\rotatebox{0}{\includegraphics[width=4.8cm,height=2.5cm,angle=0]{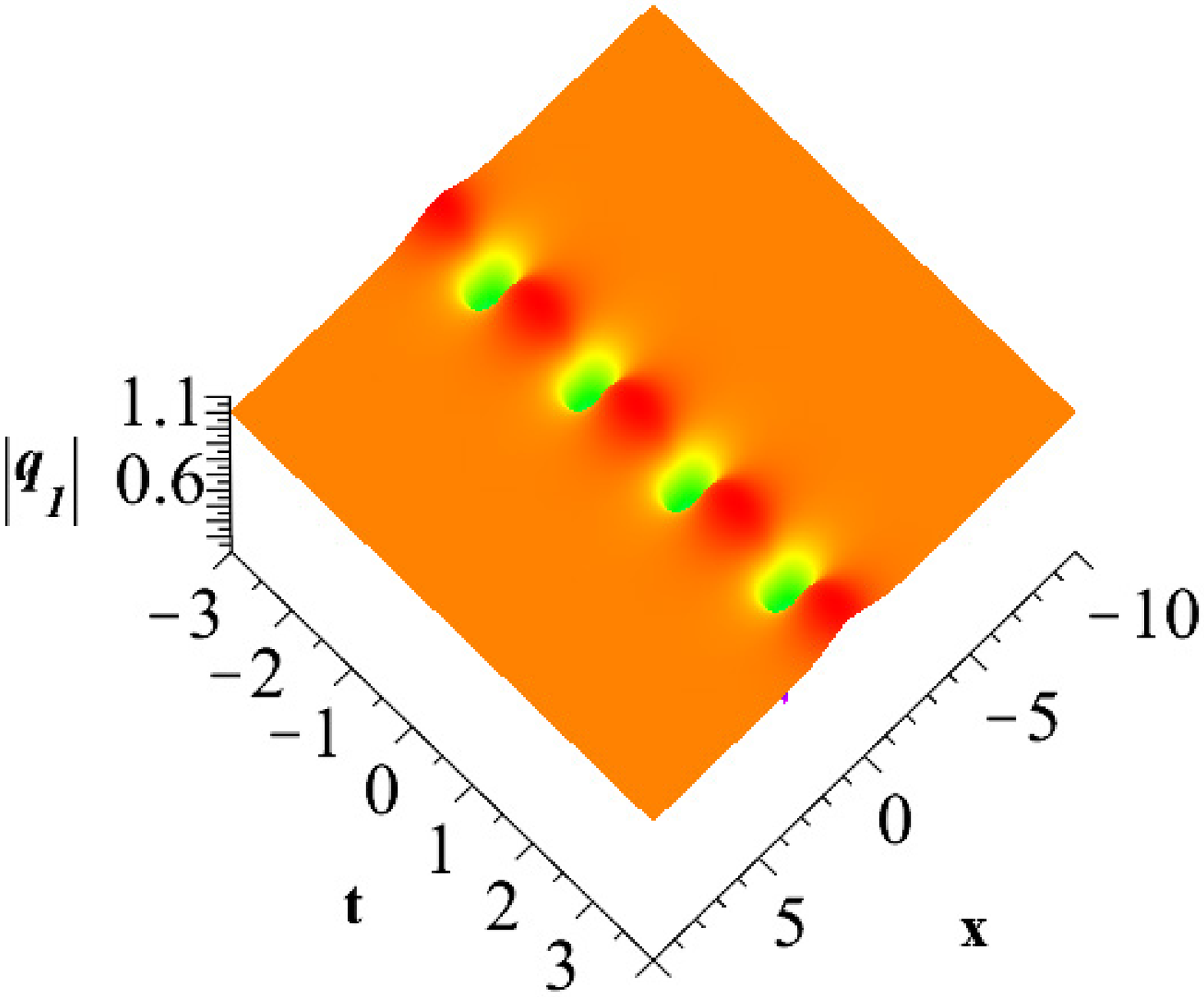}}}
{\rotatebox{0}{\includegraphics[width=4.8cm,height=2.5cm,angle=0]{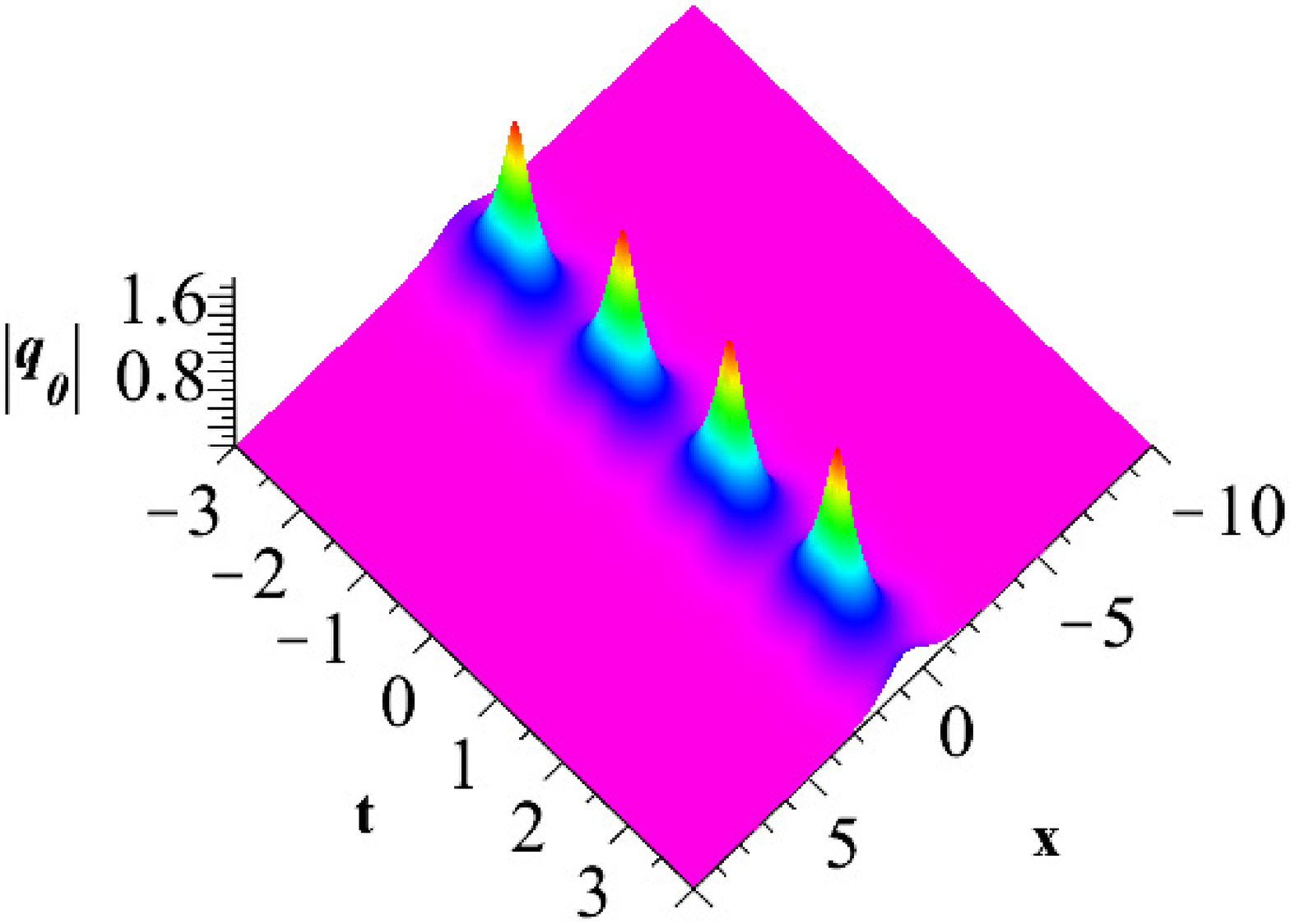}}}
{\rotatebox{0}{\includegraphics[width=4.8cm,height=2.5cm,angle=0]{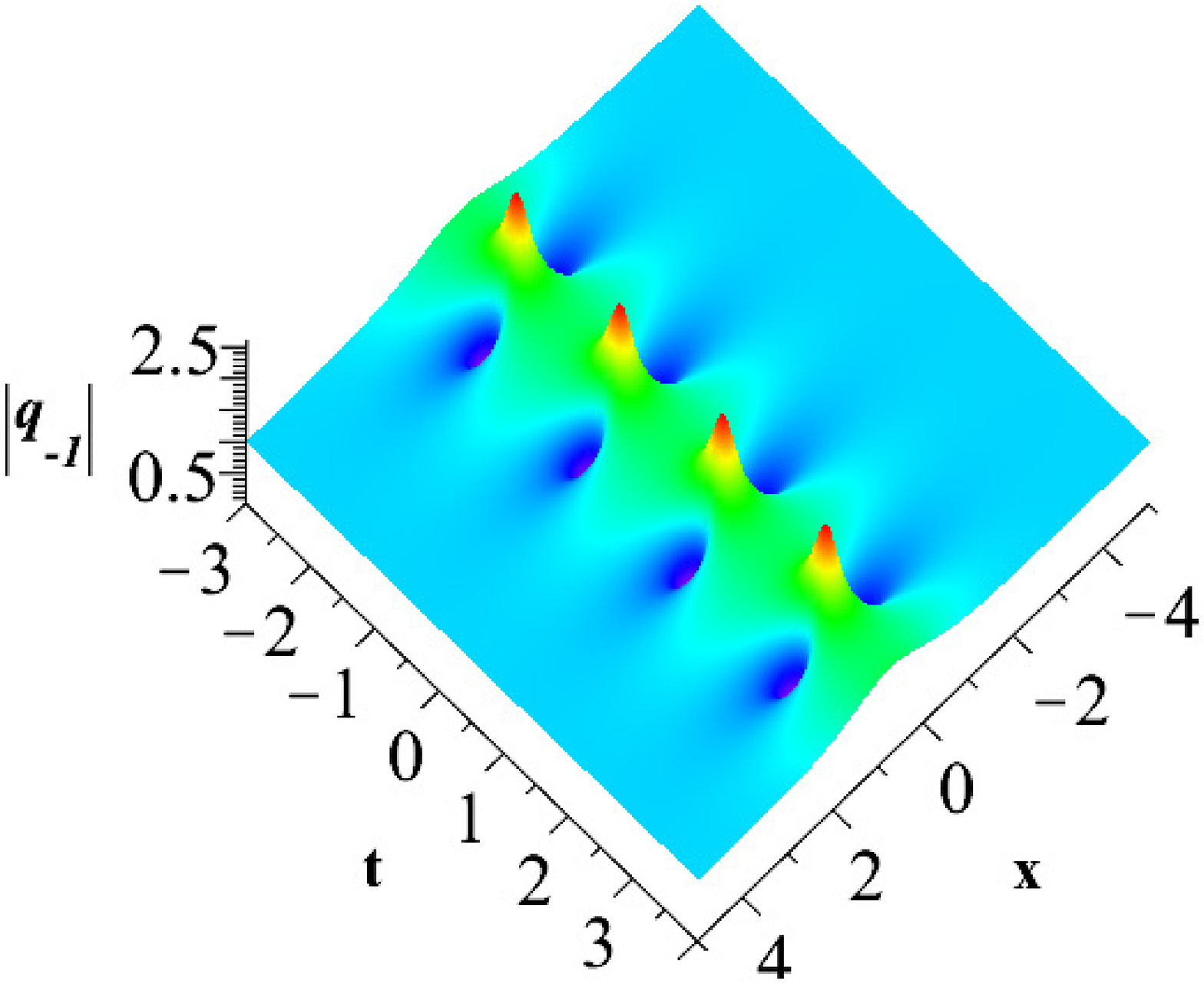}}}

$\qquad\qquad\textbf{(a)}\qquad\qquad\qquad\qquad\qquad\textbf{(b)}
\qquad\qquad\qquad\qquad\qquad\qquad\textbf{(c)}$\\

\noindent
{\rotatebox{0}{\includegraphics[width=4.8cm,height=2.5cm,angle=0]{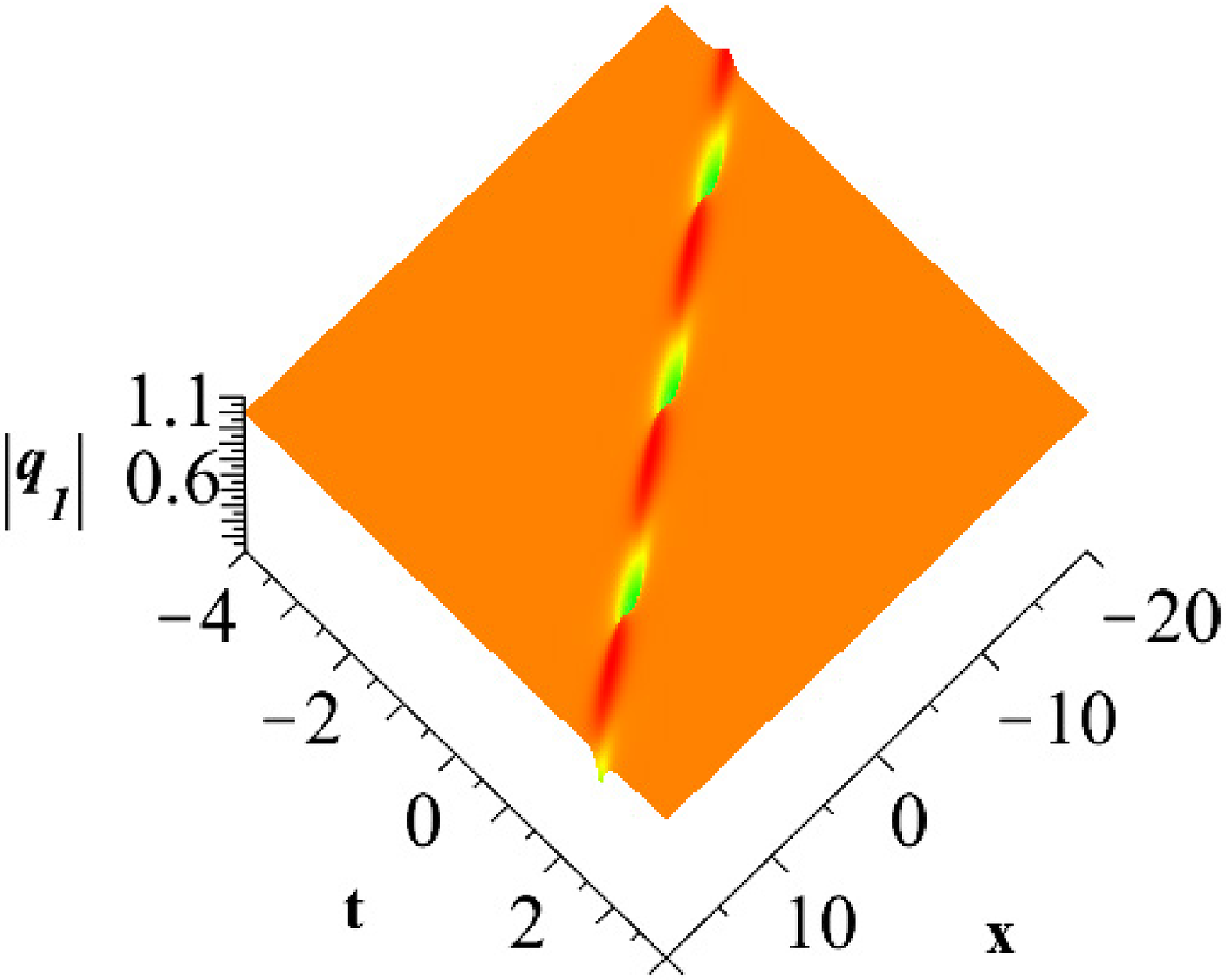}}}
{\rotatebox{0}{\includegraphics[width=4.8cm,height=2.5cm,angle=0]{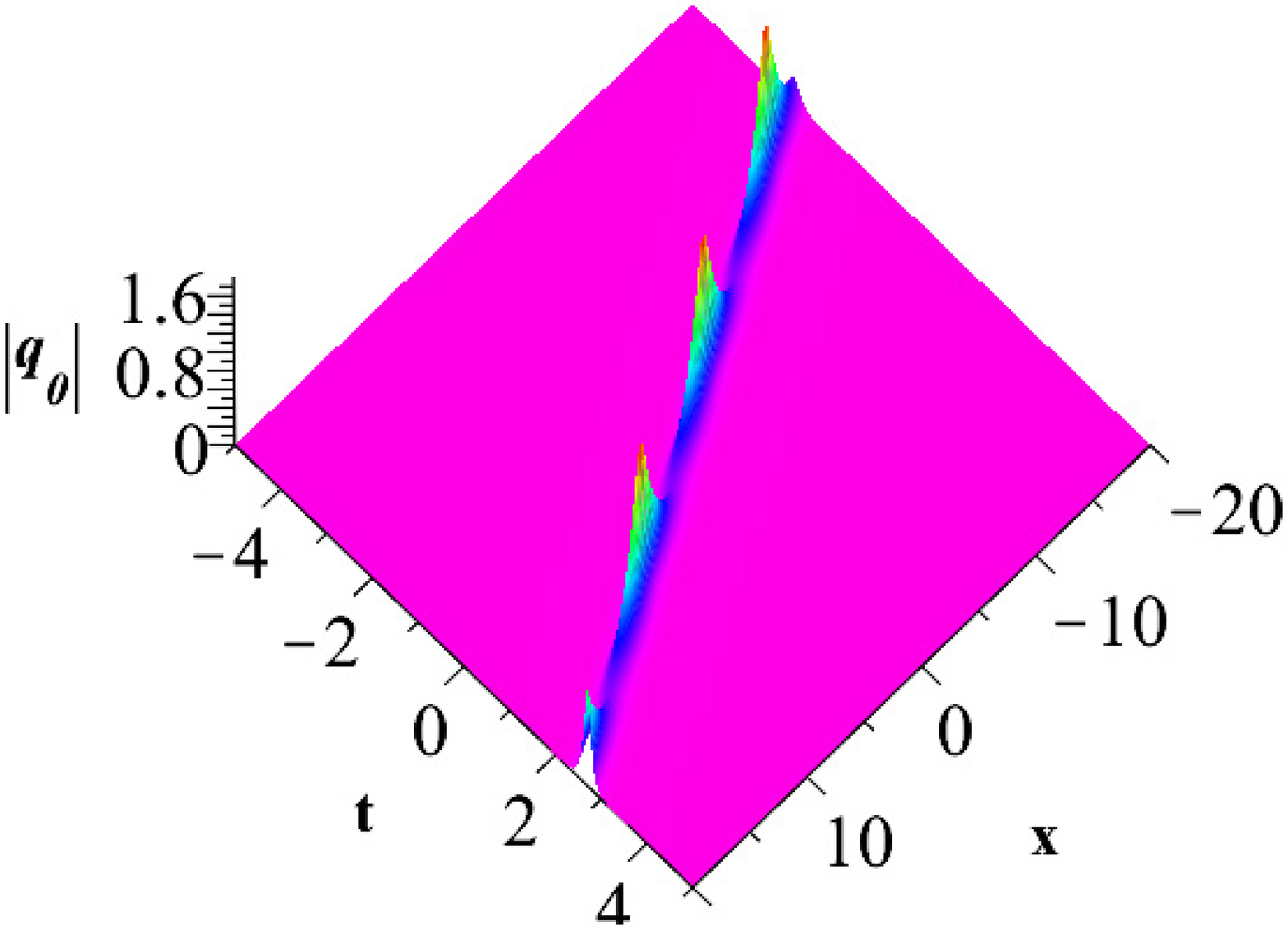}}}
{\rotatebox{0}{\includegraphics[width=4.8cm,height=2.5cm,angle=0]{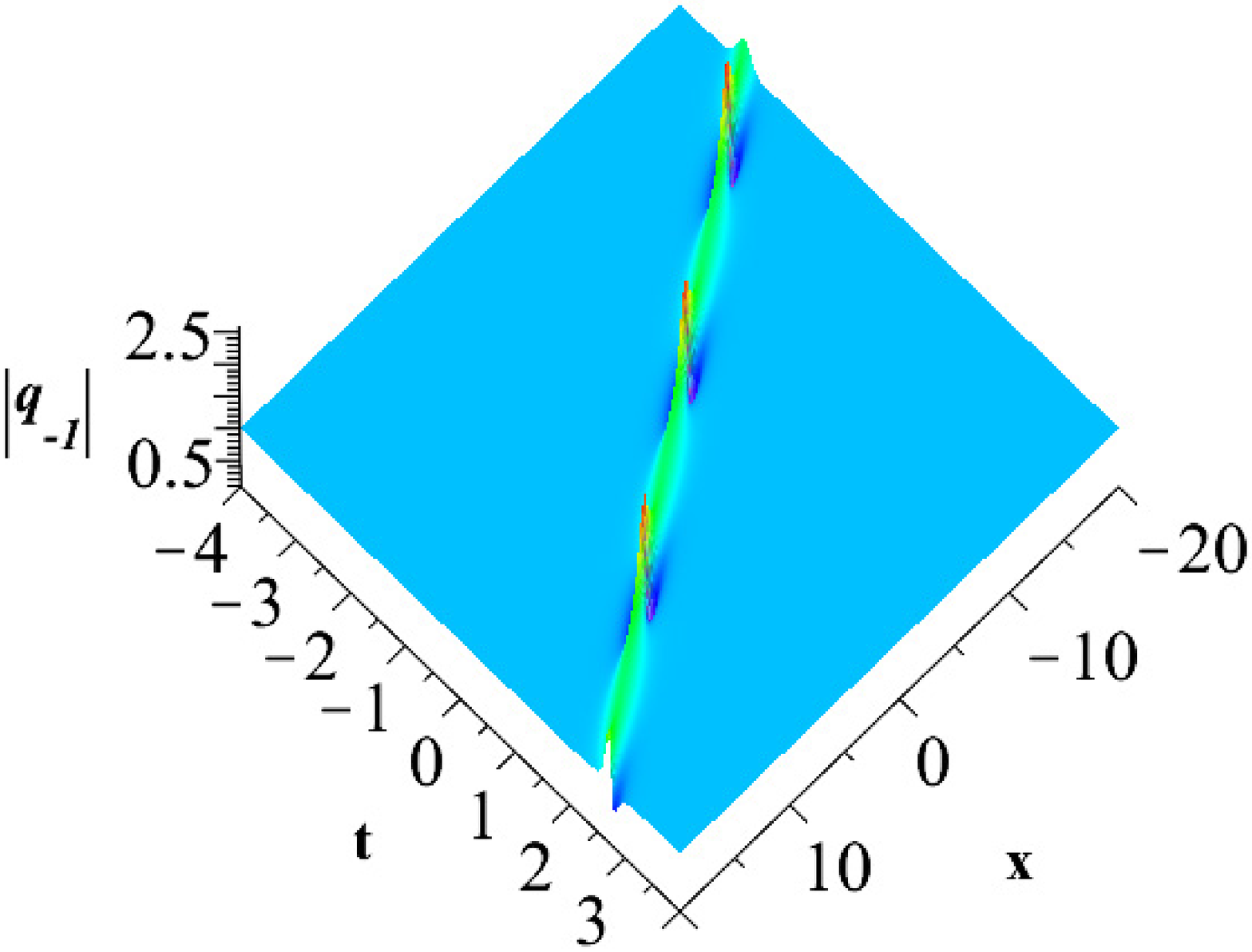}}}

\noindent { \small \textbf{Figure 10.} (Color online) Breather wave via solution \eqref{SSF-6} with parameters
$\alpha=1, \mathcal {Q}_{+}=\mathcal {I}_{2}, \zeta_{1}=2i, k_{0}=1,\gamma_{1}=1,\gamma_{0}=2,\gamma_{-1}=4$.
$(\textbf{a},\textbf{b},\textbf{c})$: $\beta=0.1$;
$(\textbf{d},\textbf{e},\textbf{f})$: $\beta=1$;\\}

\noindent
{\rotatebox{0}{\includegraphics[width=4.8cm,height=2.5cm,angle=0]{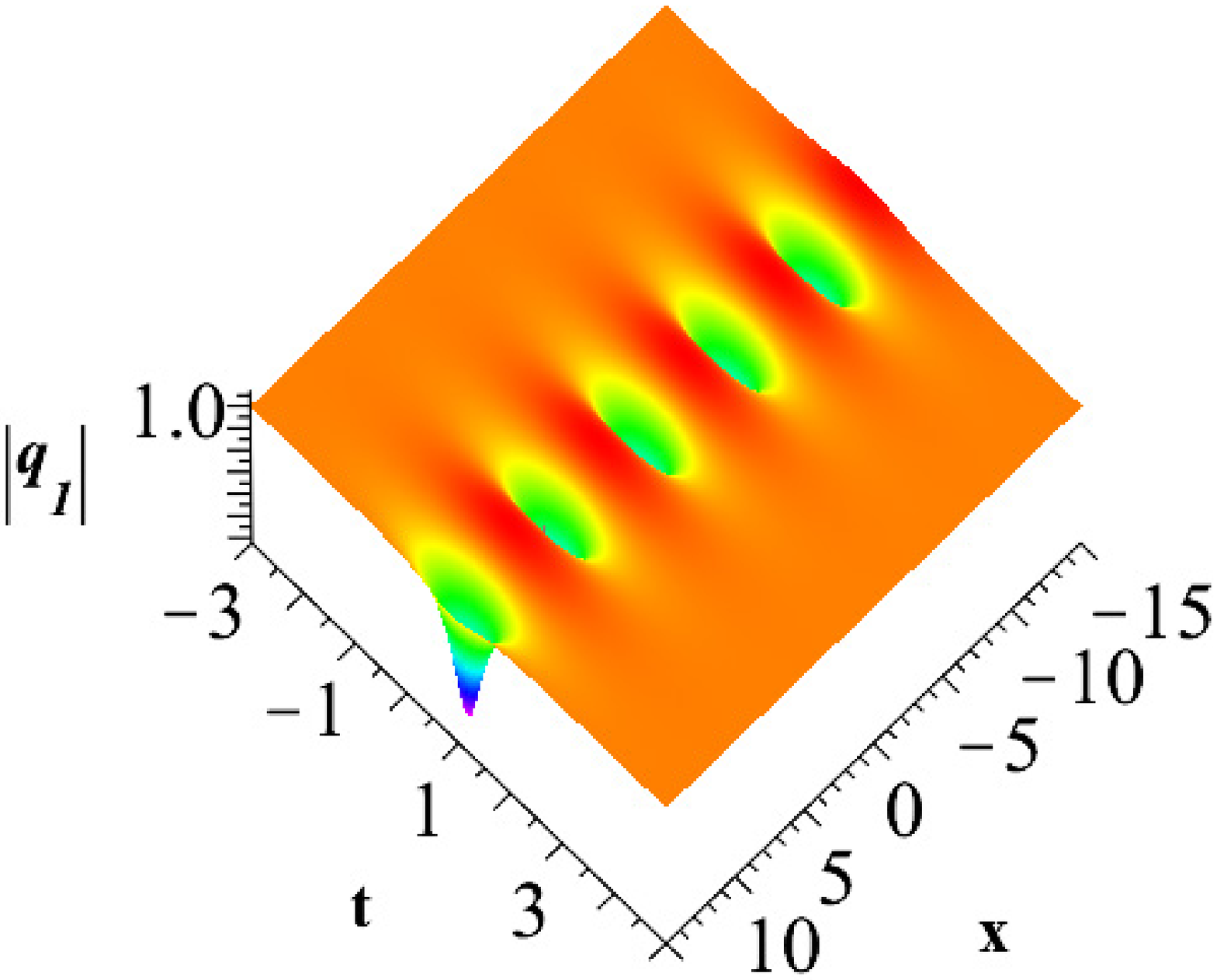}}}
{\rotatebox{0}{\includegraphics[width=4.8cm,height=2.5cm,angle=0]{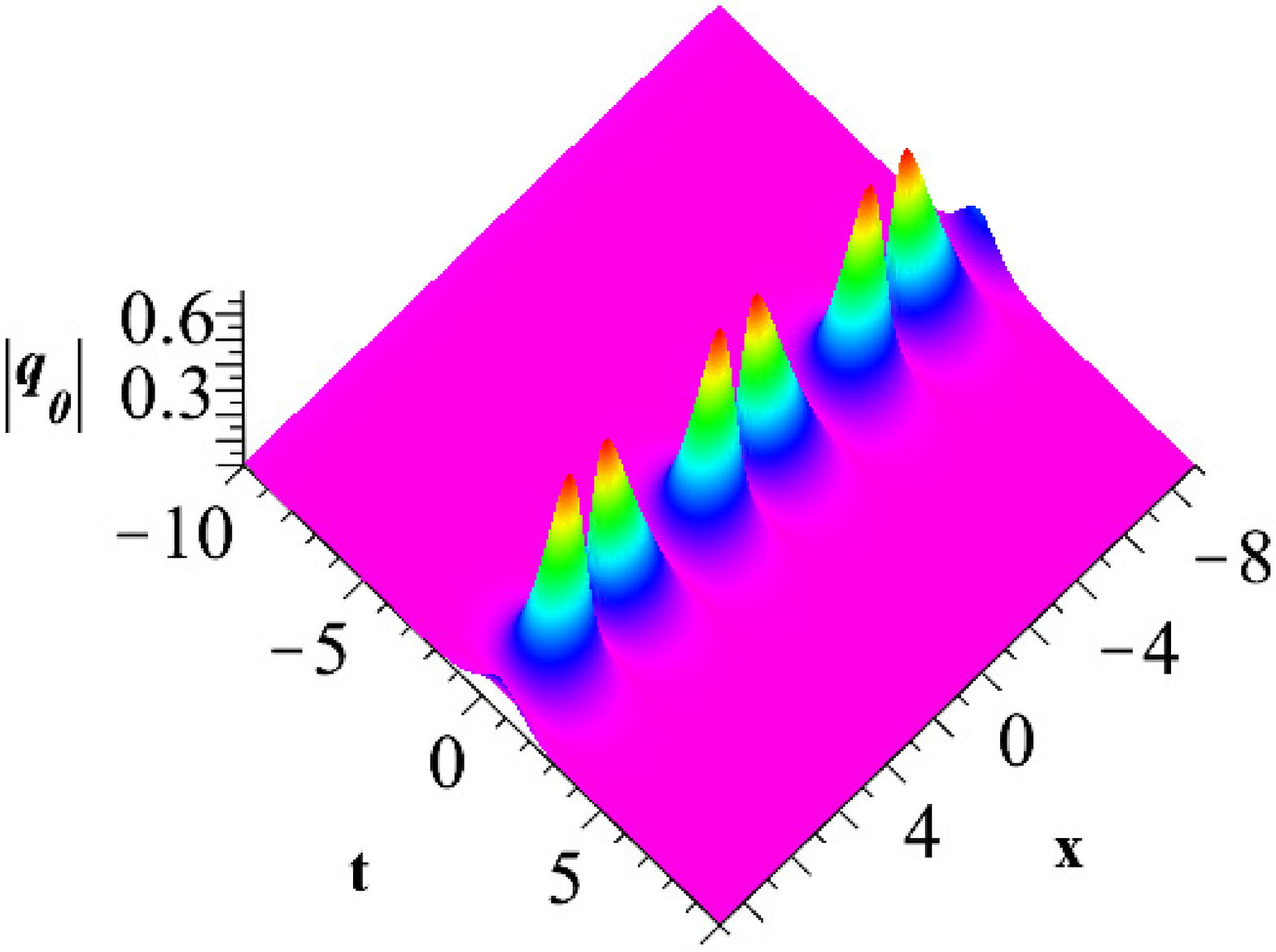}}}
{\rotatebox{0}{\includegraphics[width=4.8cm,height=2.5cm,angle=0]{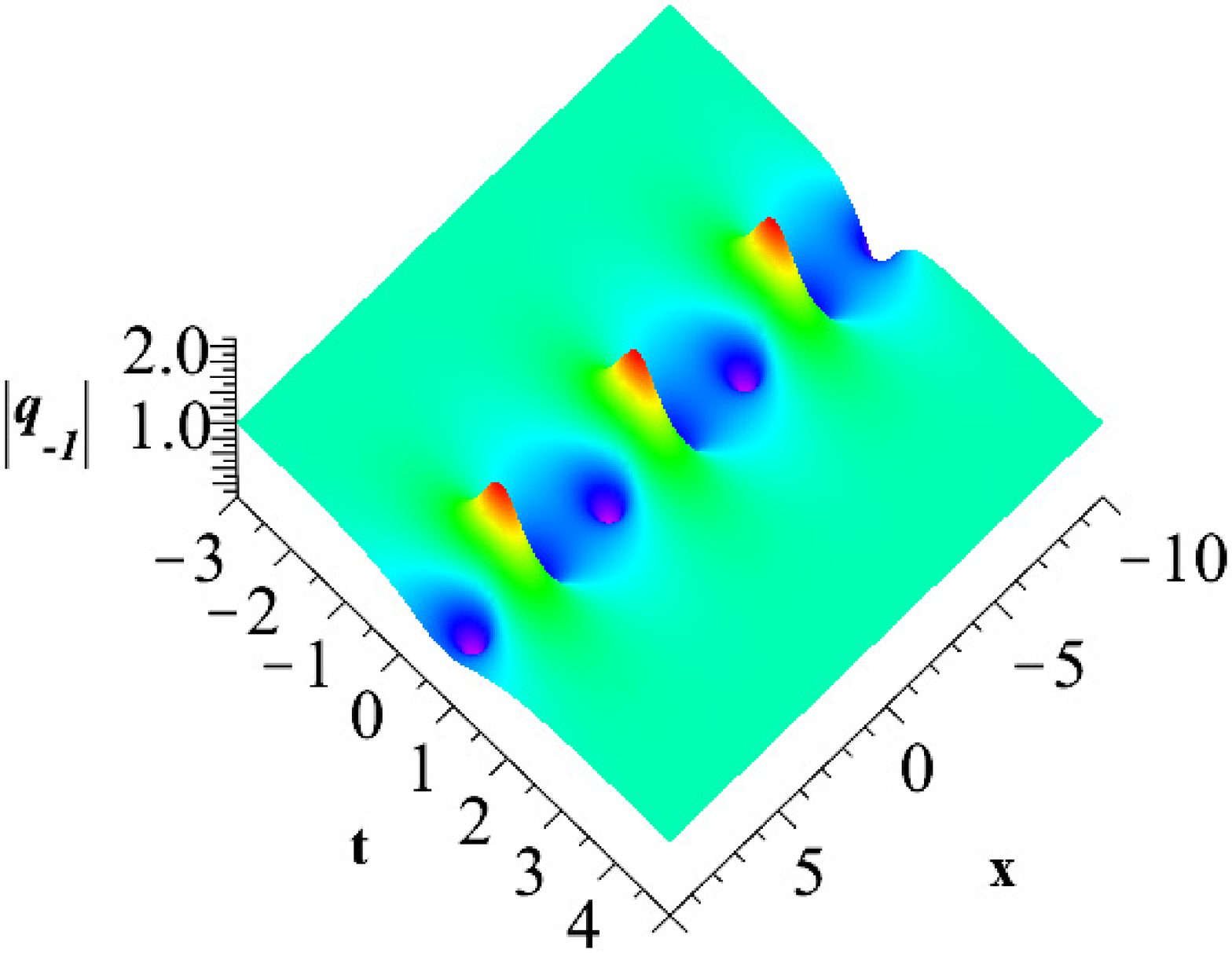}}}

$\qquad\qquad\textbf{(a)}\qquad\qquad\qquad\qquad\qquad\textbf{(b)}
\qquad\qquad\qquad\qquad\qquad\qquad\textbf{(c)}$\\

\noindent { \small \textbf{Figure 11.} (Color online) Breather wave via solution \eqref{SSF-6} with parameters
$\alpha=1, \beta=0.1, \mathcal {Q}_{+}=\mathcal {I}_{2}, \zeta_{1}=\frac{1}{2}+\frac{\sqrt{3}}{2}i, k_{0}=1,\gamma_{1}=i,\gamma_{0}=2,\gamma_{-1}=-4i$.\\}

%\noindent
%{\rotatebox{0}{\includegraphics[width=4.8cm,height=2.5cm,angle=0]{12-1.eps}}}
%{\rotatebox{0}{\includegraphics[width=4.8cm,height=2.5cm,angle=0]{12-2.eps}}}
%{\rotatebox{0}{\includegraphics[width=4.8cm,height=2.5cm,angle=0]{12-3.eps}}}
%
%$\qquad\qquad\textbf{(a)}\qquad\qquad\qquad\qquad\qquad\textbf{(b)}
%\qquad\qquad\qquad\qquad\qquad\qquad\textbf{(c)}$\\
%
%
%\noindent { \small \textbf{Figure 12.} (Color online) Periodic wave via solution \eqref{SSF-6} with parameters
%Plotted is Breather wave via solution \eqref{SSF-6} for the values of parameters in Fig.11 except $\alpha=0$, $\beta=1$.\\}
%

\section{Conclusions and discussions}
In this paper,
the IST with NZBCs at infinity is developed for the general coupled Hirota system \eqref{mNLSS1} with higher-order effects,
%we have developed the IST for the matrix Hirota system \eqref{mNLSS1} with NZBCs,
and we have provided
that the problem is significantly more complicated than the scalar case.
Moreover, the exact solutions of the the general coupled Hirota system are presented.
In particular,
we have found that these solutions provided in this paper possess a rich family of soliton solutions.
We expect the content of this work to be useful in characterizing recent experiments in BEC \cite{HC-2004,YDJJ-2004}
and nonlinear optics \cite{MJBP-2004,MJBP-2004,TSVA-2010}.
More importantly, these new
rational solutions show the potential rich dynamics in breather wave solutions, and promote our understanding of breather wave phenomena.
In addition, the matrix Hirota system \eqref{mNLSS1} we investigated in this work is fairly more general
as it admits the free constants $\alpha$, $\beta$.
Consequently, the solutions of the integrable spin-1
Gross-Pitaevskii equations and the modified matrix Korteweg-de Vries equation can be respectively constructed by reducing the solutions of the
the matrix Hirota system \eqref{mNLSS1}.

%\section*{Acknowledgements}
%\hspace{0.3cm}%We express our sincere thanks to all the persons who have provided valuable suggestions to this paper.
%This work is supported by the National Natural Science Foundation of China under Grant
%No.11871180.

\section*{Appendix: Trace formula}
%By trace formula we mean an expression for the analytic entry/entries of the scattering matrices $a(z)$ and $\bar{a}(z)$
%in terms of scattering data (in the scalar case: discrete eigenvalues and reflection coefficient, see for instance \cite{}).
As reported in \cite{VEZAB-1972,GB-1983,LD-1987}, the
trace formula also offers a relationship between the scattering data and the asymptotic phase difference of the potential under NZBCs.
In what follows we will
construct the trace formula for $\det a(z)$, from which $\det \bar{a}(z)$ can be obtained by symmetry, in the $\sigma=1$ case.
This will also offer a weak
version of the $\theta$-condition, establishing a relationship between the asymptotic phases of $\det Q_{+}$ and $\det Q_{+}$ and the spectral data.
Following a similar way, the $\sigma=1$ case can also be obtained.

\noindent
\textbf{Simple zeros}:
Let us investigate the case where all discrete eigenvalues are simple zeros of $\det a(z)$.
In particular, the corresponding norming constants are of rank one.
$z=z_{n}$ and $z=-k_{0}^2/z_{n}^{\ast}$ are the simple zeros of $\det a(z)$ (are analytic in $D^{+}$),
and $z=z_{n}^{\ast}$ and $z=-k_{0}^2/z_{n}$ are the simple zeros of $\det \bar{a}(z)$ (are analytic in $D^{-}$).
Consequently, we give
\begin{align}\label{TRCEf-1}
&\tilde{\alpha}^{+}(z)=\det a(z)\prod_{n=1}^{\mathcal {N}}\frac{\left(z-z^{\ast}_{n}\right)\left(z+k_{0}^2/z_{n}\right)}{\left(z-z_{n}\right)\left(z+k_{0}^2/z_{n}^{\ast}\right)},~~z\in D^{+},\notag\\
&\tilde{\alpha}^{-}(z)=\det \bar{a}(z)\prod_{n=1}^{\mathcal {N}}\frac{(z-z_{n})\left(z+k_{0}^2/z_{n}^{\ast}\right)}{\left(z-z_{n}^{\ast}\right)\left(z+k_{0}^2/z_{n}\right)},~~~z\in D^{-},
\end{align}
where $\tilde{\alpha}^{\pm}(z)$ are analytic in $D^{\pm}$, respectively, and they do not have zeros.
Furthermore, it follows from \eqref{ASAS-8} that $\alpha^{\pm}(z)\rightarrow1$ as $z\rightarrow\infty$
in the suitable region, and from \eqref{FSYMM-13} and \eqref{FSYMM-14} for $\sigma=-1$ it follows that
\begin{equation}\label{TRCEf-2}
\tilde{\alpha}^{+}(z)\tilde{\alpha}^{-}(z)=\det\left(\mathcal {I}_{m}+\rho^{\dag}(z^{\ast})\rho(z)\right)^{-1},~~z\in\Sigma.
\end{equation}
%The latter is nothing but a RHP for $\tilde{\alpha}^{\pm}(z)$ across the contour $\Sigma$,
Using the Cauchy projectors reported in Section 3.1 the formal solution of the RH problem can be written as
\begin{equation}\label{TRCEf-3}
\log\tilde{\alpha}^{\pm}(z)=\mp\frac{1}{2\pi i}\int_{\Sigma}\log\det\left(\mathcal {I}_{m}+\rho^{\dag}(\zeta^{\ast})\rho(\zeta)\right)\frac{d\zeta}{\zeta-z},~~z\in D^{\pm},
\end{equation}
So, a weak form of the trace formula is also obtained
{\small\begin{equation*}\label{TRCEf-4}
\det a(z)=\exp\left\{-\frac{1}{2\pi i}\int_{\Sigma}\log\det\left(\mathcal {I}_{2}+\rho^{\dag}(\zeta^{\ast})\rho(\zeta)\right)\frac{d\zeta}{\zeta-z}\right\}\prod_{n=1}^{\mathcal {N}}\frac{(z-z_{n})\left(z+k_{0}^2/z_{n}^{\ast}\right)}{\left(z-z_{n}^{\ast}\right)\left(z+k_{0}^2/z_{n}\right)}.
\end{equation*}}
In view of \eqref{ASAS-9} one can compute the behavior of $\det a(z)$ as $z\rightarrow 0$
\begin{equation*}\label{TRCEf-5}
\det a(z)\sim\frac{1}{k_{0}^{2m}}\det \mathcal {Q}_{+}\det\mathcal {Q}^{\dag}_{-},~~z\rightarrow 0.
\end{equation*}
It follows from above expressions that
\begin{equation*}\label{TRCEf-6}
\det \mathcal {Q}_{+}\det \mathcal {Q}^{\dag}_{-}=k_{0}^{2m}\exp\left\{-\frac{1}{2\pi i}\int_{\Sigma}\log\det\left(\mathcal {I}_{m}+\rho^{\dag}(\zeta^{\ast})\rho(\zeta)\right)\frac{d\zeta}{\zeta}\right\}\prod_{n=1}^{\mathcal {N}}e^{4i\delta_{n}},
\end{equation*}
where $\delta_{n}$ represents the phase of the discrete eigenvalue $z_{n}$ ($z_{n}=|z_{n}|e^{i\delta_{n}}$).

From the constraint \eqref{LP-4} on the boundary conditions, we find
\begin{equation*}\label{TRCEf-7}
\det \mathcal {Q}_{+}=k_{0}^{m}e^{i\theta_{+}},~~\det \mathcal {Q}_{-}=k_{0}^{m}e^{i\theta_{-}},
\end{equation*}
then a weak form of the $\theta$-condition can be derived
\begin{equation*}\label{TRCEf-8}
\theta_{+}-\theta_{-}=\frac{1}{2\pi}\int_{\Sigma}\log \det\left(\mathcal {I}_{m}+\rho^{\dag}(\zeta^{\ast})\rho(\zeta)\right)\frac{d\zeta}{\zeta}
+4\sum_{n=1}^{\mathcal {N}} \delta_{n}.
\end{equation*}

\noindent
\textbf{Double zeros}:  In this part, we investigate the case in which all discrete eigenvalues are double zeros.
Similar to the simple zeros, we first introduce
\begin{align*}\label{DZE-1}
&\tilde{\alpha}^{+}(z)=\det a(z)\prod_{n=1}^{\mathcal {N}}\frac{\left(z-z_{n}^{\ast}\right)^2\left(z+k_{0}^2/z_{n}\right)^2}{\left(z-z_{n}\right)^2\left(z+k_{0}^2/z_{n}^{\ast}\right)^2},~~z\in D^{+},\notag\\
&\tilde{\alpha}^{-}(z)=\det a(z)\prod_{n=1}^{\mathcal {N}}\frac{\left(z-z_{n}\right)^2\left(z+k_{0}^2/z_{n}^{\ast}\right)^2}{\left(z-z_{n}^{\ast}\right)^2\left(z+k_{0}^2/z_{n}\right)^2},~~z\in D^{-},
\end{align*}
where $\tilde{\alpha}^{\pm}$ do not have zeros. Proceeding as before we obtain
{\small\begin{equation*}\label{DZE-2}
\det a(z)=
\exp\left\{-\frac{1}{2\pi i}\int_{\Sigma}\log \det\left(\mathcal {I}_{m}+\rho^{\dag}(\zeta^{\ast})\rho(\zeta)\right)\frac{d\zeta}{\zeta-z}\right\}
\prod_{n=1}^{\mathcal {N}}\frac{(z-z_{n})^2\left(z+k_{0}^2/z_{n}^{\ast}\right)^2}{(z-z_{n}^{\ast})^2\left(z+k_{0}^2/z_{n}\right)^2},
\end{equation*}}
and
\begin{equation*}\label{DZE-3}
\theta_{+}-\theta_{-}=\frac{1}{2\pi}\int_{\Sigma_{o}}\log \det\left(\mathcal {I}_{m}+\rho^{\dag}(\zeta^{\ast})\rho(\zeta)\right)\frac{d\zeta}{\zeta}
-8\sum_{n=1}^{\mathcal {N}}\delta_{n}.
\end{equation*}
%Obviously, one can easily combine the two cases to account for both simple and double zeros.
Let $\{z_{n}\}_{n=1}^{\mathcal {N}_{1}}$ be all the simple zeros of $\det a(z)$,
and Let $\{\check{z}_{n}\}_{n=1}^{\mathcal {N}_{2}}$ be all the simple zeros of $\det a(z)$.
Then we can write the trace formula and the $\theta$-condition as
\begin{align*}\label{DZE-4}
&\det a(z)=\exp\left\{-\frac{1}{2\pi i}\int_{\Sigma}\log\det\left(\mathcal {I}_{m}+\rho^{\dag}(\zeta^{\ast})\rho(\zeta)\right)\frac{d\zeta}{\zeta-z}\right\}\notag\\
&\times\prod_{n=1}^{\mathcal {N}_{1}}\frac{(z-z_{n})^2\left(z+k_{0}^2/z_{n}^{\ast}\right)^2}{(z-z_{n}^{\ast})^2\left(z+k_{0}^2/z_{n}\right)^2}
\prod_{n=1}^{\mathcal {N}_{2}}\frac{(z-\check{z}_{n})^2\left(z+k_{0}^2/\check{z}_{n}^{\ast}\right)^2}{(z-\check{z}_{n}^{\ast})^2\left(z+k_{0}^2/\check{z}_{n}\right)^2},\notag\\
&\theta_{+}-\theta_{-}=\frac{1}{2\pi}\int_{\Sigma}\log\det\left(\mathcal {I}_{m}+\rho^{\dag}(\zeta^{\ast})\rho(\zeta)\right)\frac{d\zeta}{\zeta}
-4\sum_{n=1}^{\mathcal {N}_{1}}\delta_{n}-8\sum_{n=1}^{\mathcal {N}_{2}}\check{\delta}_{n},
\end{align*}
where
$z_{n}=|z_{n}|e^{i\delta_{n}}$ and $\check{z}_{n}=|\check{z}_{n}|e^{i\check{\delta}_{n}}$.

\section*{Acknowledgments}
This work is supported by the National Natural Science Foundation of China under Grant Nos. 12201622 and 11975306.
\\

\noindent
\textbf{Compliance with ethical standards}\\

\noindent
\textbf{Conflict of interest} The authors declare that they have no conflict
of interest.

%\end{CJK*}
%\end{document}

% Converted from Microsoft Word to LaTeX
% by Chikrii SoftLab Word2TeX converter (version 2.4)
% Copyright (C) 1999-2001 Kirill A. Chikrii, Anna V. Chikrii
% Copyright (C) 1999-2001 Chikrii SoftLab.
% All rights reserved.
% http://www.word2tex.com/
% mailto: info@word2tex.com, support@word2tex.com

% Warning: You are using UNREGISTERED Chikrii SoftLab Word2TeX!
%          In UNREGISTERED mode some restrictions will apply.
%          For more information please visit http://www.word2tex.com/
% YOU CAN USE THIS FILE WITH THE SOLE PURPOSE OF EVALUATING Word2TeX.

%\documentclass [12pt]{article}

%\begin{document}


\begin{thebibliography}{99}
\bibitem{GCS-1967}
Gardner, C.S., Greene, J.M., Kruskal, M.D., Miura, R.M.:  Method for solving the Korteweg-de Vries equation.
Phys. Rev. Lett. \textbf{19}, 1095-1097 (1967)
\bibitem{Lax-1968}
Lax, P.D.:  Integrals of nonlinear equations of evolution and solitary waves. Commun. Pure Appl. Math. \textbf{21}, 467-490 (1968)
\bibitem{AS-1972}
Shabat, A., Zakharov, V.:  Exact theory of two-dimensional self-focusing and one-dimensional self-modulation
of waves in nonlinear media. Sov Phys JETP. \textbf{34}, 62 (1972)

\bibitem{MJA-1973}
Ablowitz, M.J., Kaup, D.J., Newell, A.C., Segur, H.: Nonlinear-evolution equations of physical significance.
Phys. Rev. Lett. \textbf{31}, 125-127 (1973)
\bibitem{MJA-1974}
Ablowitz, M.J., Kaup, D.J., Newell, A.C., Segur, H.: The inverse scattering transform-Fourier analysis for
nonlinear problems. Stud. Appl. Math. \textbf{53}, 249-315 (1974)



\bibitem{MJADJ-1973}
Ablowitz, M.J., Kaup, D.J., Newell, A.C., Segur, H.: Method for solving the sine-Gordon equation.
Phys. Rev. Lett. \textbf{30}, 1262-1264 (1973)
\bibitem{MW-1973}
Wadati, M.: The modified Korteweg-de Vries equation. J. Phys. Soc. Jpn \textbf{34}, 1289-1296 (1973)
\bibitem{MW-1982}
Wadati, M., Ohkuma, K.: Multiple-pole solutions of the modified Korteweg-de Vries equation. J. Phys. Soc. Jpn \textbf{51}, 2029-2035 (1982)
\bibitem{MJADB-1983}
Ablowitz, M.J., Yaacov, D.B., Fokas, A.:  On the inverse scattering transform for the Kadomtsev-Petviashvili
equation. Stud. Appl. Math. \textbf{69}, 135-143 (1983)
\bibitem{ACVS-2006}
Constantin, A., Gerdjikov, V.S., Ivanov, R.I.: Inverse scattering transform for the Camassa-Holm equation.
Inverse Prob. \textbf{22}, 2197 (2006)
\bibitem{ACBU-1983}
Constantin, A., Ivanov, R.I., Lenells, J.: Inverse scattering transform for the Degasperis-Procesi equation.
Nonlinearity  \textbf{23}, 2559 (2010)
\bibitem{AFMB-1983}
Fokas, A.S., Ablowitz, M.J.: The inverse scattering transform for the Benjamin-Ono equation pivot to multidimensional problems.
Stud. Appl. Math. \textbf{68}, 1-10 (1983)


\bibitem{BFMJ-2006}
Prinari, B., Ablowitz, M.J., Biondini, G.:
Inverse scattering transform for the vector nonlinear Schr\"{o}dinger equation with nonvanishing boundary conditions.
J. Math. Phys. \textbf{47}, 063508 (2006)
\bibitem{BFMJ-20061}
Prinari, B., Ortiz, A.K., van der Mee, C., Marek Grabowski, M.: Inverse scattering transform and solitons for square matrix nonlinear Schr\"{o}dinger equations.
Stud. Appl. Math. \textbf{141}, 308-352 (2018)
\bibitem{BFMJ-20062}
Biondini, G., Lottes, J., Mantzavinos, D.:
Inverse scattering transform for the focusing nonlinear Schr\"{o}dinger equation with counterpropagating flows.
Stud. Appl. Math. \textbf{146}, 371-439 (2021)
\bibitem{VEZAB-1972}
Zakharov, V.E., Shabatm A.B.: Exact theory of two-dimensional self-focusing and one-dimensional self-modulation of waves in nonlinear media.
Sov Phys-JETP. \textbf{34}, 62-69 (1972)
\bibitem{AMJ-2007}
Ablowitz, M.J., Biondini, G., Prinari, B.:
Inverse scattering transform for the integrable discrete nonlinear Schr\"{o}dinger equation with nonvanishing boundary conditions.
Inverse Prob. \textbf{23}, 1711-1758 (2007)
\bibitem{NOMJ}
Ablowitz, M.J., Musslimani, Z.H.: Inverse scattering transform for the integrable nonlocal nonlinear Schr\"{o}dinger equation.
Nonlinearity \textbf{29}, 215 (2016)
\bibitem{FBFLS}
Feng, B.F., Luo, X.D., Ablowitz, M.J., Musslimani, Z.H.:  General soliton solution to a nonlocal nonlinear Schr\"{o}dinger equation with zero and nonzero boundary conditions. Nonlinearity \textbf{31}, 5385 (2018)
\bibitem{YJKLS}
Yang, J.: Nonlinear Waves in Integrable and Nonintegrable Systems. SIAM 2010.
\bibitem{BFFBG-2011}
Prinari, B., Biondini, G., Trubatch, A.D.: Inverse scattering transform for the multi-component nonlinear Schr\"{o}dinger equation with nonzero boundary
conditions. Stud. Appl. Math. \textbf{126}, 245-302 (2011)
\bibitem{BFFBG-2013}
Demontis, F., Prinari, B., van der Mee, C., Vitale, F.:  The inverse scattering transform for the defocusing nonlinear Schr\"{o}dinger equations with nonzero boundary conditions. Stud. Appl. Math. \textbf{131}, 1-40 (2013)
\bibitem{GB-19831}
Biondini, G.,  Kova\v{c}i\v{c}, G.: Inverse scattering transform for the focusing nonlinear Schr\"{o}dinger equation with nonzero boundary conditions.
J. Math. Phys. \textbf{55}, 031506 (2014)
\bibitem{BP-2015}
Prinari, B., Vitale, F.: Inverse scattering transform for the focusing nonlinear Schr\"{o}dinger equation with one-sided nonzero boundary condition.
Cont Math. \textbf{651}, 157-194 (2015)
\bibitem{XZJLW-2015}
Bilman, D., Trogdon, T.:  On numerical inverse scattering for the Korteweg-de Vries equation with discontinuous step-like data.
Nonlinearity \textbf{33}, 2211 (2020)
\bibitem{DF-2014}
Demontis, F., Prinari, B., van der Mee, C., Vitale, F.: The inverse scattering transform for the focusing nonlinear Schr\"{o}dinger equation with asymmetric
boundary conditions. J. Math. Phys. \textbf{55}, 101505 (2014)
\bibitem{GB-2016}
Biondini, G., Fagerstrom, E., Prinari, B.:  Inverse scattering transform for the defocusing nonlinear Schr\"{o}dinger equation with fully asymmetric non-zero
boundary conditions. Physica D \textbf{333}, 117-136 (2016)
\bibitem{GB-1983}
Pichler, M., Biondini, G.: On the focusing nonlinear Schr\"{o}dinger equation with non-zero boundary conditions and double poles.
IMA J. Appl. Math. \textbf{82}, 131-151 (2017)
\bibitem{IST-2017}
Prinari, B., Demontis, F., Li, S., Horikis, T.P.: Inverse scattering transform and soliton solutions for square matrix
nonlinear Schr\"{o}dinger equations with non-zero boundary conditions. Physica D \textbf{368}, 22-49 (2018)
\bibitem{IST-JNMP}
Ortiz, A.K., Prinari, B.: Inverse scattering transform and solitons for square matrix nonlinear Schr\"{o}dinger equations with mixed sign reductions and nonzero boundary conditions. J. Nonlinear Math. Phys. \textbf{27}, 130-161 (2020)
\bibitem{yzy-cnsns}
Zhang, G., Chen, S., Yan, Z.: Focusing and defocusing Hirota equations with non-zero
boundary conditions: Inverse scattering transforms and
soliton solutions. Commun. Nonlinear Sci. Numer. Simulat. \textbf{80}, 104927 (2020)
%\bibitem{yzy-2020}
%G. Zhang, Z. Yan, Focusing and defocusing mKdV equations with nonzero boundary conditions: inverse
%scattering transforms and soliton interactions. Physica D 410 (2020) 132521.
\bibitem{yzy-2020pd}
Zhang, G., Yan, Z.: Inverse scattering transforms and soliton solutions of focusing and defocusing
nonlocal mKdV equations with non-zero boundary conditions. Physica D \textbf{402}, 132170 (2020)
\bibitem{GXG}
Liu H, Shen J, Geng X.: Inverse scattering transformation for the N-component
focusing nonlinear Schr\"{o}dinger equation with nonzero
boundary conditions. Lett. Math. Phys. \textbf{113},  23 (2023)
\bibitem{wxbJMMA}
Wang, X.B., Han, B.: Inverse scattering transform of an extended nonlinear Schr\"{o}dinger
equation with nonzero boundary conditions and its multisoliton solutions. J. Math. Anal. Appl. \textbf{487}, 123968 (2020)

\bibitem{Hirota-PRE}
Park, Q.H., Shin, H.J.: Painlev\'{e} analysis of the coupled nonlinear Schr\"{o}dinger equation for polarized optical waves in an isotropic medium.
Phys. Rev. E. \textbf{59}, 2373 (1999)
\bibitem{WSSWL-2020}
Xu, H.X., Yang, Z.Y., Zhao, L.C., Duan, L., Yang, W.L.:  Breathers and solitons on two different backgrounds in a generalized
coupled Hirota system with four wave mixing. Phys. Lett. A \textbf{382}, 1738-1744 (2018)
%Sun W S, Wang L 2020 Solitons, breathers and rogue waves of the coupled Hirota system with $4\times4$ Lax pair
%\emph{Commun. Nonlinear Sci. Numer. Simulat.} 82 105055

\bibitem{YZY-2017}
Yan, Z.:  An initial-boundary value problem for the integrable spin-1 Gross-
Pitaevskii equations with a $4\times4$ Lax pair on the half-line. Chaos \textbf{27}, 053117 (2017)
\bibitem{GXGG-2017}
Geng X, Wang K, Chen M.: Long-time asymptotics for the Spin-1 Gross-Pitaevskii equation. 
Commun. Math. Phys. \textbf{382}, 585-611 (2021)
\bibitem{YZY-20171}
Ieda, J-ichi., Miyakawa, T., Wadati, M.: Exact analysis of soliton dynamics in Spinor Bose-Einstein condensates.
Phys. Rev. Lett. \textbf{93}, 194102 (2004)
\bibitem{MU-2017}
Uchiyama, M., Ieda, J-ichi., Wadati, M.:  Dark solitons in F=1 Spinor Bose-Einstein condensate. J. Phys. Soc. Jpn. \textbf{75}, 064002 (2006)
\bibitem{VS-2009}
Gerdjikov, V.S., Kostov, N.A., Valchev, T.I.: Solutions of multi-component NLS models and Spinor Bose-Einstein condensates.
Physica D \textbf{238}, 1306-1310 (2009)
\bibitem{wxb-2018}
Wang, X.B., Han, B.: Riemann-Hilbert problem and multi-soliton solutions of the integrable Spin-1
Gross-Pitaevskii equations. Z. Naturforsch A \textbf{74}, 139-145 (2018)


\bibitem{BFDA-2012}
Baronio, F., Degasperis, A., Conforti, M., Wabnitz, S.:
Solutions of the vector nonlinear Schr\"{o}dinger equations: Evidence for deterministic rogue waves.
Phys. Rev. Lett. \textbf{109}, 044102 (2012)
\bibitem{CSLY-2013}
Chen, S., Song, L.Y.:  Rogue waves in coupled Hirota systems.
Phys. Rev. E \textbf{87}, 032910 (2013)
\bibitem{DA-2013}
Degasperis, A., Lombardo, S.: Rational solitons of wave resonant-interaction models. Phys. Rev. E 88, 052914 (2013)
\bibitem{DA-2007}
Degasperis, A., Lombardo, S.:
Multicomponent integrable wave equations I. Darboux-dressing transformation.
J. Phys. A: Math. Theor. \textbf{40}, 961-977 (2007)
\bibitem{GAA-CMP}
Biondini, G., Kraus, G., Prinari, B.:
The three-component defocusing nonlinear
Schr\"{o}dinger equation with nonzero boundary conditions. Commun. Math. Phys.
\textbf{348}, 475-533 (2016)
\bibitem{BF-2013}
Baronio, F., Conforti, M., Degasperis, A., Lombardo, S.:
Rogue waves emerging from the resonant interaction of three waves. Phys. Rev. Lett. \textbf{111}, 114101 (2013)




\bibitem{AB111-2013}
Akhmediev, N., Ankiewicz, A., Soto-Crespo, J.M.: Rogue waves and rational solutions of the nonlinear Schr\"{o}dinger equation.
Phys. Rev. E \textbf{80}, 026601 (2017)
\bibitem{AB111-1986}
Akhmediev, N., Korneev, V.I.: Modulation instability and periodic solutions of the
nonlinear Schr\"{o}dinger equation. Theor. Math. Phys. \textbf{69}, 1089 (1996)
\bibitem{AB111-2014}
Ankiewicz, A., Akhmediev, N.: Rogue wave solutions for the infinite integrable nonlinear Schr\"{o}dinger equation hierarchy.
Phys. Rev. E \textbf{96}, 012219 (2017)
\bibitem{YJKLS-2014}
Pelinovsky, D., Yang, J.: Instabilities of multihump vector solitons in coupled nonlinear Schr\"{o}dinger equations.
Stud. Appl. Math. \textbf{115}, 109-137 (2005)
%\bibitem{YZYzgq}
%G. Zhang, Z. Yan , The n-component nonlinear Schr\"{o}dinger equations: dark-bright mixed n-and higher-order solitons and breathers, and dynamics,
%Proc. R. Soc. A 474 (2018) 20170688.
\bibitem{YZYzgq-2}
Zhang, G., Yan, Z.: Inverse scattering transforms and soliton solutions of focusing and defocusing nonlocal mKdV equations
with non-zero boundary conditions. Physica D \textbf{402}, 132170 (2020)
\bibitem{mwxx-1}
Ma, W.X.: The inverse scattering transform and soliton solutions of a combined modified Korteweg-de Vries equation.
J. Math. Anal. Appl. \textbf{471}, 796-811 (2019)
\bibitem{mwxx-20}
Ma, W.X.:  Inverse scattering and soliton solutions of nonlocal reverse-spacetime nonlinear Schr\"{o}dinger equations.
Proc. Amer. Math. Soc. \textbf{149}, 251-263 (2021)
\bibitem{mwxx-2}
Ma, W.X., Huang, Y., Wang, F.: Inverse scattering transforms and soliton
solutions of nonlocal reverse-space nonlinear
Schr\"{o}dinger hierarchies.
Stud. Appl. Math. \textbf{145}, 563-585 (2020)
\bibitem{JNS}
Zhang, G.Q., Yan, Z.:
The derivative nonlinear Schr\"{o}dinger equation with zero/nonzero boundary conditions: Inverse scattering transforms and n-double-pole solutions.
J. Nonlinear Sci. \textbf{30}, 3089-3127 (2020)
%\bibitem{wxxxb-2}
%Wang XB, Tian SF, Zhang TT. Characteristics of the breather and rogue waves in a (2+1)-dimensional nonlinear Schr\"{o}dinger equation.
%\emph{Proc Am Math Soc.} 2018; 146: 3353-3365.
\bibitem{wxxxb-5}
Wang, X.B., Tian, S.F., Feng, L.L., Zhang, T.T.: On quasi-periodic waves and rogue waves to the (4+1)-dimensional nonlinear Fokas
equation. J. Math. Phys. \textbf{59}, 073505 (2020)
%\bibitem{wxxxb-JPSJ}
%Wang XB, Han B.  Vector nonlinear waves in a two-component Bose-Einstein condensate system.
%\emph{J Phys Soc Jpn.} 2020; 89: 124003.
%\bibitem{wxbb-3}
%X.B. Wang, B. Han, The three-component coupled nonlinear Schr\"{o}dinger equation: Rogue waves on a multi-soliton background and dynamics,
%Europhys. Lett. 126 (2018) 15001.
\bibitem{wxbb-4}
Wang, X.B., Han, B.: A Riemann-Hilbert approach to a generalized nonlinear Schr\"{o}dinger equation
on the quarter plane. Math. Phys. Anal. Geom. 23, 25 (2020)
\bibitem{tsff-3}
Tian, S.F.: Initial-boundary value problems of the coupled modified Korteweg-de Vries equation on the half-line via the Fokas method.
J. Phys. A: Math. Theor. \textbf{50}, 395204 (2017)
\bibitem{tsff-ams}
Tian, S.F and Zhang, T.T.: Long-time asymptotic behavior for the Gerdjikov-Ivanov
type of derivative nonlinear Schr\"{o}dinger equation with time-periodic boundary condition.
Proc. Amer. Math. Soc. \textbf{146}, 1713-1729 (2018)
\bibitem{tsff-SAMP}
Tian, S.F., Zhang, H.Q.: On the integrability of a generalized variable-coefficient forced Korteweg-de Vries equation in fluids.
Stud. Appl. Math. \textbf{132}, 212-246 (2014)
\bibitem{wdss-4}
Wang, D.S., Guo, B., Xu, X.: Long-time asymptotics of the focusing Kundu-Eckhaus equation with nonzero boundary conditions.
J. Differential Equations \textbf{266}, 5209-5253 (2019)
%\bibitem{jdexx-4}
%Long-time asymptotics for the nonlocal nonlinear
%Schr\"{o}dinger equation with step-like initial data, J. Differential Equations 270 (2021) 694-724.
%\bibitem{jdexx-4}
%Liu H, Geng X and Xue B 2018 The Deift-Zhou steepest descent method to long-time
%asymptotics for the Sasa-Satsuma equation J. Differential Equations 265 (2018) 5984-6008.
\bibitem{xjfeg-1}
Xu, J., Fan, E.G.: Long-time asymptotics for the Fokas-Lenells equation with decaying initial
value problem: Without solitons. J. Differential Equations. \textbf{259}, 1098-1148 (2015)
\bibitem{xjfeg-2}
Xu, J., Fan, E.G and Chen, Y.: Long-time asymptotic for the derivative nonlinear
Schr\"{o}dinger equation with step-like initial value. Math. Phys. Anal. Geom. \textbf{16}, 253-288 (2015)




\bibitem{MJBP-2004}
Ablowitz, M.J., Prinari, B., Trubatch, A.D.: Discrete and Continuous Nonlinear Schr\"{o}dinger Systems.
London Mathematical Society Lecture Note Series, 302, Cambridge University Press, 2004.




\bibitem{HC-2004}
Hamner, C., Chang, J.J., Engels, P and Hoefer, M.A.: Generation of dark-bright soliton trains in
superfluid-superfluid counterflow. Phys. Rev. Lett. \textbf{106}, 065302 (2011)
\bibitem{YDJJ-2004}
Yan, D., Chang, J.J., Hamner, C., Hoefer, M., Kevrekidis, P.G., Engels, P., Achilleos, V., Frantzeskakis, D.J
and Cuevas, J.: Beating dark-dark Solitons in Bose-Einstein condensates. J Phys B. \textbf{45}, 115301 (2012)
\bibitem{MJBP-2004}
Conti, C., Fratalocchi, A., Peccianti, M., Ruocco, G and Trillo, S.: Observation of a gradient catastrophe generating solitons.
Phys. Rev. Lett. \textbf{102}, 083902 (2009)
\bibitem{MJBP-2004}
Fratalocchi, A., Conti, C., Ruocco, G and Trillo, S.: Free-energy transition in a gas of noninteracting
nonlinear wave particles. Phys. Rev. Lett. \textbf{101}, 044101 (2008)
\bibitem{TSVA-2010}
Trillo, S and Valiani, A.: Hydrodynamic instability of multiple four-wave mixing.
Opt Lett. \textbf{35}, 3967 (2010)

\bibitem{LD-1987}
Faddeev, L.D., Takhtajan, L.A.: Hamiltonian Methods in the Theory of Solitons. Springer, Berlin, 1987.




\end{thebibliography}
\end{document}